\documentclass[11pt]{article}
\usepackage{changepage}
\usepackage{graphicx} 
\usepackage{mathtools}
\usepackage{amssymb}
\usepackage{fullpage}
\usepackage{amsthm}
\usepackage{thmtools} 
\usepackage{xcolor}
\usepackage{tikz}
\usepackage{float}
\usetikzlibrary{trees}
\usetikzlibrary{circuits.logic.US, shapes.geometric, positioning}
\usetikzlibrary{arrows.meta}
\usepackage{amsmath}
\usepackage{booktabs}
\usepackage{algorithm}
\usepackage{algpseudocode}
\usepackage[referable]{threeparttablex}
\usepackage{base}
\usepackage{macros}
\usepackage{minimacros}
\usepackage{booktabs,tabularx}

\usepackage{esvect}

\usepackage{array}
\usepackage{longtable}
\usepackage{adjustbox}

\theoremstyle{plain}

\newcommand{\Input}{\Require}
\newcommand{\Output}{\Ensure}

\makeatletter
\newcommand{\vast}{\bBigg@{4}}
\newcommand{\Vast}{\bBigg@{5}}
\makeatother

\newcounter{proofcase}

\newcommand{\ord}{\text{ord}}

\newcommand{\hr}{\mathcal{H}_{\text{recon}}}
\newcommand{\hp}{\mathcal{H}_{\text{pit}}}
\newcommand{\hre}{\mathcal{H}^{\text{ext}}_{\text{recon}}}
\newcommand{\epsprod}{\epsilon_{\text{pit}, \text{prod}}}
\newcommand{\epsst}{\epsilon_{\text{pit}, \text{s2}}}
\newcommand{\epsrcn}{\epsilon_{\text{recon}}}

\newcommand{\roabp}{\mathsf{roABP}}
\newcommand{\ROABP}{\textsf{roABP}}
\newcommand{\aoroabp}{\mathsf{ao-roABP}}
\newcommand{\abp}{\mathsf{ABP}}
\newcommand{\mlrkf}{\mathsf{mlRkF}}

\newcommand{\e}{\boldsymbol{e}}
\newcommand{\x}{\boldsymbol{x}}
\newcommand{\y}{\boldsymbol{y}}
\newcommand{\z}{\boldsymbol{z}}
\newcommand{\bu}{\boldsymbol{u}}
\newcommand{\bv}{\boldsymbol{v}}
\newcommand{\bw}{\boldsymbol{w}}
\newcommand{\ba}{\boldsymbol{a}}
\newcommand{\bb}{\boldsymbol{b}}

\newcommand{\balp}{\boldsymbol{\alpha}}

\newcommand{\0}{\boldsymbol{0}}

\newcommand{\rof}{\mathsf{ROF}}
\newcommand{\prof}{\mathsf{PROF}}
\newcommand{\prop}{\mathsf{PROP}}

\newcommand{\rtf}{\mathsf{R2F}}

\newcommand{\rkf}{\mathsf{RkF}}

\newcommand{\fread}{\text{Read}}

\newcommand{\op}{\text{op}}

\newcommand{\tl}{\text{L}}
\newcommand{\tr}{\text{R}}
\newcommand{\rop}{\mathsf{ROP}}

\newcommand{\defsymbol}{\hfill $\lozenge$}

\newcommand{\gla}{\text{GL}^{\text{aff}}_n(\mathbb{F})}

\newcommand{\charac}{\operatorname{char}}

\title{PIT and Reconstruction for Sums of Powers}
\author{Nimrod Kaplan\thanks{This research was funded by the European Union (ERC, EACTP, 101142020). Views and opinions expressed are however those of the author(s) only and do not necessarily reflect those of the European Union or the European Research Council Executive Agency. Neither the European Union nor the granting authority can be held responsible for them.} \and Amir Shpilka\footnotemark[1] \and Yann Tal\footnotemark[1]}

\date{}

\begin{document}

\maketitle

\pagenumbering{arabic}
\setcounter{page}{1} 

\begin{abstract}
We study deterministic black-box polynomial identity testing (PIT) and reconstruction for
polynomials of the form
\[
    f(x)=\sum_{i=1}^r \alpha_i f_i(x)^d,
\]
where each $f_i$ belongs to some underlying polynomial class $\cC$.
This representation generalizes Waring decomposition, several natural small-depth circuit models, and polynomial decompositions arising in moment-based learning problems.

Our main results are three lifting theorems showing that, when the outer power is sufficiently large, PIT and reconstruction for the inner class can be lifted to sums of powers. When the power degree is small, we can perform these tasks directly in some of the cases using closure properties of the underlying models. In particular we obtain the following results:


\begin{itemize}
    \item When the inner polynomials have bounded-individual-degrees and are sparse, we obtain deterministic reconstruction for every $d$: the original power-sum decomposition is recovered at high degree, while at low degree the polynomial is reconstructed nonproperly by sparse interpolation.

    \item For sums of powers of same-order and any-order $\roabp$s, the same proper/nonproper dichotomy gives deterministic reconstruction for every powering degree.

    \item Explicit quasipolynomial-size hitting sets for every powering degree for several depth-$5$ and depth-$6$ circuit families, as well for the models mentioned above.

\end{itemize}


Our results for high powering degree rely on the polynomial ABC theorem, which to the best of our knowledge is the first application of this theorem in the context of PIT. 


\medskip
\noindent\emph{Remark}: Some of the results for sparse inner polynomials appeared in a preliminary manuscript posted online \cite{ST26}.  The present paper subsumes and substantially extends that manuscript.

\end{abstract}

\newpage

\tableofcontents

\newpage

\section{Introduction}

Power-sum decomposition asks to recover structured polynomials
$f_1,\ldots,f_r$ from black-box access to
\begin{equation}
\label{eq:intro:model}
    f(\vx)=\sum_{i=1}^{r}\alpha_i f_i(\vx)^d.
\end{equation}
This representation  
generalizes several well-studied reconstruction problems.
When the $f_i$'s are linear forms, \eqref{eq:intro:model} is a Waring
decomposition, or equivalently a symmetric tensor decomposition
\cite{ComonGKM08,BrachatCMT10,Kay12}.
Tensor decomposition has been studied extensively in theoretical computer
science and machine learning; see, e.g.,
\cite{goyal2014fourier,anandkumar2014tensor,montanari2014statistical,
hopkins2015tensor,ma2016polynomial,hopkins2016fast,BhargavaSV21,PSVTensor24,bhargavaDevansh25,kothari2025smooth}.
The case of quadratic inner polynomials (i.e., the $f_i$) is already substantially more
difficult. Sums of powers of quadratic forms arise naturally in
moment-based approaches to the identifiability and learning of mixtures of
Gaussians; see, e.g., \cite{GKS20,BHKX22} and the related Gaussian-mixture
literature
\cite{sanjeev2001learning,dasgupta2007probabilistic,anderson2014more,
regev2017learning,liu2021settling,di2023multidimensional}.

From the viewpoint of algebraic complexity, allowing the $f_i$'s to belong to richer classes gives natural polynomial
identity testing (PIT) and reconstruction problems for arithmetic circuits.
For example, sparse inner polynomials give depth-$4$ powering circuits,
while depth-$3$ or depth-$4$ inner circuits give corresponding depth-$5$
and depth-$6$ models. Taking the $f_i$'s to be formulas or algebraic
branching programs gives further exact-learning problems for succinct arithmetic computations.  
A strong motivation for studying small-depth classes comes from depth
reduction theorems. These results show how to simulate general arithmetic circuits by shallow circuits, making depth-$5$ powering circuits, depth-$3$ and depth-$4$ models central for lower bounds and PIT
\cite{AgrawalV08,Koiran12,Tavenas15,GuptaKKS16chasm}. 

 
The reconstruction problem is difficult even in some of these special
cases. Indeed, computing tensor rank is NP-hard over the rationals and
NP-complete over finite fields \cite{Hastad90}. Accordingly, efficient
reconstruction algorithms have typically been sought under structural
promises or in restricted parameter regimes.
Prior work obtains algorithms for \emph{generic} or \emph{non-degenerate} powers of low-degree forms \cite{GKS20,BHKX22}, for \emph{random} arithmetic formulas \cite{GKL11,GKQ14}, and for several worst-case restricted models, such as read-once formulas and multilinear tensors of bounded or small growing rank \cite{sv14_rec_rof,BhargavaSV21,PSVTensor24,bhargavaDevansh25}.  

\medskip
In this paper we study the case in which the outer power $d$ is large.
More generally, let $\mathcal C$ be a class of polynomials and write
\[
\sum^{r}\!\bigwedge^{d}\!\mathcal C
:=
\left\{
    \sum_{i=1}^{r}\alpha_i f_i^d : f_i\in\mathcal C
\right\}.
\]

We show that in many cases, an identity-testing or reconstruction algorithm for $\mathcal C$ can be used to obtain such an algorithm for $\Sigma^{r}\!\wedge^{d}\!\mathcal C$, when the outer power is sufficiently large.  The main algebraic tool is the ABC theorem for function fields of Stothers and Mason and its extension by Vaserstein and Wheland \cite{stothers1981polynomial,mason,VW2023}.  These results imply that a sufficiently short sum of sufficiently high powers of pairwise non-associate univariate polynomials cannot vanish.  This implies linear independence of the powered polynomials and is the basis for our PIT algorithms. It also implies that such representations are unique (up to permutations), which is a property that is important for many reconstruction approaches.


The lifting theorems apply when the powering degree $d$ is large.  For small $d$, in several of our applications the resulting polynomial belongs to a  somewhat larger class for which PIT or reconstruction is already known.  Combining the two cases gives reconstruction for every $d$ for sparse bounded-individual-degree polynomials and for same-order or any-order $\roabp$s.  It also gives PIT for every $d$ for several depth-$5$ and depth-$6$ circuit classes.  We obtain additional results in the high-power regime for several other inner classes.


\subsection{Our results}
\label{sec:intro:results}

We may assume that the representation in \eqref{eq:intro:model} is minimal: i.e., that no $f_i$ is zero and no two are associates. An \emph{interpolating set} for $\cC$ is an indexed set $H$ such that
$f(H)=g(H)$ implies $f=g$ for every $f,g\in\cC$.
This is equivalent to $H$ being a hitting set for $\Sigma^2\cC$. In particular, every nonzero polynomial computable in $\cC$ has a nonzero evaluation vector, and non-associate polynomials have non-proportional evaluation vectors. A non-adaptive \emph{reconstruction set} is a fixed set of points such that a polynomial computable in $\cC$ can be recovered efficiently from its values on that set.

We summarize our three lifting theorems next. The formal versions include more precise technical assumptions and dependence. 

\begin{theorem}[Lifting theorems, informal]
\label{thm:intro:lifting}
Let
\[
    f=\sum_{i=1}^{r}\alpha_if_i^d
\]
be a minimal representation.
\begin{enumerate}[label=\textup{(\roman*)},leftmargin=2.4em]
    \item \emph{\textbf{PIT lifting (informal version of \Cref{thm:bb pit for high powers}).}}\label{item:intro-main-pit}
    Suppose that the $f_i$'s have \emph{total degree} at most $\Delta$ and that
    the inner class has an interpolating set of size $s$.  If 
    \[
        d+1\ge (r-1)^{2},
    \]
    then $f$ has an explicit hitting set of size at most
    \[
        (s-1)\Delta d+1.
    \]

    \item \emph{\textbf{Total-degree reconstruction (informal version of \Cref{thm:gen recon algo}).}} \label{item:intro-main-total}
    Suppose that $\deg(f_i)\le\Delta$ and the inner class has an efficient
    deterministic non-adaptive reconstruction algorithm.  If
    \[
        d>(r+1)^4\Delta,
    \]
    then one can recover the representation, up to permutation and scalar
    rescaling of the summands, in time polynomial in the parameters of the
    inner reconstruction problem.

    \item \emph{\textbf{Individual-degree reconstruction (informal version of \Cref{thm:small_individual_degree_reconstruction}).}}
    Suppose that \(\operatorname{ideg}(f_i)\leq\delta\) for every \(i\in[r]\),
    that the underlying class $\cC$ admits an efficient deterministic non-adaptive
    reconstruction algorithm, and that there is an efficient hitting set for
    $\Sigma^2\Pi^2 \cC$.
If
    \[
        d>(r+1)^4r\delta,
    \]
    then one can recover the representation deterministically and
    non-adaptively in time
    \[
        n^{O(r)}\cdot
        \poly
        \bigl(r,d,\delta,\text{inner class parameters}\bigr).
    \]
\end{enumerate}
\end{theorem}

\medskip

Note that the total-degree theorem
has only polynomial dependence on $n$, but it requires powering degree growing with the total degree $\Delta$.  The individual-degree theorem replaces the dependence on the total degree $\Delta$ with the individual degree $\delta$, at the price of an $n^{O(r)}$ overhead.  



\subsubsection{Applications of the lifting theorem for reconstruction}

We next give several applications of \Cref{thm:intro:lifting}. The main application is for reconstructing sums of powers of sparse polynomials (see \Cref{sec:sparse}) and more generally, sums of powers of read once algebraic branching programs, $\roabp$s for short (see \Cref{sec:roabp}). We summarize both results in the following statement. 

\begin{theorem}[Main reconstruction results, informal]
\label{thm:intro:all-degree}
The following deterministic black-box algorithms exist.
\begin{enumerate}[label=\textup{(\alph*)},leftmargin=2.4em]
    \item \label{tm:intro:rec-sparse} \emph{\textbf{Informal version of \cref{thm:sum_pow_sparse_poly_recon}.}} Suppose that every $f_i$ is $s$-sparse and has individual degree
    at most $\delta$.  For $d>(r+1)^4r\delta$, the algorithm properly recovers
    the summands.  For $d\le (r+1)^4r\delta$, it reconstructs $f$ as a sparse
    polynomial with at most
    \[
        r\binom{s+d-1}{d} \leq(s+r^5\delta)^{O(r^5\delta)}
    \]
    monomials. In particular, regardless of $d$, for $r=\delta=O(1)$  this gives polynomial-time reconstruction, and for $r,\delta=\poly(\log n)$  the running time is quasipolynomial.

    \item \label{tm:intro:rec-roabp} \emph{\textbf{{Informal version of \cref{thm:main-rec-abp}}.}} 
    Suppose that every $f_i$ is a width-$w$, individual-degree-$\delta$ $\roabp$, and either all $f_i$'s have the same known variable order or every $f_i$ is an any-order $\roabp$.  For $d>(r+1)^4r\delta$, the algorithm properly recovers the summands. For $d\le (r+1)^4r\delta$, it reconstructs $f$ as one $\roabp$ of width at most
    \[
        r \binom{w+d-1}{d}\le r\min\{w^d,(d+1)^{w-1}\} \leq r\min \{w^{O(r^5\delta)},{(r\delta+1)}^{O(w)}\}
    \]
    and individual degree at most $d\delta$. In particular, if $r,\delta = \poly(\log n)$ then the running time is quasipolynomial \[
    \poly(d)\cdot \exp(\poly(\log n,\log w)).
    \]
    
    If, moreover, the inner polynomials are any-order $\roabp$s and $r,\delta,w=O(1)$, then the reconstruction algorithm runs in polynomial time in $n,d$.

    Furthermore, in the case $d>(r+1)^4\Delta$ the same algorithm also works when the inner $\roabp$s have arbitrary orders.
\end{enumerate}
\end{theorem}


\subsubsection{Applications of the lifting theorem for PIT}

We next present our main PIT results. They are obtained in a different manner when $d+1\ge (r-1)^2$ and in the complement case. Thus, for each of the classes in \Cref{tab:roabp-hitting-sets-overview} and \Cref{tab:intro-hitting-set-sizes-small-depth}, we obtain an explicit hitting set for every
powering degree $d$.  The two columns of the tables give the bounds obtained in the complementary high- and low-power regimes. 

For sums of powers of $\roabp$s we obtain the following PIT results. We state a slightly informal version of \Cref{thm:main-pit-abp}.

\begin{theorem}[{PIT for sums of powers of $\roabp$s, informal version of \Cref{thm:main-pit-abp}}]
\label{thm:intro:pit-abp}
There are explicit hitting sets for
    $\Sigma^r \wedge^d \roabp(w,\delta)$ with the following parameters as summarized in \Cref{tab:roabp-hitting-sets-overview}.
\begin{table*}[h]
\centering

\setlength{\arraycolsep}{12pt}
\[
\begin{array}{lll} 
    \toprule
    \multirow{2}{*}{\textbf{\(\roabp\) type}} &
    \multicolumn{2}{c}{
        \hspace{-3cm}\textbf{Hitting set size}
    }
    \\
    &
    \textbf{\(d+1 \geq (r-1)^2\)}
    &
    \textbf{\(d+1 < (r-1)^2\)}
    \\
    \midrule

    \text{arbitrary orders}
        &
        \displaystyle
        d\cdot
        \left(n\delta w\right)^{
            O\left(\log(n\delta w)\right)
        }
        &
        \displaystyle
        \left(n\delta w\right)^{
            2^{O(r)}\log(n\delta w)
        }
        \\[3.5ex]

    \text{same variable order}
        &
        d\cdot\poly\left(
            \delta,
            (nw)^{O(\log n)}
        \right)
        &
        O\left(
            \delta\cdot
            (nw)^{O(r^2\log n)}
        \right)
        \\[3.5ex]

    \text{any-order}
        &
        d\cdot
        \poly\left(
            n\delta,\,
            w^{\log\log w}
        \right)
        &
        \poly\left(
            n\delta,\,
            w^{O\left(r^2(\log r+\log\log w)\right)}
        \right)
        \\

    \bottomrule
\end{array}
\]
\caption{Hitting-set sizes for sums of powers of $\roabp$s (See \Cref{tab:roabp-hitting-set-sizes} for full details)}
\label{tab:roabp-hitting-sets-overview}
\end{table*}

In particular, for summands with arbitrary orders the hitting set is of \emph{quasipolynomial size}  whenever $r=O(\log\log n)$. If we further know that the summands have the same order (or indeed are in any order) then it suffices that $r=\poly(\log n)$.
\end{theorem}


Next, we consider bounded depth circuits. Many of the strongest known PIT results for small-depth circuits require a bounded or moderately growing top fan-in. Depending on the setting, we may allow polynomially large top fan-in.  



\begin{theorem}[PIT for bounded depth powering circuits, informal version of \Cref{thm:main-pit-small-depth}]
\label{thm:intro:pit-bounded-depth}

    We can explicitly construct the hitting set in \Cref{tab:intro-hitting-set-sizes-small-depth}.
    \begin{table}[h]
\centering
\setlength{\arraycolsep}{12pt}
\[
\begin{array}{lll} 
    \toprule
    \multirow{2}{*}{\textbf{Circuit Model}} &
    \multicolumn{2}{c}{\hspace{-3cm}\textbf{Hitting set size } T} 
     \\
    & \textbf{\(d+1 \geq (r-1)^2\)} &
      \textbf{\(d+1 < (r-1)^2\)}\\
    \midrule

    \Sigma^r\!\wedge^d\!\Sigma^s\!\wedge^\Delta\!\Sigma
        &
        d \cdot \poly(n,\Delta,s)
        &
        \poly\left(n,(s\Delta + r)^{r^2}\right)
\\

    \textbf{quasi-poly regime}
        & \text{always}
        & r=\poly(\log n)
        \\[2ex]

    \Sigma^r\!\wedge^d\!\Sigma^s\Pi^\Delta\Sigma
        &
        d \cdot \poly(n)(\Delta+1)^{2s+1}
        &
        \poly(n)(d\Delta+1)^{rs^{r^2}}
        \\

    \textbf{quasi-poly regime}
        & s=\poly(\log n)
        & r=O(1),\,s=\poly(\log n)
        \\[2ex]

    \Sigma^r\!\wedge^d\!\Sigma^s
    \Pi^\Delta\Sigma^m\!\wedge^\delta
        &
        d
        \left(\Delta m\delta\right)^{
            \tilde{O}\left(
                s \log\log(\Delta m \delta)
            \right)
        }
        &
        \left( \Delta m\delta \right)^{
             s^{O(r^2)}\cdot\log\log(\Delta m \delta )
             
        }
        \\

\textbf{quasi-poly regime}
        & s=\poly(\log n)
        & r=O(1),\,s=\poly(\log n)
        \\[2ex]

    \Sigma^r\!\wedge^d\!\Sigma^s
    \Pi^\Delta\Sigma^m\Pi^\delta
        &
        d 
        \left(\Delta m\right)^{
            \tilde{O}\left(
                \delta^2s\cdot\log(\Delta m)
            \right)
        }
        &
        \left( \Delta m\delta \right)^{
            O\left(
                \delta^2 s^{O(r^2)}\cdot\log(\Delta m\delta)
            \right)
        }
        \\

    \textbf{quasi-poly regime}
        & \delta,s=\poly(\log n)
        & r=O(1),\,\delta,s=\poly(\log n)
        \\

    \bottomrule
\end{array}
\]
\caption{Hitting sets sizes for small depth powering circuits
(see \Cref{tab:hitting-set-sizes-small-depth} for full details)}
\label{tab:intro-hitting-set-sizes-small-depth}

\end{table}
\end{theorem}




\subsubsection{More results}

Other applications of our lifting theorems include high-power PIT for sums of powers of read-once formulas,
preprocessed read-once formulas, bounded-read multilinear formulas, and several affine-orbit classes.  

We also obtain proper reconstruction algorithms in the high-power regime when the inner class is preprocessed read-once polynomials.

\subsection{Related work}
\label{sec:intro:related}

\paragraph{Algebraic lifting theorems.}

The closest work in spirit is the lifting result of Bhargava, Saraf, and Volkovich \cite{BhargavaSV26}, which transfers PIT algorithms between certain circuit classes
and gives PIT for constant-occur depth-$4$ circuits, and constant-occur depth-$5$ circuits, where the fan-in of the inner sum gate is small (this is the parameter $s$ in \Cref{tab:intro-hitting-set-sizes-small-depth}).

The lifting theorem of \cite{BhargavaSV26} relies on the property that each variable appears a small number of times (constant-occur) so there is rank concentration at a small set of variables. 
Our PIT lifting theorem is different and does not require constant-occur. Instead it relies on the fact that sufficiently high powers of pairwise non-associate polynomials are linearly independent.

\paragraph{Sums of powers of linear forms.}
When the inner polynomials are linear forms, our reconstruction problem is
Waring reconstruction, or equivalently symmetric tensor decomposition.
The algorithmic study of this problem goes back to Sylvester's work on
binary forms~\cite{sylvester1851lx}. Kayal gave a randomized polynomial-time
algorithm for reconstructing
\[
    f=\sum_{i=1}^{r}\alpha_i\ell_i^d
\]
when the linear forms are linearly independent~\cite{Kayal11}. His subsequent work on affine projections gave reconstruction algorithms for \emph{random}
depth-$3$ powering circuits in a broader parameter regime~\cite{Kay12}.
Further algorithms for random sums of powers and related polynomial equivalence problems were given by García-Marco, Koiran, and Pecatte~\cite{GKP17}.

Koiran and Skomra studied reconstruction for sums of cubes of linearly independent linear forms~\cite{KoiranSkomra21}. Koiran and Saha subsequently gave black-box reconstruction algorithms for degree three and higher~\cite{KoiranSaha23}. These results concern decompositions into
linearly independent forms.

A different line studies \emph{worst-case} reconstruction parameterized by the
rank. Bhargava, Saraf, and Volkovich gave polynomial-time algorithms for
constant-rank tensors (set-multilinear depth-$3$ circuits) and symmetric tensors~\cite{BhargavaSV21}.
Peleg, Shpilka, and Volk extended this to superconstant rank, obtaining
fixed-parameter reconstruction algorithms for Waring and tensor
decompositions~\cite{PSVTensor24}. Bhargava and Shringi later obtained a
deterministic $2^{r^{O(1)}}\poly(n,d)$-time algorithm for worst-case
tensor decomposition over $\mathbb R$ and $\mathbb C$
\cite{bhargavaDevansh25}.

\paragraph{Sums of powers of higher-degree forms.}
For inner polynomials of degree larger than one, Garg, Kayal, and Saha, using techniques from algebraic complexity, gave a polynomial-time reconstruction algorithm under a \emph{non-degeneracy} condition that holds for \emph{generic} instances~\cite{GKS20}. Bafna, Hsieh, Kothari, and Xu developed polynomial-time algorithms in \emph{generic} and \emph{smoothed} regimes, including cubic powers of quadratic forms, and allowed inverse-polynomial \emph{noise}~\cite{BHKX22}. Chandra, Garg, Kayal, Mittal, and Sinha further developed the lower-bounds-to-learning framework in noisy and smoothed
settings, under suitable singular-value assumptions~\cite{ChandraGKMS24}.

We stress that these results and ours address different regimes. The generic and smoothed algorithms can handle small outer powers and substantially larger top fan-in. Our algorithms require a sufficiently large outer power, or polylogarithmic rank and individual degree, but are \emph{deterministic} and work in the \emph{worst case}, without smoothness, genericity or non-degeneracy assumptions. 

\paragraph{Reconstruction of arithmetic circuits.}
\sloppy Arithmetic-circuit reconstruction has been studied for many restricted models. Sparse-polynomial (i.e., $\Sigma\Pi$ circuits) interpolation was studied in 
\cite{Ben-OrT88,GrigorievKS90,KlivansSpielman01}.

Beimel et al.\ gave improper reconstruction algorithms for depth-$3$ powering and set-multilinear circuits, with an $\roabp$ as the output representation~\cite{BBB2000}. Klivans and Shpilka gave reconstruction algorithms for classes whose spaces of partial derivatives have polynomially bounded dimension
\cite{klivansShpilkaRecon06}.

A reconstruction algorithm for
$\Sigma^{2}\Pi\Sigma$ circuits over finite fields was given in \cite{Depth3ReconShpilka09}, and Karnin and Shpilka extended this to arbitrary constant top fan-in over small finite fields \cite{KarninS09rec}. Sinha gave polynomial-time reconstruction algorithms for top fan-in two over $\mathbb R,\mathbb C$, and later over large finite fields
\cite{Sinha16,Sinha22}. Saraf and Shringi obtained a quasipolynomial-time
algorithm for top fan-in three~\cite{saraf2025reconstruction}, and
Saraf, Shringi, and Varadarajan subsequently treated arbitrary constant
top fan-in~\cite{SSV26Depth3Recon}.

Under \emph{non-degeneracy} assumptions, Kayal and Saha gave polynomial-time
reconstruction algorithms for homogeneous depth-$3$ circuits with
potentially large top fan-in~\cite{KayalSaha19}. Average-case
reconstruction is known for \emph{random} multilinear formulas and more general random arithmetic formulas~\cite{GKL11,GKQ14}. Gupta, Kayal, and Lokam gave the first reconstruction algorithm for a depth-$4$ model, namely multilinear $\Sigma^2\Pi\Sigma\Pi$ circuits~\cite{GuptaKL12}.

Read-once formulas and known-order $\roabp$s also admit deterministic
reconstruction algorithms
\cite{klivansShpilkaRecon06,sv14_rec_rof,roabp_recon_adaptive_fs13}.

 \paragraph{Polynomial identity testing.}
Recent work of Limaye, Srinivasan, and Tavenas, followed by Andrews and
Forbes, gave the first subexponential-size hitting sets for general
constant-depth arithmetic circuits~\cite{limaye2025superpolynomial,andrews2022ideals}.  Much stronger bounds
are known for several restricted classes. 

For sparse polynomials, deterministic black-box PIT and interpolation go back
to the classical works of Ben-Or and Tiwari, Grigoriev, Karpinski, and Singer,
and Klivans and Spielman~\cite{Ben-OrT88,GrigorievKS90,KlivansSpielman01}.  For depth-$3$ powering circuits
$\Sigma^{r}\!\wedge^{d}\!\Sigma$, building on a result of Raz and Shpilka~\cite{RazS05}, Saxena gave a deterministic white-box PIT algorithm~\cite{saxenaDualityTrick}.  
Very recently, Bhardwaj and Saptharishi~\cite{BS26HSSmallDer} 
obtained polynomial-size hitting sets for this model.

For $\Sigma^{r}\Pi^{\Delta}\Sigma$ circuits, rank-based methods give
polynomial-size hitting sets when the top fan-in is constant
\cite{dvirShpilka_rank_bound,KayalSaraf09,KarninS11pit,SaxenaSesh12pit,saxena_seshadhri_rank_bound}.  Agrawal, Saha, Saptharishi, and
Saxena introduced the Jacobian hitting technique, which yields
polynomial-size hitting sets when the transcendence degree of the product
terms is bounded and also applies to certain higher-depth classes
\cite{AgrawalSSS16}.

\sloppy For depth-$4$ circuits, Peleg and Shpilka obtained polynomial-size hitting sets for $\Sigma^{3}\Pi\Sigma\Pi^{2}$ circuits~\cite{peleg2021polynomial}. Garg, Mendes de Oliveira, and Sengupta extended this to arbitrary constant bottom degree \cite{GargOS25a}.  Independently, Guo and Wang gave a polynomial-time algorithm under the additional assumption that one of the three summands is squarefree~\cite{GuoWang25}, and recently Guo removed this assumption \cite{Guo26}. Garg, Mendes de
Oliveira, Sengupta, Shalmon and Shpilka obtained polynomial time hitting sets for  $\Sigma^{O(1)}\Pi\Sigma\Pi^{2}$ circuits \cite{GargOSSS26}.

Dutta, Dwivedi, and Saxena obtained
quasipolynomial-size hitting sets when both the top and bottom fan-ins are
polylogarithmically bounded~\cite{DDS-depth4-pit}.  
Quasipolynomial and later polynomial-size hitting sets are also known for
multilinear depth-$4$ circuits with bounded top fan-in
\cite{karnin_et_al,SarafV18}.  As in the depth-$3$ setting, the polynomial-size results
for these general depth-$4$ subclasses require bounded top fan-in.  Dutta
et~al. also observed that known techniques give hitting sets of size
$s^{O(\log\log s)}$ for size-$s$ $\Sigma\wedge\Sigma\wedge$ circuits
\cite{DDS-depth4-pit}.

Forbes constructed hitting sets of size
\[
    (nd)^{O(\delta\log r)}
\]
for $\Sigma^{r}\!\wedge^{d}\!\Sigma\Pi^{\delta}$ circuits
\cite{ForbesDepth4PIT}. 

Raz and Shpilka~\cite{RazS05} gave a polynomial-time white-box PIT algorithm for $\roabp$.
In the black-box model, a line of work of Forbes and Shpilka, Forbes, Saptharishi, and Shpilka, Gurjar,
Korwar, and Saxena, and Guo and Gurjar~\cite{roabp_recon_adaptive_fs13,forbesSapthaShpilka_HS_for_ml_ROABP,GurjarKS17,guo-gurjar-roabps-pit} resulted in hitting sets of size
\[
   \poly\big(\delta,(nw)^{\frac{\log n}{\max\left(1,\log\log n - \log \log (w)\right)}}\big), 
\]
for width $w$ $\roabp$.

\subsubsection{Relation to \cite{BDKSS26}}

In a companion paper, Bhat, Dutta, Shringi and the first two authors~\cite{BDKSS26} study the border of sums of powers
$\Sigma^{k}\wedge^{e}\mathcal D$, with possibly unequal exponents,
using the same Wronskian construction. Their inner polynomials have
low total degree, have a small space of partial derivatives, or are
$\roabp$s in a common order or in every order. For these classes they
give debordering and hitting sets, and for inner polynomials of low
total degree also a non-proper reconstruction algorithm, which outputs
a depth-$6$ formula.

Both papers construct a low-order differential operator that
annihilates the input. In~\cite{BDKSS26}, they work with the shifted polynomial $f(y+zx)$. They construct an annihilating operator with suitable coefficient bounds and leading coefficient nonzero at $z=0$. This operator determines the polynomial from its first few Taylor coefficients. They construct such an operator also for border inputs, without requiring uniqueness. For sufficiently large powers, we instead show that a suitably normalized operator is
unique, and use its leading coefficient and solution space to recover
the inner polynomials themselves.

Our results differ from those of~\cite{BDKSS26} in three ways. First,
our reconstruction is proper for sufficiently large powers, and it
also covers $\roabp$s, and sparse inner polynomials with bounds in
terms of their sparsity and individual degree rather than their total
degree. 
Second, for large powers our bounds are polynomial in the
number of summands. For example, for sums of $r$ powers of $s$-sparse
polynomials of degree $\Delta$, we reconstruct in time
$\poly(n,r,s,d,\Delta)$ and give hitting sets of size
$\poly(n,s,d,\Delta)$, whereas the bounds of~\cite{BDKSS26} have
exponents depending on $r$ and $\Delta$. Third, our hitting sets also
apply to $\roabp$s in different orders, to small-depth circuits and
bounded-read formulas, and over fields of large positive
characteristic. The results of~\cite{BDKSS26} apply throughout to border inputs and allow unequal, unrestricted exponents.

\subsection{Proof overview}
\label{sec:intro:overview}

We first describe the univariate argument used in all three theorems, in the high power regime, and then explain the two different ways in which we lift it back to the multivariate setting.

\paragraph{High-powers and PIT.}
Let $H=\{\vu_1,\ldots,\vu_h\}$ be an interpolating set for the inner
class, and let $M(t)$ be a polynomial curve of degree at most $h-1$ passing
through the points of $H$.  If $A$ and $B$ are non-associate members of the
inner class, then $A\circ M$ and $B\circ M$ remain non-associate. Consequently, the projected inner polynomials are pairwise non-associate univariates of degree at most $\Delta(h-1)$. By the polynomial ABC theorem \cite{stothers1981polynomial,mason,VW2023} a sufficiently short sum of sufficiently high powers of the projected polynomials cannot vanish.  Since
\[
    \deg(f\circ M)\le d\Delta(h-1),
\]
interpolation along the curve gives the hitting set in
\Cref{thm:intro:lifting}\ref{item:intro-main-pit}.

\paragraph{Univariate reconstruction.}
Using different methods, both reconstruction theorems reduce to recovering
\[
    F(t)=\Sigma_{i=1}^{r}\alpha_i F_i(t)^d
\]
for pairwise non-associate univariate $F_i$'s of bounded degree.  The main idea is to compute a differential operator
\[
    L=Q_0(t)+Q_1(t)\partial+\cdots+Q_r(t)\partial^r
\]
that annihilates $F$ and the individual powers $F_i^d$.  This operator is obtained by considering the Wronskian of the powers of the inner polynomials. The ABC theorem implies that its kernel is exactly\footnote{In positive characteristic we have to restrict the degrees of the polynomials that we consider in the kernel.}
\[
    \operatorname{span}\{F_1^d,\ldots,F_r^d\}.
\]
We then prove that every irreducible factor of each $F_i$ divides the leading coefficient of $L$, $Q_r(t)$. 
After factoring \(Q_r(t)\), it remains to determine which factors occur in each \(F_i\) and with what multiplicities. 

To achieve this we organize the possible choices in a rooted tree: the root represents the empty product, and each edge multiplies the current product by one irreducible factor, with repetitions accounting for multiplicities (the depth of the tree is bounded by the total degree of the inner polynomials). We explore this tree by depth-first search (DFS), using linear algebra to detect and discard branches that cannot lead to a previously unrecovered summand. We show that only polynomially many nodes are visited. Finally, we recover the coefficients \(\alpha_i\) by solving a linear system obtained from evaluations on a hitting set for the model.

When reconstructing different univariate projections of the multivariate
input, we may obtain unordered summands with unrelated scalar
normalizations. We align them through a common anchor $\vu$ at which the
values $\alpha_i f_i(\vu)^d$ are nonzero and pairwise distinct. For every evaluation point $\bv$ we consider the 
line $\vu+t(\bv-\vu)$. At $t=0$, the line gives the alignment value, while at $t=1$, it gives an evaluation of the corresponding inner polynomial. 

The process described above may fail at some evaluation points that make the polynomials associate. We overcome this by introducing the notion of \emph{robust reconstruction sets} which allow reconstruction even if some evaluation points are missing (in a similar fashion to erasure decoding in coding theory). We note that a simple extension of the above can also handle corrupted evaluations, but in our case we can spot when an evaluation point is not a good anchor.

\paragraph{From total degree to individual degree.}
For the total-degree theorem, the line
$\vu+t(\bv-\vu)$ preserves the bound $\deg(f_i)\le\Delta$, so the preceding
argument applies directly.

For the individual-degree theorem, moving every coordinate on one line would
produce degree $n\delta$. We instead apply \cite[Lemma 4.3]{BhargavaSV26} (see \Cref{lem:swli}) that gives a set $S$ of at most $r-1$ variables such that, after a suitable specialization $\vu$ of the other variables,
\[
    f_1(\vx_S,\vu),\ldots,f_r(\vx_S,\vu)
\]
remain nonzero and pairwise non-associate.

We use this lemma in the following way.
Given a target evaluation point $\vw$ in a robust reconstruction set, we move from $\vu$ to $\vw$ by replacing the
coordinates of $\vu$ by those of $\vw$ one at a time.  At each step only the
variables in $S$ and one interpolation variable $t$ (corresponding to the modified coordinate) are nonconstant, so every inner polynomial has total degree at most $(|S|+1)\delta\le r\delta$.  At $t=0$, the decomposition agrees with the already aligned preceding hybrid, while at $t=1$ it yields the next hybrid.  

As before there can be technical difficulties arising from the failure of a hybrid to serve as a good anchor for the next hybrid. However this is captured by an identity from a related class and we construct the robust reconstruction set so that, for most  $\vw$, none of these witnesses vanishes at any hybrid stage.  

The $n^{O(r)}$ overhead comes from enumerating the candidate sets $S$ of at most $r-1$ variables.  

\paragraph{The low-power case.}
The overview above holds when the powering degree is high. To obtain results for the low power regime we use simple closure properties.  For example, a power of an $s$-sparse polynomial has at most $\binom{s+d-1}{d}$ monomials and a power of a common-order $\roabp$ is a wider $\roabp$ in the same order. Similar closure properties hold for small depth models.  Together, the high- and low-power algorithms give results for every powering degree $d$.

\subsection{AI disclosure}

We have used ChatGPT models for proofreading and improving the writing. ChatGPT~6 also improved one of our arguments:
Our original proof bounded the $\roabp$ complexity of the $d$-th power of a $\roabp$ of width $w$ by first concatenating $d$ copies of an $\roabp$ computing
$f$ into a $d$-pass $\abp$, and then applying a conversion lemma
of \cite{AndersonFSSV18}. This yielded an $\roabp$ of width $w^{2d}$.
The construction suggested by ChatGPT (\Cref{lem:power-roabp}) gives an $\roabp$ of width $\binom{w+d-1}{d}$.


\subsection{Notations} \label{sec:notations}

For an integer $r\in \N$ we denote $[r]=\{1,\ldots,r\}$ and $[[r]]=\{0,1,\ldots,r\}$. Also denote $[0] = \emptyset$. We use the symbol~$\coloneqq$ to denote definitions. 
We shall use $\Delta$ to denote total degree and $\delta$ to denote individual degree upper bound.

For a polynomial class $\cC$ and number $\Delta \in \N$, we denote by $\cC_{\Delta}$ the class of polynomials from $\cC$ of degree at most $\Delta$. We also denote the subset of polynomials of individual degree at most $\delta$ by \[
\mathcal C_{\Delta,\delta}
:=
\{f\in\cC_\Delta: \deg_{x_j}f\le\delta\text{ for every }j\}.
\]

We say that two nonzero polynomials $f,g\in\F[\vx]$ are associate if there exists a nonzero scalar $\alpha\in\F$ such that $g(\vx)=\alpha f(\vx)$, and non-associate otherwise. We say that a set of polynomials is non-associate if no two polynomials in the set are associate. This is equivalent to saying that the polynomials in the set are pairwise linearly independent.

In what follows, and especially in \autoref{sec:main_theorems}, we frequently use the expression $\binom{r}{2}$, sometimes also in the case $r=1$. For brevity, in this case we abuse notation and set $\binom{1}{2}=1$.

For a set of points $S = \{\boldsymbol{\alpha}_1, \dots, \boldsymbol{\alpha}_m\} \subseteq \mathbb{F}^n$, we denote by $f(S) = \{f(\boldsymbol{\alpha}) \mid \boldsymbol{\alpha} \in S\}$ the set of evaluations of $f$ on $S$. 

For a vector $\bv \in \F^n$ and a set \(S\subseteq[n]\), we denote \(\bv_S=(v_j)_{j\in S}\) and \(\bv_{-S}=(v_j)_{j\notin S}\).


\subsection{Conventions}

To help shorten the presentation we rely on the following conventions and do not refer to them in our proofs. 

\paragraph{Computational model and bit complexity.}
\label{par:computational-model}

We analyze the running times of our algorithms by counting the number of operations for the algebraic steps. The corresponding bit complexity depends on the underlying field and on the representations of the input polynomials.  
\begin{center}\it
    Throughout the paper, \(B_{\text{in}}\) denotes an upper bound on the\\ representation size (in bits) of the input polynomials.
\end{center} If a circuit representation of such a polynomial has $s$ edges, the representation's formal degree\footnote{The formal degree is defined recursively: every input has degree \(1\), addition gates take the maximum of the input degrees, and multiplication gates take their sum. 
} is $D$ and
each scalar in it requires at most \(\beta\) bits, then \(B_{\text{in}}=O(D s\beta)\).
For example, a \(\Sigma^r\Pi^d\Sigma\) circuit has \(O(ndr)\)
edges and therefore total bit complexity \(O(ndrD\beta)\).
We count a power gate \(\wedge^d\) as having \(d\) edges. Under
this convention, a \(\Sigma^r\wedge^d\Sigma\) circuit also has
edge complexity \(O(ndr)\). Over a finite field \(\F\),
each scalar can be represented using \(O(\log|\F|)\) bits,
yielding \(B_{\text{in}}=O(Ds\log|\F|)\).

To account for bit complexity, we must also consider the bit complexity of the points in the hitting sets and reconstruction sets we use. All sets used in this work are constructed using hitting-set generators and polynomial maps whose outputs have bit complexity polynomial in the parameters of the relevant class. 
\begin{center}\it
    Throughout the paper,  $B_{\text{eval}}$ denotes the bit complexity of points\\ in the input hitting/reconstruction sets.
\end{center}

With these conventions, evaluating an input polynomial \(P\) at a point of bit complexity at most \(B_{\text{eval}}\) produces a scalar of bit complexity \(O(B_{\text{in}}B_{\text{eval}})\).

\begin{center}
    \it Unless bit complexity is explicitly discussed, all running-time bounds\\ suppress factors polynomial in \(B_{\text{in}}\)  and \(B_{\text{eval}}\) .
\end{center}

\paragraph{The underlying field.} Our algorithm require polynomially many evaluation points from a field. If the field is too small we may not have such point. We therefor use the following convention
\begin{center}\it
    Throughout the paper, whenever the field $|\F|$ is too small for the\\ required construction, we move to an extension field $\mathbb{E}$\\ of size polynomial in the required parameters.
\end{center}

\subsection{Organization}

The paper is organized as follows. The main algorithm for reconstructing sums of powers of univariate polynomials is given in \Cref{sec:uni-reconstruct}. The rest of \Cref{sec:univariate-rec} gives the required tools to prove its correctness. We then prove the main lifting theorems, and show how to construct robust hitting and reconstruction sets in \Cref{sec:main_theorems}.

The remaining sections are organized by the underlying circuit class: \Cref{sec:sparse} is dedicated to sparse polynomials,
\Cref{sec:roabp} studies \(\roabp\)s,
\Cref{sec:dmall-depth} treats small-depth circuits, and
\Cref{sec:rof} concerns bounded-read formulas.

\section{Reconstruction of Univariate Sums of Powers}\label{sec:univariate-rec}


In this section we give a reconstruction algorithm for sums of powers of univariate polynomials satisfying the degree condition of \Cref{thm:intro:lifting}\ref{item:intro-main-total}.

\begin{theorem}[univariate reconstruction]\label{thm:univariate-reconstruct}
Let $r,d,\Delta\in \N$ with $d>(r+1)^4\Delta$.
Let $\F=\Q$, or let $\F=\F_q$ be a finite field of characteristic $p>2rd\Delta$.

Suppose we are given black-box access to a nonzero univariate polynomial $f\in\F[x]$ that admits a minimal representation
\begin{equation}\label{eq:univariate-input}
f(x)=\sum_{i=1}^{r'} \alpha_i\,f_i(x)^d,
\qquad
 \alpha_i\neq 0, \qquad \deg(f_i)\le \Delta, \qquad r'\le r,
\end{equation}
where each $f_i$ is monic, and the polynomials $\{f_i\}_{i=1}^{r'}$ are non-associate.
Then there is a deterministic algorithm that  outputs, in time $\poly(r,\Delta,d,p)$, 
a set $\{(\alpha'_i,f'_i)\}_{i=1}^{r'}$ such that, for some permutation $\pi$,
\[
f(x)=\sum_{i=1}^{r'} \alpha'_i\,f'_i(x)^d
\qquad\text{and}\qquad
f'_i=f_{\pi(i)},\ \ \alpha'_i=\alpha_{\pi(i)}.
\]
\end{theorem}

\begin{remark} \label{remark:small_r_wlog}
    We note that we only have an upper bound on the number of summands, $r$, and in reality, $f$ can be represented as a sum of $r'<r$ terms. This distinction does not significantly affect our algorithm, as we will simply try each $r'\in[1,r]$. Thus, for simplicity, we assume that $r$ is the exact number of summands, and we continue to denote it by $r$. From now on, until we present the algorithm itself, we assume that $f$ cannot be represented as in \eqref{eq:univariate-input} with fewer than $r$ terms. Minimality also implies that every $\alpha_i$ is nonzero and that the polynomials $f_i^d$ are linearly independent.
\end{remark}

\medskip

The rest of this section is devoted to proving \Cref{thm:univariate-reconstruct}. 

\medskip

As $f(x)$ is a univariate polynomial with $\deg(f)\leq d\cdot \Delta$, we can interpolate its coefficients to obtain a representation 
\[
f(x)=\sum_{i=0}^{d\Delta}\beta_ix^i.
\]
Hence, we may assume that this representation of $f(x)$ is given as input to the algorithm.

As described in \Cref{sec:intro:overview}, our approach is to learn a differential operator $L$ of order $r$ such that $L(f)=0$. To define the operator we shall use the Wronskian matrix that we define next. 

\subsection{Wronskian and Differential Operators over Finite Fields}\label{sec:wronski}
We recall the definition of the \emph{Wronskian}.\footnote{The Wronskian can be defined for any collection of functions that are sufficiently differentiable on an interval; here we consider only polynomials.}
Let $g_1,\ldots,g_n\in \F[x]$. Their Wronskian is the determinant of the $n\times n$ matrix whose $(i,j)$-entry, for $i\in[[n-1]]$ and $j\in[n]$, is $\nabla^{(i)} g_j$, where
\[
\nabla^{(i)} := \frac{d^i}{dx^i}.
\]
\begin{equation}\label{eq:wronskian-g}
W(g_1,\ldots,g_n)\coloneq 
\det\!\begin{pmatrix}
    g_1(x) & g_2(x) & \cdots & g_n(x)\\
    \nabla g_1(x) & \nabla g_2(x) & \cdots & \nabla g_n(x)\\
    \vdots & \vdots & \ddots & \vdots\\
    \nabla^{(n-1)} g_1(x) & \nabla^{(n-1)} g_2(x) & \cdots & \nabla^{(n-1)} g_n(x)
\end{pmatrix}\in \F[x] .
\end{equation}
We treat $\nabla$ as a linear operator on the space of polynomials, so it is well defined over any field.

The Wronskian is a classical object in the theory of differential equations (see, e.g., \cite[\S3.4]{ODE}), but it also plays an important role for polynomials over finite fields. The key fact is that if the characteristic of the field is sufficiently large, then the Wronskian vanishes if and only if the polynomials are linearly dependent. For completeness, we provide the proof of the following theorem in \autoref{app:wronskian}. 

\begin{restatable}{theorem}{thmwronskilinind}\label{thm:wronskian-lin-ind}
Let $g_1, \dots, g_n$ be polynomials over a field $\F$ of characteristic $p$.  
If the maximum degree of any $g_i$ is $d$ and $p=0$ or $p > d$, then  
$g_1, \dots, g_n$ are linearly dependent over $\F$ if and only if their Wronskian
$W(g_1, \dots, g_n)$ is identically zero.    
\end{restatable}

\medskip

The next theorem, an analogue of the fundamental theorem of linear homogeneous ODEs for polynomials over arbitrary fields, states that the solution space of an operator of order $r$ has dimension at most $r$.

\begin{theorem}\label{thm:ode}
Let $\F$ be a field of characteristic $p$, and let\footnote{Since $x$ and $\nabla$ do not commute, we denote the noncommutative polynomial ring of differential operators with coefficients in $\F[x]$ by $\F[x]\langle\nabla\rangle$.}
\[
A \;=\; \sum_{i=0}^{r} Q_i(x)\,\nabla^{i} \;\in\; \F[x]\langle\nabla\rangle
\]
be a nonzero operator of order $r$, with $Q_r\not\equiv 0$.
Fix $D\in\N$ and assume $p=0$ or $p> D$.
Then the space
\[
\mathcal S_{\le D}\;=\;\{\,g\in\F[x]:\ \deg g\le D,\ A(g)=0\,\}
\]
has $\dim_\F \mathcal S_{\le D}\le r$. 
\end{theorem}

\begin{proof}\
If $r>D$ then the claim is obvious since $\dim_\F \mathcal S_{\le D}\le D+1\leq r$.
We may extend scalars to an algebraic closure, since this does not change the dimension of the kernel. Hence, we can pick $a\in\overline{\F}$ with $Q_r(a)\neq 0$. Define the translated operator
\[
\tilde{A} \;=\; \sum_{i=0}^{r}  \tilde{Q}_i(x)\,\nabla^i,
\qquad  \tilde{Q}_i(x):=Q_i(x+a),
\]
and the translated polynomial $\tilde{g}(x):=g(x+a)$. 
Since ordinary derivatives commute with translation, we have
$A(g)=0$ if and only if $\tilde{A}(\tilde{g})=0$.
Write $\tilde{g}(x)=\sum_{i=0}^{D} c_i x^i$, where $c_0,\ldots,c_D\in\F$ are indeterminates. Then
\[
\nabla^k \tilde{g}(0)=\frac{d^k \tilde{g}}{dx^k}(0) \;=\; k!\, c_k \qquad (0\le k\le D).
\]
Consider the identities $\nabla^t\left(\tilde{A}(\tilde{g})\right)(0)=0$ for $t=0,1,\ldots,D-r$. 
By Leibniz’s rule
\[
0 \;=\; \nabla^t\left(\tilde{A}(\tilde{g})\right)(0)
 \;=\; \sum_{i=0}^{r} \sum_{j=0}^{t} \binom{t}{j}\,\nabla^{t-j} {\tilde{Q}}_i(0)\,\nabla^{i+j} {\tilde{g}}(0),
\]
hence, using $\nabla^{i+j} {\tilde{g}}(0)=(i+j)!\,c_{i+j}$,
\begin{equation}\label{eq:triangular-recurrence-corrected}
\sum_{i=0}^{r} \sum_{j=0}^{t} \binom{t}{j}\, \nabla^{t-j} {\tilde{Q}}_i(0)\,(i+j)!\, c_{i+j} \;=\; 0,
\qquad t=0,1,\ldots,D-r.
\end{equation}
In \eqref{eq:triangular-recurrence-corrected}, the term with $(i,j)=(r,t)$ equals
\[
\binom{t}{t}\,{\tilde{Q}}_r (0)\,(r+t)!\, c_{r+t} \;=\; Q_r(a)\,(r+t)!\, c_{r+t},
\]
while all other terms involve only $c_k$ with $k\le r+t-1$. 
Because $p=0$ or $p>D$, each factorial $(r+t)!$ is invertible in $\F$ for $0\le t\le D-r$, 
and $Q_r(a)\ne 0$ by construction. Thus, the coefficient of $c_{r+t}$ is nonzero. 
Therefore, for each $t$, we can solve uniquely for $c_{r+t}$ in terms of $c_0,\ldots,c_{r+t-1}$. 
Inductively, all $c_r,\ldots,c_D$ are determined by the initial block $(c_0,\ldots,c_{r-1})\in\F^r$, 
so $\dim_\F \mathcal S_{\le D}\le r$.
\end{proof}

For a differential operator $A$ and an integer $D\ge0$, we write 
\[ 
\Ker_{\le D}(A) := \{\,g\in\F[x] \mid \deg(g)\le D,\ A(g)=0\,\}. 
\] 
We remark that the degree bound is essential in positive characteristic since without it the kernel has infinite dimension.


\subsection{The Differential Operator $L$}\label{sect:wronskian}

Consider the Wronskian of $f,f_1^d,\ldots,f_r^d$: 
\begin{equation}\label{detW}
W(x)=W(f,f_1^d,\ldots,f_r^d)\coloneq \det\!\begin{pmatrix}
    f(x) & f_1^d(x) & \cdots & f_r^d(x)\\
    \nabla f(x) & \nabla f_1^d(x) & \cdots & \nabla f_r^d(x)\\
    \vdots &\vdots & \ddots & \vdots\\
    \nabla^{r} f(x) & \nabla^{r}f_1^d(x) & \cdots & \nabla^{r}f_r^d(x)
    \end{pmatrix}.
\end{equation}

Clearly, $W(x)\equiv0$ because the first column is a linear combination of the others. Expanding the determinant along the first column yields
\begin{equation}\label{eq:W-with-M}
\sum_{i=0}^r (-1)^i \,   W_i(x) \, \nabla^{i}f(x) \equiv 0,    
\end{equation}
where $W_i(x)$ denotes the determinant of the $r\times r$ minor obtained by deleting the $i$-th row and the first column of the matrix in \eqref{detW}. 

\begin{claim}\label{Wronski}
Each polynomial $W_k(x)$, obtained from the Laplace expansion of the Wronskian, can be written as
\[
W_k(x) = (-1)^k\left(\prod_{i\in[r]} f_i^{\,d-r}\right) \tilde{P}_k(x),
\]
where $\tilde{P}_k(x)$ is a polynomial of degree at most $r^2\Delta$.
\end{claim}

\begin{proof}
Recall that $\deg(f_i)=\Delta_i\le \Delta$. Fix $k\in [[r]]$. 
For each $i\in [[r]]\setminus\{k\}$ and $j\in [r]$, we can write
\begin{equation}
    \label{eq:def-gij}
\nabla^{(i)} f_j^d \;=\; f_j^{\,d-r}\, g_{i,j}, 
\qquad \text{with}\qquad \deg(g_{i,j}) \le r\Delta_j - i \;\le\; r\Delta \, .    
\end{equation}
Hence
\[
W_k(x)
= \left(\prod_{j\in[r]} f_j^{\,d-r}\right) \det\!\big( g_{i,j}(x) \big)_{i\in [[r]]\setminus\{k\},\, j\in [r]}.
\]
Define
\begin{equation*}
\tilde{P}_k(x) \;:=\; (-1)^{k}\det\!\big( g_{i,j}(x) \big)_{i\in [[r]]\setminus\{k\},\, j\in [r]}.
\end{equation*}
Then
\[
W_k(x)=(-1)^{k}\left(\prod_{j\in[r]} f_j^{\,d-r}\right) \tilde{P}_k(x).
\]
Since each $g_{i,j}$ has degree at most $r\Delta$, $\tilde{P}_k(x)$ has degree at most $r^2\Delta$, as claimed. \qedhere
\end{proof}

Let $P=\gcd(\tilde{P}_0,\ldots,\tilde{P}_r)$ and denote $P_i=\tilde{P}_i/P$. 

\begin{corollary}\label{cor:OpL}
Assume that $p$, the characteristic of $\F$, satisfies $p=0$ or $p>d\Delta$. 
Then the differential operator 
\[
\tilde{L} \;\coloneq\; \sum_{i=0}^r P_i(x)\,\nabla^{(i)} \in \F[x]\langle\nabla\rangle
\]
has order exactly $r$, and 
\[\Ker_{\le d\Delta}(\tilde{L}) = \Span\{f_1^d,\ldots,f_r^d\}.
\]
In particular, $\tilde{L}$ annihilates $f$ and every $f_i^d$.
\end{corollary}

\begin{proof}
The minor $W_r$ is precisely \[ W(f_1^d,\ldots,f_r^d). \] Since $f_1^d,\ldots,f_r^d$ are linearly independent, \Cref{thm:wronskian-lin-ind} gives $W_r\not\equiv0$. Hence $\widetilde P_r\not\equiv0$, and therefore $P_r\not\equiv0$. Thus $\widetilde L$ has order exactly $r$.
 
From \autoref{Wronski} and \eqref{eq:W-with-M}, we have
\begin{align*}
0&= \sum_{i=0}^r (-1)^i \,  W_i(x) \, \nabla^{i}f(x) \\   
&= \left(\prod_{j\in[r]} f_j^{\,d-r}\right)\!\cdot  
  \sum_{i=0}^r \tilde{P}_i(x)\, \nabla^{i}f(x)\\
&= \left(\prod_{j\in[r]} f_j^{\,d-r}\right)\!\cdot P \cdot 
  \sum_{i=0}^r P_i(x)\, \nabla^{i}f(x)\\
&=
\left(\prod_{j\in[r]} f_j^{\,d-r}\right) \cdot P \cdot \tilde{L}(f)\,.
\end{align*}
Since $\left(\Pi_{j\in[r]} f_j^{\,d-r}\right)\!\cdot P$ is not identically zero, it follows that $\tilde{L}(f)\equiv 0$.
The same argument applies to any linear combination of $f_1^d,\ldots,f_r^d$, hence
\[
\mathrm{span}\{f_1^d,\ldots,f_r^d\}\subseteq\Ker_{\le d\Delta}{(\tilde{L})}.
\]

By \autoref{thm:ode}, the solution space of an operator of order $r$ has dimension at most $r$.  
As the $f_i^d$ are linearly independent, they form a basis of $\Ker_{\le d\Delta}{(\tilde{L})}$. 
\end{proof}



We now set up a linear system whose solution  yields the desired operator $L$ from \autoref{cor:OpL}.
Consider the following system of linear equations with unknown coefficients $\gamma_{i,j}\in\F$:
\begin{equation}\label{eq:L-lin-system}
\sum_{i=0}^{r}\!\left(\sum_{j=0}^{r^2\Delta}\gamma_{i,j}x^j\right) \nabla^{i}f(x)\;\equiv\;0\,.
\end{equation}
The system \eqref{eq:L-lin-system} consists of $\deg(f)+r^2\Delta+1$ linear equations in the unknowns $\gamma_{i,j}$, with coefficients that are linear combinations of the coefficients of~$f$.

Clearly, one solution is obtained by setting
\[
\sum_{j=0}^{r^2\Delta}\gamma_{i,j}x^j \;=\; P_i(x)\,,
\]
which corresponds exactly to the operator $\tilde L$ from \autoref{cor:OpL}.
We next show that any nontrivial solution of the form $\Sigma_{j=0}^{r}Q_j\nabla^j$ satisfying $\gcd(Q_0,\ldots,Q_r)=1$ is a scalar multiple of $\tilde{L}$. In particular, any nontrivial solution that minimizes $\deg(Q_r)$ equals a scalar multiple of $\tilde{L}$.
For this, we rely on the ABC theorem for function fields that we discuss next.

 \subsection{The ABC theorem for function fields}\label{sec:abc}

An important tool in all proofs is the ABC theorem for function fields, known as the Mason-Stothers theorem \cite{mason,stothers1981polynomial}.
We state the following strengthening  due to  Vaserstein and Wheland \cite{VW2023}. While most of the results in \cite{VW2023} are stated for fields of characteristic zero, an inspection of the proofs in   \cite[Section 3]{VW2023} reveals that they also hold for fields of large enough characteristic.\footnote{Vaserstein and Wheland were primarily interested in characteristic $0$ or fixed characteristic independent of the degrees and number of polynomials (see \cite[Section 5]{VW2023}). }

For a polynomial $h\in\F[x]$, let $\nu(h)$ denote the number of \emph{distinct} roots of $h$ over the algebraic closure $\overline{\F}$.

\begin{theorem}[\cite{VW2023}, Theorem~2.2(a)]\label{thm:ABC}
 Let $D, r\in \N$ such that $r\geq 2$. Let $\F$ be a field of characteristic $p$ such that $p=0$ or $p>(r-1)D$. Let $h_1,h_2,\ldots,h_r\in\F[x]$ satisfy
\[
h_2+\cdots+h_r = h_1,\qquad \gcd(h_2,\ldots,h_r)=1,
\]
where $\max_i\deg{h_i}\leq D$, not all $h_i$ are constant and no nonempty subsum of $h_2,\ldots,h_r$ vanishes. 
Then
\[
\deg(h_1) < (r-2)\!\left(\sum_{i=1}^r \nu(h_i)\right).
\]
\end{theorem}

In other words, \autoref{thm:ABC} asserts that if $h_1$ has ``high'' degree while each other $h_i$ has ``few'' distinct roots, then $h_1$ cannot be expressed as a linear combination of the other $h_i$’s. 
We note that we require an upper bound on $\deg(h_i)$ only in the case of positive characteristic.

\begin{corollary}\label{cor:ABC}
Let $r,\eta,\delta,e\in \N$ such that $r\geq 2$.
Let $\F$ be a field of characteristic $p$ such that $p=0$ or $p>(r-1)(\eta+e\delta)$. Let $P_1(x)^e,P_2(x)^e,\ldots,P_r(x)^e\in \F[x]$ be pairwise linearly independent polynomials, such that for all $i$, $\deg(P_i)\leq \delta$. 
Let $g_1(x),\ldots,g_r(x)\in\F[x]$ be additional polynomials, such that  for all $i$, $\deg(g_i)\leq \eta$. 
Assume that
\[
e \;\ge\; r(r-1)(\eta+1).
\]
Then the identity
\[
\sum_{i=1}^r g_i(x)\,P_i(x)^e = 0
\]
holds if and only if each $g_i(x)$ is identically zero.
\end{corollary}

\begin{proof}
The “if” direction is immediate. For the converse, suppose
\[
\sum_{i=1}^r g_i P_i^e = 0
\]
is a nontrivial identity. Passing, if necessary, to a nontrivial relation of minimal support, we may assume that every $h_i := g_i P_i^e$ in the relation is nonzero, and hence each corresponding $g_i$ is nonzero. Set
\[
H(x)=\gcd(h_1,\ldots,h_r),\qquad P(x)=\gcd(P_1,\ldots,P_r) .
\] 

\begin{claim}
There exists $G(x)\in\F[x]$ such that $H(x)=P(x)^e G(x)$ and $\deg G \le r\eta$.
\end{claim}

\begin{proof}
Clearly $P^e$ divides each $h_i$, so $P^e$ divides $H$. As $\gcd((P_1/P)^e,\ldots,(P_r/P)^e)=1$, we get that
\[
\gcd\big(g_1(P_1/P)^e,\ldots,g_r(P_r/P)^e\big)\quad \text{divides}\quad \prod_{i=1}^r g_i.
\]
Thus $G\coloneq H/P^e$ divides $\Pi_i g_i$, so
\[
\deg(G)=\deg(H/P^e)\ \le\ \sum_i \deg g_i\ \le\ r\eta.\qedhere
\]
\end{proof}

Let $S := \{i:\, g_i \not\equiv 0\}$ be the support, and choose $t\in S$ such that
\[
\nu_{\mathrm{sup}} := \max_{i\in S} \nu(P_i/P)= \nu(P_t/P) .
\]
Since the polynomials are non-associate, $\nu_{\mathrm{sup}}\ge 1$.
Define $\tilde h_i := h_i/H$. 
Then the equality $\Sigma_{i=1}^r h_i=0$ holds if and only if $\sum_{i=1}^r \tilde h_i=0$, and moreover
\[
\gcd(\tilde h_1,\ldots,\tilde h_r)=1.
\]

By \autoref{thm:ABC}, if $\Sigma_{i=1}^r \tilde h_i=0$ and not all $\tilde h_i$ vanish, then
\begin{align*}
 e\,\nu_{\mathrm{sup}}-\deg(G)&= e\nu(P_t/P)-\deg(G)\\
 &\le \deg((P_t/P)^e)-\deg(G)
 \\& \le \deg(\tilde h_t)\\
 &< (r-2)\sum_{j=1}^r \nu(\tilde h_j)\\ &\le (r-2)\sum_{j=1}^{r}\nu(g_j (P_j/P)^e)\\
 &\le (r-2)r\big(\eta+\nu_{\mathrm{sup}}\big).
\end{align*}
Since $\deg(G)\le r\eta$ we get
\[
e\leq r(r-1)\eta + r(r-2) < r(r-1)(\eta+ 1),
\]
contradicting the assumption on $e$. 
\end{proof}

We did not attempt to optimize the parameters in \autoref{cor:ABC}. In particular, for the special case where all $g_i$ are scalars a tighter result is known.

\begin{theorem} [{\cite[Corollary 3.8]{VW2023}}]\label{thm:abc_constants}
    Let $r,d ,\delta \in \N$ such that $r \geq 3$. Let $\F$ be of characteristic $p = 0$ or $p > (r-1)d\delta$. Let $f_1(x),\ldots, f_r(x) \in \F[\x]$ be pairwise non-associate polynomials of degree at most $\delta$. If $(r-1)^2 \leq d + 1$ then $f_1^d,\ldots,f_r^d$ are linearly independent.
\end{theorem}


The above theorem implies a unique representation of small sums of polynomials.

\begin{corollary} [{Uniqueness of Representation for Small Sums of High Powers}] \label{cor:uniq repres of high powers}
    Let $r,d,\delta \in \N$. Let $\F$ be of characteristic $p = 0$ or $p > (2r-1)d\delta$. Let $\{f_i\}_{i=1}^r\subseteq \F[\x]$ be polynomials satisfying the conditions in \autoref{thm:abc_constants}, with $d + 1 \geq (2r-1)^2$. Then, for any set $\{g_i\}_{i=1}^{r'} \neq \{f_i\}^r_{i=1}$ of polynomials of degree at most $\delta$ such that $r' \leq r$ and not all $g_i$s are a scalar multiple of some $f_i$, and scalars $\alpha_1,\ldots,\alpha_r,\beta_1,\ldots,\beta_{r'} \in \F\setminus\{0\}$, 
    $$
      \sum_{i=1}^r  \alpha_i f_i^d \not \equiv \sum^{r'}_{i=1} \beta_i g_i^d \, .
    $$
\end{corollary}
    
\begin{proof}
    Assume without loss of generality that the polynomials in $\{g_i\}_{i=1}^{r'}$ are pairwise non-associates and denote by $S_g$ and $S_f$ the set of polynomials in $\{g_i\}_{i=1}^{r'}$ and $\{f_i\}_{i=1}^{r}$ respectively which have an associate polynomial in the other set of polynomials. Obviously, $|S_g| = |S_f|$, assume without loss of generality, that $S_g = S_f = [k]$ for some $k \in [\Z^{+}]$ ($[0] = \emptyset$). We get
    $$
        0 \equiv \sum_{i=1}^r \alpha_i f_i^d - \sum_{i=1}^{r'} \beta_i g_i^d = \sum_{i=1}^k \gamma_i f_i^d +\sum_{i=k+1}^r \alpha_i f_i^d - \sum_{i=k+1}^{r'} \beta_i g_i^d \, ,
    $$
     for scalars $\gamma_1,\ldots,\gamma_k \in \F$. This is a linear combination of at most $2r$ polynomials, which are pairwise non-associate. According to \autoref{thm:abc_constants} this sum cannot be identically zero, thus we arrive at a contradiction.
\end{proof}

\subsection{Uniqueness of $L$}\label{sec:uniqueness}


\begin{claim}\label{cla:L-unique}
    Assume that the characteristic of the field $\F$ satisfies $p=0$ or $p>2rd\Delta$. Let $\tilde{L}\in\F[x]\langle \nabla\rangle$ be the operator from \autoref{cor:OpL}.
    Assume $d>(r+1)^4\Delta$. Then any nontrivial operator $L'=\Sigma_{i=0}^{r}Q_i(x)\nabla^i$ satisfying  $\deg(Q_i)\leq r^2\Delta$  for all $i$, $\gcd(Q_0,\ldots,Q_r)=1$, and $L'(f)=0$ must equal a scalar multiple of $\tilde{L}$.
\end{claim}
\begin{proof}
    Let $L'$ be as in the statement of the claim. We have that 
    \begin{align*}
        0\equiv L'(f) &= \sum_{j=1}^{r}\alpha_j L'(f_j^d)\\
        &=\sum_{j=1}^{r}\sum_{i=0}^{r}\alpha_j Q_i(x) \nabla^{i}(f_j^d)\\
        &=\sum_{j=1}^{r} \sum_{i=0}^{r}\tilde{Q}_{i,j} f_j^{d-i}\\
        &=\sum_{j=1}^{r} \tilde{Q}_{j} f_j^{d-r}\, ,
    \end{align*}
        where  $\tilde{Q}_{i,j}$ is defined similarly to \eqref{eq:def-gij}, and $\alpha_jL'(f_j^d)=\tilde{Q}_{j} f_j^{d-r}$.
        Observe that, as in \eqref{eq:def-gij}, \[\deg(\tilde{Q}_{i,j}f_j^{r-i})\leq  r^2\Delta + (i\Delta-i)+(r-i)\Delta \,.\] Hence, $\deg(\tilde{Q}_j)\leq  (r^2+r)\Delta$. Since the $f_i$ are   non-associate, 
        \[
        d-r > (r+1)^4\Delta-r= (r^2+2r+1)^2\Delta -r \geq (r^2+r)((r^2+r)\Delta +1)\, ,
        \]
        and $p=0$ or $p\geq 2rd\Delta> r((r^2+r)\Delta + d\Delta)$, 
        \autoref{cor:ABC} implies that $\tilde{Q}_j\equiv 0$ for all $j$. Consequently, $L'(f_j^d)=0$ for all $j$. \autoref{thm:ode} implies that the kernel of $L'$ (up to degrees $2rd\Delta$) is spanned by $f_j^d$, and in particular that $Q_r\neq 0$. Observe now that the operator $Q_r\cdot\tilde{L} - P_r\cdot L'$ has order at most $r-1$ and its kernel contains all $f_j^d$. Hence, by \autoref{thm:ode} it must be identically zero. As $\gcd(P_0,\ldots,P_r)=\gcd(Q_0,\ldots,Q_r)=1$ the claim follows. 
\end{proof}

\begin{corollary}\label{cor:unique-L}
Let $p,d$ satisfy the requirements of \autoref{cla:L-unique}. 
Then there exists a unique operator 
\[
L = \sum_{i=0}^{r} Q_i(x)\,\nabla^i
\]
satisfying $\deg Q_i \le r^2\Delta$ for all $i$, 
$\gcd(Q_0,\ldots,Q_r) = 1$, $Q_r$ monic, 
and $L(f) = 0$.
Furthermore, $L$ can be computed in time $\poly(r,d,\Delta)$.
\end{corollary}

\begin{proof}
    The claim regarding the uniqueness follows immediately from \autoref{cla:L-unique} and the fact that we chose $Q_r$ to be monic and of minimal degree.

    Since we can find, in polynomial time, a basis for the solution space of \eqref{eq:L-lin-system}, by simple diagonalization we can find a solution that minimizes $\deg(Q_r)$ such that $Q_r$ is monic. 
\end{proof}

We next show how to extract the polynomials $f_i$ given $L$. 

\subsection{Properties of the Solution Space of $L$}\label{sec:L-properties}


\begin{claim}\label{cla:no-other-d-power}
    Let $p$ satisfy the requirements of \autoref{cla:L-unique}. If $r^2 \le d+1$ then the monic $d$-th powers contained in $\Ker_{\le d\Delta}(L)$ are precisely $f_1^d,\ldots,f_r^d$.
\end{claim}

\begin{proof}
    Assume there is a polynomial $h(x)$ such that $L(h(x)^d)=0$ and $h$ is not a scalar multiple of any of the $f_j$. Since $\Span\{f_i^d\}_{i\in[r]}=\Ker_{\le d\Delta}(L)$, there are scalars $\gamma_i$ such that $\Sigma_{i}\gamma_if_i^d=h^d$. However, by choice of $d$, the assumption on $p$,  and since the polynomials $\{f_i\}\cup\{h\}$ are   non-associate, we get a contradiction to  \Cref{thm:abc_constants}. 
\end{proof}

Thus, our task is finding all polynomials that are $d$-th powers in $\Ker_{\le d\Delta}\!{(L)}$. We next show that each root of each $f_j$ is also a root of $Q_r$.

\begin{definition}[$\ord_z(g)$]
    Let $g(x)$ be a polynomial. We define \emph{$\ord_z(g)$} to be the multiplicity of $z$ as a root of $g(x)$. Similarly, for an irreducible polynomial $\phi(x)$ we define $\ord_\phi(g)$ to be the largest power $e$ such that $\phi^e|g$.
\end{definition}

\begin{claim}\label{cla:qr-zero}
    Let $p$ satisfy the requirements of \autoref{cla:L-unique}. If $d\geq r$ then the leading coefficient $Q_r(x)$ of $L=\Sigma_{i=0}^rQ_i(x)\nabla^{(i)}$ vanishes at every root $z$ of every $f_i$.
\end{claim}

\begin{proof}
Let $z\in\overline{\F}$ be a root of $f_j$, and write $m:=\ord_z(f_j^d)$. Then $r\le d\le m\le d\Delta<p$ in positive characteristic. Writing $f_j^d=(x-z)^mu(x)$ with $u(z)\ne0$, we obtain \[ \ord_z\!\left(\nabla^i(f_j^d)\right)=m-i \qquad (0\le i\le r).\]  Suppose for contradiction that $Q_r(z)\ne0$. Since $L(f_j^d)=0$, \[ Q_r\nabla^r(f_j^d) = -\Sigma_{i=0}^{r-1}Q_i\nabla^i(f_j^d). \] The left-hand side has order exactly $m-r$ at $z$, whereas every term on the right has order at least \[ m-(r-1)=m-r+1, \] 
a contradiction. 
 
\end{proof}

\begin{corollary}\label{cor:irred-factor}
    Each irreducible factor, over $\F$, of each $f_i$ is an irreducible factor of $Q_r$. Furthermore, the factorization of $Q_r$ can be computed efficiently.
\end{corollary}

\begin{proof}
    This follows immediately from \Cref{cla:qr-zero} by considering minimal polynomials of roots. The claim about efficiency follows from known factorization algorithms that we cite in \Cref{sec:factorization} below. Specifically, it follows from \Cref{thm:lll} and \Cref{thm:finite-factor}, depending on the underlying field.
\end{proof}

\begin{remark}\label{rem:slower-alg}
\autoref{cor:irred-factor} tells us that, upon factorizing the leading coefficient $Q_r(x)$, we obtain a set of $m$ irreducible factors $\phi_1,\ldots,\phi_m$ of $Q_r$. 

Hence, we can iterate over all possible multiplicity vectors $\ve$, of which there are at most
${{m+\Delta} \choose {\Delta}} \approx (r^2 \Delta)^{\Delta}$, to construct candidate polynomials
\[
g_{\boldsymbol{e}}(x) = \prod_{j \in [m]} \phi_j^{e_j}.
\]
For each such $g_{\boldsymbol{e}}$, we can apply $L$ to $g_{\boldsymbol{e}}^d$ to test whether 
$g_{\boldsymbol{e}}^d \in \Ker_{\le d\Delta}(L)$. This yields an algorithm with running time 
$\poly(n, d, (r \Delta)^{\Delta})$.

However, our goal is to design an algorithm whose running time is polynomial in all the parameters.
\end{remark}

\subsubsection{Polynomial Factorization }\label{sec:factorization}

We state two well known results on polynomial factorization. The first is the famous LLL algorithm for factorization over $\Q$. 

\begin{theorem}[Factorization over $\Q$, \cite{LLL}]\label{thm:lll}
    Let $f\in \Q[x]$ have degree $d$ and bit complexity $B_{\text{in}}$. Then, there is a deterministic algorithm that outputs all irreducible factors of $f$ whose running time is $\poly(d,B_{\text{in}})$.
\end{theorem}

For the following theorem, see, e.g. \cite[Section 9]{GathenShoup1992-deterministic-factor}.

\begin{theorem}[Factorization over Finite Fields]\label{thm:finite-factor}
    Let $\F_q$ be a field of size $q=p^k$ and characteristic $p>0$.
    Let $f(x)\in \F_q[x]$ be a polynomial of degree $d$. Then, there is a deterministic algorithm that outputs all irreducible factors of $f$ in time $\poly(p,\log q,d)$.
\end{theorem}

\subsection{Efficiently Recovering the $f_i$}\label{sec:recover-fi}

We first note that a basis for $\Ker_{\le d\Delta}(L)$ can be computed in polynomial time. 
Indeed, consider $d\Delta+1$ unknowns $a_i$ and consider the linear system of equations
\begin{equation}
    \label{eq:L-sol-space}
    \sum_{j=0}^{r}Q_j\nabla^j\!\left(\sum_{i=0}^{\Delta d}a_i x^i\right)\equiv 0 \,.
\end{equation}
Clearly, this system is solvable in time $\poly(d,\Delta)$. 
Furthermore, since $\deg(f_j^d)\leq d\Delta$, each $f_j^d$ lies in the solution space. 
As these polynomials span $\Ker_{\le d\Delta}(L)$, we are guaranteed to find a basis for $\Ker_{\le d\Delta}(L)$ in this process.
    
Let $h_1,\ldots,h_r$ be the basis found for $\Ker_{\le d\Delta}(L)$. Thus, 
\[
\Ker_{\le d\Delta}(L)=\Span\{f_1^d,f_2^d,\ldots,f_r^d\}
       =\Span\{h_1(x),h_2(x),...,h_r(x)\}, 
       \quad \forall i \in[r],\; \deg(h_i(x))\le\Delta d \,. 
\] 

Therefore, our goal is to find all $f_j^d$ given the $h_i$. 
We achieve this by the following observation: 
if some $g\in\Ker_{\le d\Delta}(L)$ has a root $z$ such that $\ord_z(g)\geq d$, then $z$ must be a root of one of the $f_j$. 

\begin{claim}\label{cla:high-mul-root}
    Let $p$ satisfy the requirements of \autoref{cla:L-unique}.
    Assume $d>2r^2\Delta$. 
    If $z$ is a root of order $t>r^2\Delta+r$ of some polynomial 
    $h\in \Span\{f_j^d :\ j\in [r]\}$, then it is a root of one of the $f_j$.
\end{claim}

\begin{proof}
    Denote $h=\Sigma_{i\in[r]}c_if_i^d$. Assume, w.l.o.g., that $c_1=1$.  
    Consider the order $(r-1)$ Wronskian 
    \[
        W(f_1^d,...,f_r^d)\in\F[x],
        \quad W:=\det(\nabla^i f_j^d)_{i\in [[r-1]],j\in [r]}\,.
    \] 
    As $c_1=1$, we can replace the first column of the matrix in the Wronskian 
    with the vector whose $i$-th entry is $(\nabla^i h)$, without affecting $W$.
    Since $\ord_z(h)=t>r$, we have $\ord_z(\nabla^i h)=t-i\geq t-(r-1)$.  
    Consequently, $z$ is a root of $W$ of order at least $t-r+1$. 
    As in the proof of  \autoref{Wronski}, the Wronskian factorizes as 
    $\Pi_{i\in[r]}f_i^{d-r+1}W'$, where $W'$ is a nonzero polynomial of degree at most $r^2\Delta$ (since the $f_i^d$ are linearly independent, $W$ is a nonzero polynomial). 
    Since 
    \[
        \ord_z(W)\geq t-r > r^2\Delta\geq  \deg(W')\,,
    \]
    we conclude that $\ord_z(\Pi_{i\in[r]}f_i^{d-r+1})>0$. 
    Hence, $z$ is a root of one of the $f_i$.
\end{proof}

The following simple corollary is at the heart of our algorithm.

\begin{corollary}\label{cor:fi-span-high-mul-roots}
Let $d$ be as in \autoref{cla:high-mul-root}.
The subspace of all polynomials that have a root at $z$ of order at least $d$ 
\[
\{h\in \Span\{f_1^d,\ldots,f_r^d\} \mid \ord_z(h)\ge d\}
\]
is spanned by all the $f_i^d$ that have $z$ as a root.     
\end{corollary}

\begin{proof}
    Assume w.l.o.g. that among $\{f_i^d\}$ only $f_1^d,...,f_k^d$ 
    have $z$ as a root, and hence it is a root of order at least $d$ at each of them. 
    Assume for contradiction that some nonzero polynomial 
    $h\in \Span\{f_{k+1}^d,...,f_r^d\}$ satisfies $\ord_z(h)\geq d$. 
    Applying \autoref{cla:high-mul-root} to $h\in \Span\{f_{k+1}^d,...,f_r^d\}$ 
    shows that $z$ must be a root of some $f_j$ for $k<j\leq r$, a contradiction.
\end{proof}

\begin{corollary}\label{cor:fi-span-high-mul-factors}
    Let $\phi\in \F[x]$ be an irreducible polynomial. Let $d$ be as in \autoref{cla:high-mul-root}. The subspace of all polynomials that have $\phi^d$ as a factor
is spanned by all the $f_i^d$ such that $\phi$ is a factor of $f_i$. 
\end{corollary}

\begin{proof}
    Apply \autoref{cor:fi-span-high-mul-roots} to any root of $\phi$.
\end{proof}

Armed with \autoref{cor:fi-span-high-mul-factors}, 
our next step is to sieve the polynomials in $\Ker_{\le d\Delta}(L)$ 
that have a factor of high multiplicity.

\begin{claim}\label{cla:basis-for-factor}
    Given an irreducible $\phi\in \F[x]$ such that $\deg(\phi)\leq \Delta$, and $e\leq d$, we can find in time $\poly(d,r,\Delta)$ 
    a basis for the space of polynomials
    \[
    \{\, g \in \Span\{h_1,\ldots,h_r\} \;\big|\; \ord_{\phi}(g)\geq e \,\}.
    \]
\end{claim}

\begin{proof}
    Let $b_1,\ldots,b_r$ and $a_0,\ldots,a_{\Delta d-e\deg(\phi)}$ be indeterminates. 
    Solve the homogeneous linear system 
    \[
        \phi^e\cdot \left(\sum_{i=0}^{\Delta d-e\deg(\phi)}a_ix^i\right) 
        = \sum_{i=1}^{r}b_i h_i\;.
    \]
    Any basis for the solution space immediately gives a basis for the requested subspace.
\end{proof}

We recover the monic polynomials $f_i^d$ by performing a DFS (with pruning) on a suitable tree. We first give the main idea of the algorithm.
Let $\{\phi_j\}_{j\in [m]}$ denote the irreducible factors of $Q_r$. 
As proved in \autoref{cor:irred-factor}, this set contains all factors of each $f_i$. 
Hence, each $f_i$ corresponds to an exponent vector $\ve_i$ defined by 
$\ve_{i,j} = \ord_{\phi_j}(f_i)$. 
Equivalently,
\[
    f_i(x) = \prod_{j\in [m]} \phi_j^{e_{i,j}}.
\]
Clearly, for each $i$, we have $\Sigma_j e_{i,j} \deg(\phi_j) = \deg(f_i) \le \Delta$.

Consider now a rooted tree of depth $\Delta$, where each node has $m$ children, 
and the edge to the $j$-th child is labeled by $\phi_j$. 
Every  root-to-leaf path in the tree corresponds to an exponent vector, counting the multiplicities of the $\phi_j$ encountered along the path. 
Label each node with the exponent vector defined by the path from the root to that node, 
and associate to each such node the subspace of polynomials in $\Ker_{\le d\Delta}(L)$ 
that have $\phi_j^d$ as a factor with multiplicity equal to the number of times 
$\phi_j$ appears along the path. 

Note, however, that this tree is highly redundant, since many distinct nodes correspond 
to the same subspace. Moreover, the tree has $m^\Delta$ leaves. 

Our algorithm performs a ``smart'' depth-first search (DFS) on this tree to recover the $f_j$ in the lexicographic order of their exponent vectors. 
It is important to note that the algorithm visits only those vertices that are guaranteed to lead to previously undiscovered polynomials, and thus visits in total 
only $\poly(r,\Delta)$ nodes overall.

\begin{theorem}\label{thm:dfs}
Let $\{\phi_1,\ldots,\phi_m\}$ be the set of monic irreducible factors of $Q_r$. 
Assume $d\geq (r+1)^4\Delta$ and the characteristic is $p=0$ or $p>2dr\Delta$.  
Then there exists a deterministic algorithm that, given a basis 
$\{h_1,\ldots,h_r\}$ for $\Ker_{\le d\Delta}(L)$, recovers all $f_i^d$ 
in time $\poly(m,r,\Delta,d)$.
\end{theorem}

We first give the pseudocode of the algorithm (\Cref{alg:dfs}) and then prove that it satisfies the claim in the theorem.

\begin{algorithm}[H] \caption{\textsc{DFS-Recover-$f_i^d$}} \label{alg:dfs} \begin{algorithmic}[1] \Require A basis $\mathcal B_{\mathbf 0}$ for $\mathcal K=\Ker_{\leq d\Delta}(L)$; monic irreducible polynomials $\phi_1,\ldots,\phi_m$ 
\Ensure The set $U=\{f_1^d,\ldots,f_r^d\}$ \State $U\gets\varnothing$ 
\Procedure{DFS}{$\mathbf e,\mathcal B_{\mathbf e}$} 
\State $M_{\mathbf e}\gets\Pi_{j=1}^m\phi_j^{e_j}$ 
\If{$\dim\Span(\mathcal B_{\mathbf e})=1$} 
\State Let $g$ be the unique monic representative of $\Span(\mathcal B_{\mathbf e})$ 
\State $U\gets U\cup\{M_{\mathbf e}^dg\}$ 
\State \Return 
\EndIf 
\If{$1\in\Span(\mathcal B_{\mathbf e})$ \textbf{and} $M_{\mathbf e}^d\notin U$} \label{step:constant-test}
\State $U\gets U\cup\{M_{\mathbf e}^d\}$ 
\EndIf 
\For{$j=1$ \textbf{to} $m$}
    \State $A_j(\mathbf e)\gets
      \{g\in\operatorname{Span}(\mathcal B_{\mathbf e}): \phi_j^d\mid g\}$
    \State $U(\mathbf e)\gets
      \{q/M_{\mathbf e}^d:q\in U,\ M_{\mathbf e}^d\mid q\}$
    \If{$A_j(\mathbf e)\not\subseteq\operatorname{Span}U(\mathbf e)$}\label{step:edge-condition}
\State Let $\mathcal A_j(\mathbf e)$ be a basis for $A_j(\mathbf e)$ 
\State $\mathcal B_{\mathbf e+\mathbf e_j} \gets \{\,g/\phi_j^d: g\in\mathcal A_j(\mathbf e)\,\}$ 
\State \Call{DFS}{ $\mathbf e+\mathbf e_j, \mathcal B_{\mathbf e+\mathbf e_j}$} 
\EndIf 
\EndFor 
\EndProcedure 
\State \Call{DFS}{$\mathbf 0,\mathcal B_{\mathbf 0}$} 
\State \Return $U$ \end{algorithmic} \end{algorithm}

\begin{proof}

The tree is composed of nodes that are accessed during recursive calls.
We naturally identify nodes in the tree by their multiplicity vectors $\ve$, where the root corresponds to the all-zero vector. 
Note that different calls may have the same multiplicity vector. 

Denote by $T_\text{DFS}$ the tree defined by the DFS algorithm. That is, there is an edge between a node that is labeled by the multiplicity vector $\ve$ to its child $\ve+\ve_j$ only if the condition in Step~\ref{step:edge-condition} holds.

For a vector $\ve=(e_1,\ldots,e_m)$, write \[ M_{\mathbf e}:=\prod_{j=1}^m\phi_j^{e_j} \] and define \[ I(\mathbf e) := \{\,i\in[r]:M_{\mathbf e}\mid f_i\,\}. \] Set \[ V(\mathbf e) := \Span\left\{ \left(\frac{f_i}{M_{\mathbf e}}\right)^d: i\in I(\mathbf e) \right\}. \] We first prove by induction on the recursion depth that, at every recursive call, \begin{equation}\label{eq:dfs-invariant} \Span(\mathcal B_{\mathbf e})=V(\mathbf e). \end{equation} 
At the root, this follows from 
\[ 
\Span(\mathcal B_{\mathbf 0}) = \Ker_{\leq d\Delta}(L) = \Span\{f_1^d,\ldots,f_r^d\}. 
\] 

Suppose that \eqref{eq:dfs-invariant} holds at $\mathbf e$. The family \[ \left\{ \left(\frac{f_i}{M_{\mathbf e}}\right)^d: i\in I(\mathbf e) \right\} \] is clearly linearly independent. Hence this family is in fact a basis of $V(\mathbf e)$. 
By \Cref{cor:fi-span-high-mul-factors}, applied to the quotient polynomials $f_i/M_{\mathbf e}$, 
\[ 
A_j(\mathbf e) = \Span\left\{ \left(\frac{f_i}{M_{\mathbf e}}\right)^d: M_{\mathbf e}\phi_j\mid f_i \right\}. 
\] 
After division by $\phi_j^d$, the basis constructed by the algorithm therefore spans 
\[ 
V(\mathbf e+\mathbf e_j). 
\] 
This proves the invariant \eqref{eq:dfs-invariant}. 

We next justify the constant test in \Cref{step:constant-test}. Suppose that $1\in V(\mathbf e)$. If no quotient $f_i/M_{\mathbf e}$ is constant, then \[ 1 = \sum_{i\in I(\mathbf e)} c_i \left(\frac{f_i}{M_{\mathbf e}}\right)^d \] is a nontrivial relation among the $d$-th powers of the pairwise non-associate polynomials \[ 1 \quad\text{and}\quad \left\{\frac{f_i}{M_{\mathbf e}}: i\in I(\mathbf e)\right\}. \] There are at most $r+1$ such polynomials, so this contradicts \Cref{thm:abc_constants}. Hence one quotient is constant. Since all polynomials involved are monic, that quotient equals $1$, and \[ f_i=M_{\mathbf e}. \] Thus the constant test adds an actual polynomial $f_i^d$ to $U$. 

If $\dim V(\mathbf e)=1$, the basis displayed above contains exactly one element, say$ \left(\frac{f_i}{M_{\mathbf e}}\right)^d$. It is monic and therefore is the unique monic representative of $V(\mathbf e)$. Consequently the algorithm adds 
\[ M_{\mathbf e}^d \left(\frac{f_i}{M_{\mathbf e}}\right)^d = f_i^d. \] 

We now prove that, before a call at $\mathbf e$ returns, every $f_i^d$ with $M_{\mathbf e}\mid f_i$ has been added to $U$. We use induction on \[ \Delta-\deg(M_{\mathbf e}). \] The one-dimensional case was handled above. Otherwise, let $f_i^d$ be compatible with $M_{\mathbf e}$ and suppose that it has not yet been recovered. If $f_i=M_{\mathbf e}$, the constant test recovers it. Otherwise, choose an irreducible factor $\phi_j$ of $f_i/M_{\mathbf e}$. Then \[ \left(\frac{f_i}{M_{\mathbf e}}\right)^d \in A_j(\mathbf e). \] 

The set $U(\mathbf e)$ consists precisely of quotient powers corresponding to already recovered polynomials compatible with $M_{\mathbf e}$. Since the quotient powers form a basis of $V(\mathbf e)$, the unrecovered vector \[ \left(\frac{f_i}{M_{\mathbf e}}\right)^d \] does not belong to $\Span U(\mathbf e)$. Therefore \[ A_j(\mathbf e)\not\subseteq\Span U(\mathbf e), \] and the algorithm makes the recursive call at $\mathbf e+\mathbf e_j$. Since \[ \deg(M_{\mathbf e+\mathbf e_j}) > \deg(M_{\mathbf e}), \] the induction hypothesis implies that this call recovers $f_i^d$ before it returns. Hence every compatible polynomial is recovered before the current call returns. In particular, the root call recovers all $f_i^d$. 

We now bound the number of recursive calls. By the condition in Line~\ref{step:edge-condition}, each non-root call starts with an unrecovered polynomial compatible with \(M_{\ve}\). By correctness, this polynomial is recovered before the call returns. In particular, each leaf of \(T_{\mathrm{DFS}}\) recovers at least one new polynomial, so the tree has at most \(r\) leaves. Moreover, each edge increases \(\deg(M_{\mathbf e})\) by at least one, and this degree is always at most \(\Delta\). Thus the tree has depth at most \(\Delta\), and therefore at most \(r\Delta\) non-root nodes.

Each call requires \(\poly(m,r,d,\Delta)\) field operations, so the total number of field operations is also \(\poly(m,r,d,\Delta)\).
\end{proof}

\subsection{The Univariate Reconstruction Algorithm}\label{sec:uni-reconstruct}

We now give the full reconstruction algorithm in the univariate case. In what follows we assume that the characteristic of $\F$ is either $p=0$ or $p>2rd\Delta$. We further assume that $d>(r+1)^4\Delta$.  \Cref{thm:univariate-reconstruct} is an immediate consequence of \Cref{thm:uni-reconstruct} below.

\begin{algorithm}[H]
\caption{\textsc{Recover Univariate Sums of Powers}}
\label{alg:recover-depth4-uni}
\begin{algorithmic}[1]
\Require Black-box access to $f(x)\in\F[x]$; parameters $r,d,\Delta$.
\Ensure Representation $f(x)=\Sigma_{i=1}^r \alpha_i\,f_i(x)^d$ with monic $f_i$.
\State \textbf{Interpolation.} Since $\deg f\le d\Delta$, query $f$ at $d\Delta+1$ points and interpolate $f(x)=\Sigma_{i=0}^{\Delta d}\beta_i x^i$.\label{alg:uni:step:interpolate}
\State \textbf{Perfect-power test.} Decide if $f=\alpha g^d$ for some scalar $\alpha$ and polynomial $g$  with $\deg(g)\le\Delta$ (e.g., via repeated $\gcd(f,\nabla f)$ / squarefree decomposition). If yes, rescale $g$ to be monic and \Return $\{(\alpha_1,f_1)\}=\{(\alpha,g)\}$.\label{alg:uni:step:perfect-power}
\State \label{alg:uni:step:find-L}\textbf{Annihilating operator.} For $s=1,2,\ldots,r$, solve the homogeneous system \[ \sum_{i=0}^{s} \left( \sum_{j=0}^{s^2\Delta}\gamma_{i,j}x^j \right) \nabla^if(x) \equiv0. \] Let $r'$ be the first value of $s$ for which a nonzero solution exists. Choose any nonzero solution $(R_0,\ldots,R_{r'})$, divide all its entries by $\gcd(R_0,\ldots,R_{r'})$, and rescale so that $R_{r'}$ is monic. Set \[ L:=\sum_{i=0}^{r'}R_i(x)\nabla^i. \]
\State \textbf{Factor.}
       Using \autoref{thm:lll} or \autoref{thm:finite-factor}, compute the set $\{\phi_j\}$ of all irreducible factors of degree at most $\Delta$ of $R_{r'}(x)$. By \autoref{cor:irred-factor}, $\{\phi_j\}$ contains all irreducible factors of the $f_i$’s.\label{alg:uni:step:factor}
\State \textbf{Solution space.}
       Compute a basis $\{h_1,\ldots,h_{r'}\}$ of $V=\Ker_{\le d\Delta}(L)$ by solving \eqref{eq:L-sol-space}.\label{alg:uni:step:solution-space}
\State \textbf{Recover the $f_i^d$.}
       Run \Cref{alg:dfs} on input $(\{h_i\},\{\phi_j\})$ to obtain $U=\{f_1^d,\ldots,f_{r'}^d\}$.\label{alg:uni:step:dfs}
\State \textbf{Recover the coefficients $\alpha_i$.}
       Solve the linear system (with unknown $\alpha_i$) $f(x)=\Sigma_{i=1}^{r'}\alpha_if_i^d$.
\State For each monic $f_i^d\in U$, compute the unique monic polynomial $f_i$.
\State \Return $\{(\alpha_i,f_i)\}_{i=1}^{r'}$.
\end{algorithmic}
\end{algorithm}

\begin{theorem}[Correctness and  complexity of the univariate reconstruction]\label{thm:uni-reconstruct}
Let $r,d,\Delta\in \N$ with $d>(r+1)^4\Delta$.
Let $\F$ be a field of characteristic $p=0$ or $p>2rd\Delta$.

Suppose a nonzero univariate polynomial $f\in\F[x]$ admits a minimal representation
\[
f(x)=\sum_{i=1}^{r'} \alpha_i\,f_i(x)^d,
\qquad
\deg(f_i)\le \Delta, \qquad r'\le r,
\]
where each $f_i$ is monic, and the polynomials $\{f_i\}_{i=1}^{r'}$ are   non-associate. 
Then \Cref{alg:recover-depth4-uni} outputs, in time $\poly(r,\Delta,d,p)$, 
a set $\{(\alpha'_i,f'_i)\}_{i=1}^{r'}$ such that, for some permutation $\pi$,
\[
f(x)=\sum_{i=1}^{r'} \alpha'_i\,f'_i(x)^d
\qquad\text{and}\qquad
f'_i=f_{\pi(i)},\ \ \alpha'_i=\alpha_{\pi(i)}.
\]
\end{theorem}

\begin{proof} Since $\deg(f)\le d\Delta$ and $|\F|>d\Delta$ in the finite-field case, $f$ can be interpolated from $d\Delta+1$ evaluations. 

If the minimal representation has one summand, the perfect-power test recovers it. Assume therefore that its minimal top fan-in is $r'\ge2$. 

For $s=r'$, \Cref{cor:OpL} provides a nonzero solution to the annihilator system with coefficient degrees at most ${r'}^2\Delta$. 
For $s<r'$, no nonzero solution exists: otherwise, as it would give an operator of order at most $s<r'$ and coefficient degrees at most ${r'}^2\Delta$ annihilating $f$, contradicting \Cref{cla:L-unique} applied to the minimal $r'$-term representation. Thus the first successful value of $s$ is exactly $r'$.  \Cref{cla:L-unique} implies that we have recovered the canonical operator $L$. 

By \Cref{cor:irred-factor}, every irreducible factor of every $f_i$ appears among the factors of the coefficient of $\nabla^{r'}$ in $L$. Equation \eqref{eq:L-sol-space} computes a basis for \[ \Ker_{\le d\Delta}(L) = \Span\{f_1^d,\ldots,f_{r'}^d\}. \] Then \Cref{thm:dfs} recovers the set \[ U=\{f_1^d,\ldots,f_{r'}^d\}. \] The polynomials $f_1^d,\ldots,f_{r'}^d$ are linearly independent by minimality of the original representation. 
Therefore the system \[ f=\sum_{i=1}^{r'}\alpha_if_i^d \] has a unique solution, and it recovers all coefficients $\alpha_i$. Factoring each recovered power and dividing its factor multiplicities by $d$ gives the unique monic polynomial $f_i$. Finally, uniqueness of the resulting representation, up to a permutation, follows from \Cref{cor:uniq repres of high powers}. All interpolation, linear-algebra, gcd, and factorization steps have polynomial bit complexity under the stated computational assumptions. 
\end{proof}


\section{The Lifting Theorems} \label{sec:main_theorems}



In this section we state and prove the formal versions of the different claims in \Cref{thm:intro:lifting}. We start with the PIT lifting claim and then give the two reconstruction results.

\subsection{PIT for High Powers} \label{subsec:main_thm_pit}

Here we give the formal statement and proof of \Cref{thm:intro:lifting}\ref{item:intro-main-pit}.

\medskip

Let $P = \Sigma^r_{i=1} \left(T_i\right)^d$ be a circuit where each $T_i \in \cC$ for a polynomial class $\cC$, and assume that the polynomials $\{T_i\}_{i=1}^r$ are pairwise non-associates. To apply the ABC Theorem for high powers (via \autoref{thm:abc_constants}), it suffices to project these $T_i$-s to univariate polynomials while preserving their pairwise non-associate property. We note that any interpolating set (that is, a hitting set for $\Sigma^2 \cC$), distinguishes any two non-associate polynomials. By composing the polynomials with an appropriate curve through such an interpolating set we achieve the sought after reduction. This approach is formalized in the following theorem.

\begin{theorem}[Black-box PIT for High Powers, formal version of \Cref{thm:intro:lifting}\ref{item:intro-main-pit}] \label{thm:bb pit for high powers}
    Let $r,d, \Delta \in \N$ such that $d + 1 \geq (r-1)^2$ and let $\cC$ be some polynomial class which is closed under scalar multiplications. Assume there exists a hitting set $\cH$ for $\Sigma^2 \cC_{\Delta}$ of size $s$ and assume $\F$ is a field of characteristic $p = 0$ or $p > (r-1) d \Delta s$. 

    Order the elements of \(\mathcal H=\{\mathbf u_i\}_{i=1}^s\) arbitrarily, and choose pairwise distinct elements \(a_1,\ldots,a_s\in\mathbb F\). Let \(H(t):\mathbb F\to\mathbb F^n\) be the unique polynomial map whose coordinates have degree at most \(s-1\) and which satisfies \(H(a_i)=\mathbf u_i\) for every \(i\in[s]\).
     
    
    Then, for any $S \subseteq \F$, any nonzero polynomial $P \in \Sigma^r \wedge^d \cC_{\Delta}$ vanishes on at most $(s-1) \cdot \Delta \cdot d$ points from the ordered set $H(S)$.\footnote{Note that since this set is ordered, even if $H(\alpha_i)= H(\beta_i) = c$ for $\alpha_i\neq\beta_i \in S$ the two values $(\alpha_i,c)$ and $(\beta_i,c)$ are distinct and are counted separately.}
\end{theorem}

\begin{proof}
    Let $0 \not\equiv P \in \Sigma^r \wedge^d \cC_{\Delta}$. We prove the theorem by induction on the top fan-in $r$.

    For $r = 1$, we have $P = T^d \in \wedge^d \cC_{\Delta}$. Since $\cH$ is a hitting set for $\Sigma^2 \cC_{\Delta}$ it follows that $T \circ H(T) \not \equiv 0$ which implies $P \circ H(T) \not \equiv 0$. Thus $P \circ H$ is a nonzero univariate polynomial of degree at most $(s-1) \cdot d \cdot \Delta$. This implies the case $r=1$.
    
    For $r \geq 2$, we need the following claim.

    \begin{claim} \label{claim:M_t_non_associates} 
    Let $A,B \in \cC_{\Delta}$ be nonassociates. Then $A \circ H \not\sim B \circ H$. 
    \end{claim} 
    
    \begin{proof} 
    Indeed, for every $\alpha \in \F$, the fact that $A \not\sim B$ implies that $A - \alpha B \not\equiv 0$. Since $A - \alpha B \in \Sigma^2 \cC_{\Delta}$, we obtain $\big(A - \alpha B\big) \circ H \not\equiv 0$. 
    \end{proof}

    Let $P = \sum_{i=1}^r \alpha_i T_i^d$, where $T_i \in \cC_{\Delta}$. We may assume that the $T_i$'s are pairwise nonassociates, since otherwise we can reduce the top fan-in. 

    For $r = 2$, we get \[\alpha_1 T_1^d \circ H \equiv - \alpha_2T_2^d\circ H  \iff -\sqrt[d]{\frac{\alpha_1}{\alpha_2}} T_1 \equiv T_2 \iff  T_1 \circ H \sim T_2 \circ H  .\] By \cref{claim:M_t_non_associates} this is impossible.
    
    For $r \ge 3$,  \autoref{claim:M_t_non_associates} implies that  the univariate polynomials $T_i \circ H$, for $i \in [r]$, are pairwise nonassociates. Therefore, by \autoref{thm:abc_constants}, the univariate polynomial $P \circ H$ is nonzero and has degree at most $(s-1) \cdot d \cdot \Delta$. This proves the claim.
\end{proof}

\begin{remark}
    We note that the above theorem only requires the characteristic bound for applying \cref{thm:abc_constants}, i.e. for the case where $r \geq 3$. Thus, for $r \leq 2$ it works over any field. This will become useful in future sections.
\end{remark}

\subsection{Reconstruction for High Powers} \label{subsec:main_thm_recon} 

In this section, we prove the reconstruction lifting theorem \Cref{thm:intro:lifting}\ref{item:intro-main-total}. 
For this we shall need the notion of a non-adaptive reconstruction algorithm, which is captured by the notion of a reconstruction set that we define next.

\begin{definition}[{Reconstruction Set}] \label{def:recon_set}
Let $\mathcal{C}$ be a class of algebraic circuits. A finite, indexed set $\mathcal{S} \subseteq \F^n$ is called a \emph{reconstruction set} for $\mathcal{C}$ if there exists a reconstruction algorithm with the following
property: for every polynomial $f \in \mathcal{C}$, given only the
indexed evaluations set
\begin{equation*}
\left\{\big(\balp, f(\balp)\big) : \balp \in \mathcal{S}\right\},
\end{equation*}
the algorithm outputs a circuit $C'$ from the class $\mathcal{C}$ that
computes $f$.
\end{definition}

Note that such a set is indeed equivalent to the notion of a non-adaptive reconstruction algorithm.


\medskip

For our algorithm we shall require the slightly stronger property of a  robust reconstruction set.

\begin{definition}[{Robust Reconstruction Set}] \label{def:robust_recon_set}
Let $\mathcal{C}$ be a class of algebraic circuits, and let
$0 \leq \epsilon < 1$. A finite set $\mathcal{S} \subseteq \F^n$ is called
an \emph{$\epsilon$-robust reconstruction set} for $\mathcal{C}$ if every
subset $\mathcal{S}' \subseteq \mathcal{S}$ satisfying
\begin{equation*}
|\mathcal{S}'|
\geq
 (1-\epsilon)|\mathcal{S}| 
\end{equation*}
is a reconstruction set for $\mathcal{C}$.
\end{definition}

By passing a curve through a reconstruction set and evaluating it on many points we obtain a robust reconstruction set. We give the formal claim and the simple proof in \Cref{subsec:recon_to_robust}.

\medskip

Our algorithm relies on two specialized sets, $\hp$ and $\hr$, that satisfy the properties given in the next two definitions.

\begin{definition}[Properties of $\hp$] \label{def:hp}
    Let $r, d, \Delta \in \mathbb{N}$ and let $\cC$ be a polynomial class. The (multi-) set $\hp(r,d,\Delta)$ used in \autoref{alg:meta_recon} to reconstruct a sum of $r$ high powers of polynomials from $\cC_{\Delta}$ must satisfy the following properties:
    \begin{enumerate}
        \item \label{item:hs prod prop of hp}  Let $0 \not \equiv g \in \Pi^r\cC_{\Delta}$, then $g(\bu) = 0$ on at most an $0 \leq \epsprod < 1$ fraction of points from $\hp(r,d,\Delta)$.
        \item \label{item:hs s2 prop of hp}  Let $0 \not \equiv g \in \Sigma^2 \wedge^d \cC_{\Delta}$, then $g(\bu) = 0$ on at most an $0 \leq \epsst < 1$ fraction of points from $\hp(r,d,\Delta)$.
        \item \label{item:size req of hp} The error parameters satisfy $\epsprod + \binom{r}{2}\epsst < 1$, and the size of the set satisfies 
        \[
            |\hp(r,d,\Delta)| \geq \frac{1}{1 - \epsprod - \binom{r}{2}\epsst}.
        \]
    \end{enumerate}
    When $r$, $d$ and $\Delta$ are clear from context we simply write $\hp$.
    \defsymbol
\end{definition}

\begin{definition} [{Properties of $\hr$ and $s_{\text{recon}}$}] \label{def:hr}
    Let $\Delta, r \in \mathbb{N}$ and let $\cC$ be a polynomial class. The set $\hr(r,\Delta)$ used in \autoref{alg:meta_recon} to reconstruct a sum of $r$ high powers of polynomials from $\cC_{\Delta}$ must satisfy the following properties:
    \begin{enumerate}
        \item \label{item:recon prop of hr}  For $0 \leq \epsilon_{\text{recon}} < 1$, $\hr(r,\Delta)$ is an $\epsilon_{\text{recon}}$-robust reconstruction set for polynomials in $\cC_{\Delta}$.
        \item \label{item:hs s2 prop of hr} Let $0 \not \equiv g \in \Sigma^2\cC_{\Delta}$, then $g(\bv) = 0$ on at most an $\frac{\epsrcn}{{r \choose 2}}$ fraction of the points in $\hr(r,\Delta)$.
    \end{enumerate}
        When $r$ and $\Delta$ are clear from context we simply write $\hr$.

        We define the parameter $s_{\text{recon}}$ as:
    \[
        s_{\text{recon}} \coloneqq \lceil(1-\epsrcn)|\hr(r,\Delta)|\rceil.
    \]
    Note that any $s_{\text{recon}}$  points from $\hr(r,\Delta)$ give a reconstruction set.
    \defsymbol
\end{definition}

\begin{remark} \label{rem:hr_for_r_eq_one}
    In the case $r=1$ the above properties are simplified:
    \begin{itemize}
        \item[$\hp$:] In \autoref{def:hp} we only require that $\hp$ is a hitting set for $\cC^d_{\Delta}$.
        \item[$\hr$:] In \autoref{def:hr} we omit property \ref{item:hs s2 prop of hr}.
    \end{itemize}
\end{remark}


\Cref{alg:meta_recon} is our reconstruction algorithm. We give its analysis in \Cref{thm:gen recon algo}.

\begin{algorithm}[!t]
\caption{$\Sigma^r \wedge^d \cC_{\Delta}$ Reconstruction}
\label{alg:meta_recon}
\begin{algorithmic}[1]
\Input{Sets $\hp, \hr$; A blackbox access to $P = \Sigma_{i=1}^r \alpha_i F_i^d$ such that $F_i \in  \cC_{\Delta}$. Parameters $r,d,\Delta$.}
\Output{$\{\lambda_i, H_i\}_{i =1}^r$, such that $P = \Sigma_{i=1}^r \lambda_i H_i^d$ and $\{H_i\}_{=1}^r \subseteq \cC_\Delta$.}
\ForAll{$(\bu,\bv) \in \hp \times \hr$} \Comment{Univariate reconstruction for each $(\bu,\bv)$}
\State \label{line:puv} Let $$
        q_{\bu,\bv}(t)\coloneqq P(\bu + t (\bv - \bu)), \quad \text{and} \quad q_{i,\bu,\bv}(t) \coloneqq F_i(\bu + t(\bv - \bu))
        $$
\State Run \cref{alg:recover-depth4-uni} on $q_{\bu,\bv}(t)$
    \If{the algorithm did not abort}
\State        Let $\{(\lambda_{i,\bu,\bv}, h_{i,\bu,\bv}(t))\}^{r_{\bu,\bv}}_{i=1}$ be its output
    \EndIf
\EndFor
\If {$r \notin \{r_{\bu,\bv}\}_{\bu,\bv}$} \label{line:meta_recon_minimal_r_check}
\State abort
\EndIf
\ForAll{$\bu \in \hp$} \Comment{Find good $\bv$'s per $\bu$}
    \State\label{line:cv_u def} $$\mathcal{V}_{\bu} \leftarrow \Big\{ \bv \in \hr \mid r_{\bu,\bv} = r\; \text{and} \; \{\lambda_{i,\bu,\bv} h_{i,\bu,\bv}(0)^d\}_{i=1}^r \; \text{are nonzero and distinct}  \Big\}$$
\EndFor
\State Find $\bu \in \hp$ such that $\left| \mathcal{V}_{\bu} \right| \geq s_{\text{recon}}$ \label{line:gen good u} \Comment{Find $\bu$ with sufficiently many good $\bv$'s}
\State Let $\bv' \in \mathcal{V}_{\bu}$
\ForAll{$\bv' \neq\bv \in \mathcal{V}_{\bu}$} \Comment{Align the mixed outputs across different $\bv$'s}
    \State\label{line:gen alignment}Reorder $\{(\lambda_{i,\bu,\bv}, h_{i,\bu,\bv}(t))\}^{r_{\bu,\bv}}_{i=1}$ so that $$
        \lambda_{i,\bu,\bv} h_{i,\bu,\bv}(0)^d = \lambda_{i,\bu,\bv'} h_{i,\bu,\bv'}(0)^d
    $$
\EndFor
\State\label{line:gen lambda i} Set $\{\mathcal{S}_i\}_{i=1}^r \leftarrow \{\emptyset\}_{i=1}^r$ and $\lambda_i = \lambda_{i,\bu,\bv'} h_{i,\bu,\bv'}(0)^d$
\ForAll{$\bv \in \mathcal{V}_{\bu}$ and $i \in [r]$} \Comment{Scaling the outputs to guarantee reconstruction}
    \State Add $\left(\bv, \frac{h_{i,\bu,\bv}(1)}{h_{i,\bu,\bv}(0)}\right)$ to $\mathcal{S}_i$ \label{line:gen S_i}
\EndFor
\ForAll{$i \in [r]$} \Comment{Reconstructing every $H_i$}
\State Use $\mathcal{S}_i$ and the reconstruction algorithm from Property~\ref{item:recon prop of hr} in \autoref{def:hr} to reconstruct $H_i$ \label{line:gen recon}
\EndFor
\State \Return The set $\{(\lambda_{i}, H_{i})\}_{i=1}^r$
\end{algorithmic}
\end{algorithm}

\begin{theorem} \label{thm:gen recon algo}
Let $r,d,\Delta \in \mathbb{N}$ with $d > (r+1)^4\Delta$, let $\F$ be a field of characteristic $p$ such that $p = 0$ or $p > 2 \Delta r d$ and assume $\cC$ is closed under scalar multiplication. 
Let $T_{\text{univ}}$ be the running time of the algorithm mentioned in \autoref{thm:uni-reconstruct} and let $T_{\text{recon}}$ be the running time of the reconstruction algorithm mentioned in \autoref{def:hr}~\eqref{item:recon prop of hr}. Then, \autoref{alg:meta_recon} runs in time $\text{poly}(|\hp|, |\hr|, T_{\text{univ}}, T_{\text{recon}})$ and its output satisfies 
\[
    \sum^r_{i=1} \lambda_i H_i(\x)^d = P(\x).
\]
\end{theorem}

\begin{proof}
    We may assume without loss of generality, as in \cref{remark:small_r_wlog}, that the top fan-in $r$ is minimal. That is, no term $F_i$ computes the zero polynomial, and for every distinct pair $i \neq j \in [r]$, the polynomials $F_i$ and $F_j$ are pairwise non-associates.

    The claimed running time follows from the fact that each step either executes one of the reconstruction algorithms, that is \autoref{alg:recover-depth4-uni} and the one for $\cC$, performs basic operations on the points in $\hp$ and $\hr$, or evaluates \(P\) at these points. 
    We next prove the correctness of the algorithm.

    The proof is composed of several steps. First, in \autoref{claim:gen exists good u}, we show that there exists at least one ``good'' point $\bu \in \hp$ for $P$, ensuring that such a point is found in line~\ref{line:gen good u}. Then, in \autoref{claim:gen good u big v set} we show that for any good point $\bu$, the associated set $\cV_{\bu}$ contains enough points that enable a robust reconstruction algorithm for $\cC$. Finally, we show that the alignment procedure in line~\ref{line:gen alignment} works. Consequently, by leveraging the evaluations of the polynomials $h_{i,\bu,\bv'}$, we can recover the target terms $H_i$.

    \begin{claim} \label{claim:gen exists good u}
    If $\hp$ satisfies \autoref{def:hp}, then there exists a point $\bu \in \hp$ such that:
    \begin{enumerate}
        \item \label{item:good_u_nonzero_prop} $F_i(\bu) \neq 0$ for all $i \in [r]$,
        and
        \item \label{item:good_u_distinct_values} the values $\gamma_i \coloneqq \alpha_i F_i(\bu)^d$ are pairwise distinct.
    \end{enumerate}
    We refer to any point $\bu \in \hp$ satisfying these conditions as \emph{good}.
\end{claim}

\begin{proof}
    By \autoref{def:hp}~\eqref{item:hs prod prop of hp} and the fact that each $F_i \not\equiv 0$, it follows that there are at most $\epsprod|\hp|$ points $\bu \in \hp$ for which there exists an index $i \in [r]$ such that $F_i(\bu) = 0$.

    Next, consider the $\binom{r}{2}$ polynomials defined by
    \[
        P_{i,j} \coloneqq \alpha_i F_i^d - \alpha_j F_j^d.
    \]
    Since the polynomials $F_i$ are pairwise non-associates, each $P_{i,j}$ is nonzero. By a union bound and \autoref{def:hp}~\eqref{item:hs s2 prop of hp}, for all but at most $\binom{r}{2}\epsst|\hp|$ points in $\hp$, none of the polynomials $P_{i,j}$ vanishes.

    Combining these two bounds via a union bound and applying the size bound from Property~\eqref{item:size req of hp}, we conclude that the number of good points in $\hp$ is at least
    \[
        \left(1 - \epsprod - \binom{r}{2}\epsst\right) |\hp| \geq 1. \qedhere
    \]
\end{proof}

Fix a good point $\bu$. We next show that the set $\cV_{\bu}$ is sufficiently large for reconstruction, and that the reconstructed univariate polynomials can be aligned correctly.

\begin{claim} \label{claim:gen good u big v set}
    If $\hr$ satisfies \autoref{def:hr}, then for every good point $\bu \in \hp$, we have $\big|\mathcal{V}_{\bu}\big| \geq s_{\text{recon}}$. Moreover, for every $\bv \in \cV_{\bu}$,
    \[
        \{\lambda_{i,\bu,\bv} h_{i,\bu,\bv}(0)^d\}_{i=1}^r = \{\alpha_{i} F_i(\bu)^d \}_{i=1}^r.
    \]
\end{claim}

\begin{proof}
    For $1 \leq j < i \leq r$, define the functions
    \[
        g_{i,j}(\x) = F_i(\bu) F_j(\x) - F_j(\bu)F_i(\x).
    \]
    Since $\bu$ is a good point, we have $F_i(\bu) \neq 0$ and $F_j(\bu) \neq 0$. Furthermore, because the polynomials $\{F_{\zeta}\}_{\zeta=1}^r$ are pairwise non-associates, it follows that $g_{i,j} \not\equiv 0$. 

    By \autoref{def:hr}~\eqref{item:hs s2 prop of hr}, each non-zero $g_{i,j}(\bv)$ vanishes on at most  $\frac{\epsilon_{\text{recon}}}{\binom{r}{2}}$ fraction of the points $\bv \in \hr$. By the union bound, for a $(1 - \epsilon_{\text{recon}})$ fraction of points in $\hr$, none of the $g_{i,j}$ vanish. Let $\bv \in \hr$ be such a point. We show that any such $\bv$ contributes an evaluation point for the reconstruction algorithm that follows (\Cref{line:gen recon}).

    \begin{observation} \label{obs:meta_recon_proj_keep_lin_indp}
        The univariate polynomials $\{q_{i,\bu,\bv}(t)\}_{i=1}^r$ (defined on \Cref{line:puv}) are pairwise non-associates.
    \end{observation}

    \begin{proof}
        Assume for the sake of contradiction that there exist indices $1 \leq j < i \leq r$ and a scalar $\alpha \in \mathbb{F}$ such that $q_{i,\bu,\bv}(t) = \alpha \cdot q_{j,\bu,\bv}(t)$. Then, by evaluating $q_{i,\bu,\bv}(t) = \alpha \cdot q_{j,\bu,\bv}(t)$ at $t=0$ and at $t=1$ we get
        \[
            F_i(\bu) = \alpha F_j(\bu) \quad \text{and} \quad F_i(\bv) = \alpha F_j(\bv) \quad \Rightarrow \quad \alpha F_i(\bu)F_j(\bv) = \alpha F_j(\bu) F_i(\bv) \, .
        \]
        This implies that $g_{i,j}(\bv) = 0$, contradicting our choice of $\bv$.
    \end{proof}

    Combining this observation with \autoref{thm:abc_constants} and the minimality of the top fan-in $r$, we get that the minimum top fan-in of $q_{\bu,\bv}$ (as defined on \Cref{line:puv}) is exactly $r$. Since \autoref{cor:uniq repres of high powers} guarantees that the representation 
    \[
        q_{\bu,\bv}(t) = \Sigma^r_{i=1} \alpha_i q_{i,\bu,\bv}(t)^d
    \]
    is unique, it follows that, up to a permutation of the indices, 
    \begin{equation} \label{eq:uniq_representation_h_vs_p}
        \lambda_{i,\bu,\bv} h_{i,\bu,\bv}(t)^d = \alpha_i q_{i,\bu,\bv}(t)^d. 
    \end{equation}
    By the properties of the good point $\bu$ established in \autoref{claim:gen exists good u}, the values 
    \[
        \{\lambda_{i,\bu,\bv} h_{i,\bu,\bv}(0)^d\}_{i=1}^r = \{\alpha_{i} q_{i,\bu,\bv}(0)^d\}_{i=1}^r = \{\alpha_{i} F_i(\bu)^d\}_{i=1}^r
    \]
    are non-zero and pairwise distinct. Consequently, every such point $\bv$ enters the set $\cV_{\bu}$ in line~\eqref{line:cv_u def}. This concludes the argument for the size of $\cV_{\bu}$.
\end{proof}

First, observe that by the second statement in \autoref{claim:gen good u big v set},
$$
    \lambda_i = \lambda_{i,\bu,\bv'} h_{i,\bu,\bv'}(0)^d = \alpha_i F_i(\bu)^d \, .
$$ 
This ensures that the definition of $\lambda_{i}$ in line~\eqref{line:gen lambda i} is well-defined and independent of the specific representative $\bv'$.

By the second statements of both \autoref{claim:gen exists good u} and \autoref{claim:gen good u big v set}, the alignment procedure in line~\ref{line:gen alignment} correctly aligns the $r$ different univariate polynomials across different $\bv$'s. Thus, after reordering, using equation~\eqref{eq:uniq_representation_h_vs_p} we get that for every $\bv \in \cV_{\bu}$, 
\[
    \lambda_{i,\bu,\bv} h_{i,\bu,\bv}(t)^d = \alpha_i q_{i,\bu,\bv}(t)^d.
\]

Let $\tilde{h}_{i,\bu,\bv}(t) \coloneqq \frac{h_{i,\bu,\bv}(t)}{h_{i,\bu,\bv}(0)}$, we obtain:
\[
    \tilde{h}_{i,\bu,\bv}(t)^d = \left( \frac{h_{i,\bu,\bv}(t)}{h_{i,\bu,\bv}(0)} \right)^d = \left( \frac{q_{i,\bu,\bv}(t)}{F_i(\bu)} \right)^d.
\]
Observing that $\frac{q_{i,\bu,\bv}(0)}{F_i(\bu)} = 1 = \tilde{h}_{i,\bu,\bv}(0)$, it follows that $\tilde{h}_{i,\bu,\bv}(t) = \frac{q_{i,\bu,\bv}(t)}{F_i(\bu)}$. This follows from the fact that it is the unique $d$-th root of the polynomial $\left(\frac{q_{i,\bu,\bv}(t)}{F_i(\bu)}\right)^d$ that evaluates to $1$ at $t=0$, as any other $d$-th root evaluates to some $d$-th root of unity. In particular, evaluating at $t=1$ gives:
\[
    \tilde{h}_{i,\bu,\bv}(1) = \frac{q_{i,\bu,\bv}(1)}{F_i(\bu)} = \frac{F_i(\bv)}{F_i(\bu)}.
\]
This is also the value inserted into $\cS_i$ in line~\eqref{line:gen S_i} for all $\bv \in \cV_{\bu}$.

By \autoref{def:hr}~\eqref{item:recon prop of hr}, alongside this evaluation and the fact that $|\cV_{\bu}| \geq s_{\text{recon}}$, the algorithm successfully reconstructs the normalized polynomial $H_i \equiv \frac{1}{F_i(\bu)}\cdot F_i$ in line~\ref{line:gen recon}.

Finally, for any $\bv \in \cV_{\bu}$, we have $\lambda_i = \lambda_{i,\bu,\bv} h_{i,\bu,\bv}(0)^d = \alpha_i F_i(\bu)^d$. Thus, 
\[
    \lambda_i H_i(\x)^d = \alpha_i F_i(\bu)^d \cdot \left( \frac{F_i(\x)}{F_i(\bu)}\right)^d = \alpha_i F_i(\x)^d, 
\]
which concludes the correctness analysis of \autoref{alg:meta_recon}.
\end{proof}


\subsection{Lifting for Small Individual Degree}
\label{subsec:recon-high-deg}

The lower bound on the powering parameter in
\autoref{thm:gen recon algo} depends on the total degree of the
inner polynomials. Consequently, even when their individual degrees
are bounded, this requirement may grow with the number of variables.
We eliminate this dependence by applying the univariate reconstruction
algorithm only to restrictions with at most \(r\) free variables.

The main idea is to retain \(r-1\) variables such that, after a suitable
substitution for the remaining variables, the inner polynomials
remain nonzero and pairwise non-associate. We next explain how we use this anchor to obtain evaluations at other points. 

Recall that in \cref{thm:gen recon algo}, an anchor point \(\bu\) is connected
to a reconstruction point \(\bv\) by the line $\bu+t(\bv-\bu)$. This substitution may affect \(n\) coordinates, and may result in a high total degree in $t$. Instead, here we work with restrictions in which only a limited number of
variables remain free. We therefore keep the \(r-1\) anchor variables
free and change the remaining coordinates one at a time, forming
a chain of hybrids. Each intermediate hybrid has at most \(r\) free
variables, and consecutive hybrids share a common restriction which is an anchor.

Throughout, let
\begin{equation*}
    P=\sum_{i=1}^r\alpha_i f_i^d,
    \qquad f_i\in\cC,
    \qquad \deg_{x_j}(f_i)\leq\delta,
\end{equation*}
and let
\begin{equation} \label{eq:definition_of_kappa}
    \Delta=n\delta,
    \qquad m=n-r+1,
    \qquad \kappa=m {r \choose 2}.
\end{equation}
We assume that \(r\) is the minimal top fan-in of such a representation.
In particular, the coefficients and inner polynomials are nonzero,
and the latter are pairwise non-associate. We also assume that the
underlying class is closed under substitutions of variables by field constants and under multiplication by field scalars.

We restrict attention to \(2\leq r\leq n\). When \(r>n\), the
variable reduction argument we introduce here offers no improvement over applying
\cref{alg:meta_recon} directly. When \(r=1\), preserving pairwise
non-association is unnecessary, and one may use \cref{alg:recover-depth4-uni} with suitable substitutions. 



\medskip

We start by defining the notion of an anchor pair.

\subsubsection{Anchor pairs}

\begin{definition}[Anchor pair]
    \label{def:anchor_pair}
    Let \(f_1,\ldots,f_r\in\F[\x]\) be nonzero and pairwise
    non-associate. An \emph{anchor pair} for this set of polynomials is a pair
    \((S,\bu)\), where \(S\subseteq[n]\), \(|S|=r-1\), and
    \(\bu_{-S}\in\F^{n-r+1}\), such that
    \begin{equation*}
        f_1(\bu_{-S},\x_S),\ldots,f_r(\bu_{-S},\x_S)
    \end{equation*}
    are nonzero and pairwise non-associate.
\end{definition}

To prove that such pairs exist and to find them efficiently, we shall need the following lemma of \cite{BhargavaSV26}.

\begin{lemma}[{\cite[Lemma~4.3]{BhargavaSV26}}]
    \label{lem:swli}
    Let \(g_1,\ldots,g_k\in\F[\x]\) be linearly independent
    over \(\F\). Then there exists \(S\subseteq[n]\) with
    \(|S|\leq k-1\) such that these polynomials remain linearly
    independent over \(\F(\x_{-S})\).
\end{lemma}

We shall need the following variant in which the polynomials are only guaranteed to be pairwise linearly independent. 

\begin{lemma}[Existence of a small variable set]
    \label{lem:recon_ext_existence_of_good_set}
    Let \(f_1,\ldots,f_r\in\F[\x]\) be pairwise linearly
    independent. Then there exists \(S\subseteq[n]\) with
    \(|S|\leq r-1\) such that the polynomials remain pairwise
    linearly independent over \(\F(\x_{-S})\).
\end{lemma}

\begin{proof}
    Let \(g_1,\ldots,g_k\) be a basis of
    \(\operatorname{Span}_{\F}\{f_1,\ldots,f_r\}\).
    By \cref{lem:swli}, there exists
    \(S\subseteq[n]\), with \(|S|\leq k-1\leq r-1\), such
    that the \(g_j\) remain linearly independent over
    \(\F(\x_{-S})\).

    Fix distinct \(i,l \in [r]\), and write
    \begin{equation*}
        f_i=\sum_{j=1}^k a_jg_j,
        \qquad f_l=\sum_{j=1}^k b_jg_j,
        \qquad a_j,b_j\in\F.
    \end{equation*}
    If \(\alpha f_i=f_l\) for some \(\alpha\in \F(\x_{-S})\), then linear independence
    over \(\F(\x_{-S})\) gives \(\alpha a_j=b_j\) for every \(j\).
    Choosing \(j\) with \(a_j\neq0\), we obtain
    \(\alpha=b_j/a_j\in\F\), contradicting the linear independence
    of \(f_i\) and \(f_l\) over \(\F\).
\end{proof}

We will use the following lemma to apply hitting sets over \(\F\) to polynomials with polynomial coefficients, i.e., to  polynomials in $\F[\y][\x]$.

\begin{lemma}
    \label{claim:hs_for_base_field_is_for_rational_field}
    Let \(\cC\subseteq\F[\z]\) be a polynomial class, and let
    \(\cH\subseteq\F^{|\z|}\) be a hitting set for \(\cC\).
    Suppose that \(g(\y,\z)\in\F[\y][\z]\) is nonzero,
    that its individual degree in each $\vy$-variable is at most
    \(\delta\), and that \(g(\ba,\z)\in\cC\) for every
    \(\ba\in\F^{|\y|}\). If \(|\F|>\delta\), then some
    \(\bb\in\cH\) satisfies \(g(\y,\bb)\not\equiv0\).

    More generally, if \(\cH\) is an \(\epsilon\)-robust hitting
    set for $\cC$, then \(g(\y,\bb)\not\equiv0\) for at least a
    \((1-\epsilon)\)-fraction of its indexed points.
\end{lemma}

\begin{proof}
    Write $g(\y,\z)$ as a polynomial in $\z$ with coefficients in $\F[\y]$, and let \(c(\y)\) be any nonzero coefficient of \(g\). Since \(|\F|>\delta\), it is well known (see \cref{thm:comb_null} below) that there exists a point \(\ba\) with
    \(c(\ba)\neq0\). Hence \(g(\ba,\z)\) is nonzero and by assumption is a member of \(\cC\). Every point \(\bb\) at which \(g(\ba,\bb)\neq0\) also satisfies \(g(\y,\bb)\not\equiv0\). Applying the ordinary or robust hitting-set property to this one specialization proves the two assertions.
\end{proof}

\begin{theorem}[Combinatorial Nullstellensatz {\cite{Alon1999CombNul}}]
    \label{thm:comb_null}
    Let \(0\neq g\in\F[\x]\), with
    \(\deg_{x_i}(g)\leq d_i\). If \(S_i\subseteq\F\) and
    \(|S_i|>d_i\) for every \(i\), then there exists $\balp \in S_1\times\cdots\times S_n$ such that \(g(\balp) \neq 0\).
\end{theorem}

To prove the existence of an anchor pair we will also need the following lemma of \cite{Kayal11}.

\begin{lemma}[{{\cite[Claim~7]{Kayal11}}}]
    \label{lem:alt_linear_independence}
    Let \(g_1,\ldots,g_k\in\F[\x]\), and let
    \(\x^{(1)},\ldots,\x^{(k)}\) be disjoint copies of \(\x\).
    Then the \(g_j\) are linearly independent over \(\F\) if
    and only if
    \begin{equation*}
        \det\bigl(g_j(\x^{(i)})\bigr)_{i,j=1}^k\not\equiv0.
    \end{equation*}
\end{lemma}

\begin{lemma}[Existence of an anchor pair]
    \label{lem:recon_ext_proj_to_r_var_HS}
    Let \(f_1,\ldots,f_r \in \cC\) be nonzero and pairwise
    non-associate, with individual degrees at most \(\delta\).
    Suppose that \(|\F| > {r \choose 2} \delta\) and that \(\cH_{\mathrm{Alt}} \subseteq \F^n\) hits \(\Pi^{{r \choose 2}}\Sigma^2\Pi^2\cC_{\Delta}\). Then some \(S \subseteq [n]\) of size \(r-1\) and some \(\bu\in\cH_{\mathrm{Alt}}\) form an anchor pair.
\end{lemma}

\begin{proof}
    Apply \cref{lem:recon_ext_existence_of_good_set}, extending the
    resulting variable set to size \(r-1\) if necessary.
    Let \(\y,\z\) be disjoint copies of \(\x_S\). For \(i<j\), define
    \begin{equation*}
        A_{ij}(\y,\z,\x_{-S}) =
        \det\left(\begin{matrix}
  f_i(\y,\x_{-S}) & f_j(\y,\x_{-S}) \\
  f_i(\z,\x_{-S}) & f_j(\z,\x_{-S})
\end{matrix}\right) = f_i(\y,\x_{-S})f_j(\z,\x_{-S})
         -f_j(\y,\x_{-S})f_i(\z,\x_{-S}).
    \end{equation*}
    By \cref{lem:alt_linear_independence},
    every \(A_{ij}\) is nonzero. Consequently,
    \begin{equation*}
        Q(\y,\z,\x_{-S})
        =\prod_{1\leq i<j\leq r}A_{ij}(\y,\z,\x_{-S})
        \not\equiv0.
    \end{equation*}
    Its individual degree in each variable from \(\y\cup\z\)
    is at most \({r \choose 2}\delta\). Every fixing of the variables in \(\y\cup\z\) gives a polynomial in $\Pi^{{r \choose 2}}\Sigma^2\Pi^2\cC_{\Delta}$. By \cref{claim:hs_for_base_field_is_for_rational_field}, some
    \(\bu\in\cH_{\mathrm{Alt}}\) satisfies
    \(Q(\y,\z,\bu_{-S})\not\equiv0\). Thus, every $A_{ij}(\y,\z,\x_{-S})$
    remains nonzero after the substitution and by \cref{lem:alt_linear_independence} all restricted polynomials $\{f_i(\x_S,\bu_{-S})\}_{i=1}^r$ are nonzero and
    pairwise non-associate.
\end{proof}

\subsubsection{The reconstruction set}

Like \cref{alg:meta_recon}, our extension relies on several specialized
sets, most of which are hitting sets for polynomial classes. We state
the requirements on these hitting sets in
\cref{thm:small_individual_degree_reconstruction}. Here, we describe
the properties required of the reconstruction set that replaces \(\hr\).

\begin{definition}[Properties of \(\hre(r,\Delta,\epsilon)\)]
    \label{def:hre}
    Let $r, \Delta \in \mathbb{N}$, let $0<\epsilon<1 $ and let $\cC$ be a polynomial class. The set $\hre(r,\Delta,\epsilon)$ used in \autoref{alg:meta_recon_ext} must satisfy the following properties:
    \begin{enumerate}
        \item\label{item:hre_pit_prop}
        It is an \(\epsilon\)-robust hitting set for
        \(\prod^{\kappa}\sum^2\prod^2\cC_{\Delta}\).

        \item\label{item:hre_recon_prop}
        It is an \(\epsilon\)-robust reconstruction set for
        \(\cC_{\Delta}\).
    \end{enumerate}
    When $r$ and $\Delta$ are clear from context we simply write $\hre(\epsilon)$.
\end{definition}

\subsubsection{Hybrids and good points}

Fix an anchor pair \((S,\bu)\) and relabel coordinates so that
\begin{equation*}
    [n]\setminus S=[m],\qquad S=\{m+1,\ldots,n\}.
\end{equation*}


For \(\bv\in\F^n\), define
\begin{align*}
    E_{i,\zeta,\bv}(\x_S)
        &\coloneqq f_i(\bv_{[\zeta]},\bu_{[\zeta+1:m]},\x_S),
        &&0\leq \zeta\leq m,\\[1.5ex]
    q_{i,\zeta,\bv}(x_{\zeta},\x_S)
        &\coloneqq f_i(\bv_{[\zeta-1]},x_{\zeta},\bu_{[\zeta+1:m]},\x_S),
        &&1\leq \zeta\leq m,\\
    P_{\zeta,\bv}(x_{\zeta},\x_S)
        &\coloneqq\sum_{i=1}^r\alpha_i q_{i,\zeta,\bv}(x_{\zeta},\x_S)^d,
        &&1\leq \zeta\leq m.
\end{align*}
We call the $q$ polynomials \emph{hybrids} and the $E$ polynomials \emph{endpoints}. Every hybrid has total degree at most \(r\delta\). Observe that every two consecutive hybrids can be restricted to the same endpoint, that is 
\begin{equation}
    \label{eq:small_degree_common_endpoints}
    q_{i,\zeta+1,\bv}(u_{\zeta+1},\x_S)=E_{i,\zeta,\bv}(\x_S)=
    q_{i,\zeta,\bv}(v_{\zeta},\x_S).
\end{equation}


\begin{definition} [Good point] \label{def:good_point}
    Call \(\bv\) \emph{good} if, for every \(0\leq  \zeta \le m\), the
    endpoint polynomials \(E_{1,\zeta,\bv},\ldots,E_{r,\zeta,\bv}\) are
    nonzero and pairwise non-associate.
\end{definition}

\begin{proposition}[Most reconstruction points are good]
    \label{prop:ext_recon_step_2_key_prop}
    If \(\hre(\epsilon)\) satisfies \cref{def:hre} and \(|\F|>\kappa\delta\), then at least \((1-\epsilon)|\hre(\epsilon)|\) of its indexed points are good.
\end{proposition}

\begin{proof}
    By \cref{def:anchor_pair} the two properties are satisfied for \(E_{1,0,\bv},\ldots,E_{r,0,\bv}\), it remains to show them for $\zeta \in [m]$.
    
    Let \(\y_S\) be a disjoint copy of \(\x_S\), and define
    \begin{equation*}
        A_{ij}(\z)
        =f_i(\z,\x_S)f_j(\z,\y_S)
         -f_j(\z,\x_S)f_i(\z,\y_S),
        \qquad \z=(z_1,\ldots,z_m).
    \end{equation*}
    Let
    \begin{equation*}
        Q_{\mathsf{Alt}}(\z)
        =\prod_{\zeta=1}^m\prod_{1\leq i<j\leq r}
            A_{ij}(\z_{[\zeta]},\bu_{[\zeta+1:m]}),
    \end{equation*}
    we call this polynomial a certificate for goodness.
    
    \cref{def:anchor_pair} gives
    \begin{equation*}
        Q_{\mathsf{Alt}}(\bu_{[m]})
        =\prod_{1\leq i<j\leq r}A_{ij}(\bu_{[m]})^m
        \not\equiv0.
    \end{equation*}

    Its individual degree in each $\x$ and $\y$ variable is at most \(\kappa\delta\). By \cref{claim:hs_for_base_field_is_for_rational_field} and the fact that $|\F| > \kappa \delta$ the robust hitting property of $\hre(\epsilon)$ applies also with the coefficients being over $\F[\x_S,\y_S]$.

    If \(Q_{\mathsf{Alt}}(\bv_{[m]})\not\equiv0\), every $A_{ij}$ of two endpoint polynomials is nonzero, \cref{lem:alt_linear_independence} then
    implies that \(\bv\) is good. The certificate is nonzero
    on at least a \((1-\epsilon)\)-fraction of the indices,
    proving the assertion.
\end{proof}

\subsubsection{The reconstruction algorithm}

In this section we give the reconstruction algorithm  (\cref{alg:meta_recon_ext}) and its analysis  (\cref{thm:small_individual_degree_reconstruction}).  Recall the definitions of $m$ and $\kappa$ from \Cref{eq:definition_of_kappa}.



\begin{algorithm}[!t]
    \caption{Reconstruction for small individual degree}
    \label{alg:meta_recon_ext}
    \small
    \begin{algorithmic}[1]
        \Input{Black-box access to \(P\); parameters \(r,d,\delta\);
        sets \(\cH_{\mathrm{Alt}}\), \(\cH_r\),
        \(\hre(\epsilon)\), $\hp^{(r)}$ and $\hr^{(r)}$.}
        \Output{\((\lambda_i^0,H_i)_{i=1}^r\) representing \(P\).}
        \State Set \(\mathsf{anchor}\gets\bot\). \Comment{Finding an anchor pair}
        \ForAll{\((S,\bu)\in\binom{[n]}{r-1}\times\cH_{\mathrm{Alt}}\)} \label{line:anchor_reconstruction}
            \State
            Run \cref{alg:meta_recon} on \(P(\bu_{-S},\x_S)\).
            \If{the algorithm does not abort}
                \State Set \(\mathsf{anchor}\gets(S,\bu)\), let \((\lambda_i^0,h_i^0)_{i=1}^r\) be an ordering of the output and exit this loop.
            \EndIf
        \EndFor
        \State\label{line:no_anchor} If \(\mathsf{anchor}=\bot\), abort.
        \State Order the coordinates such that $[n] \setminus S = [m]$ and $S = \{m+1,\ldots,n\}$; set
        \(\cE_i\gets\emptyset\) for every \(i\in[r]\).
        \ForAll{\((\gamma,\bv)\in\hre(\epsilon)\)} \Comment{Each $(\gamma,\bv)$ is called a sample}
            \State For every \(i\), set \(\ell_i\gets h_i^0\),
            \(\lambda_i\gets\lambda_i^0\), and \(\beta_i\gets1\).
            \For{\(\zeta=1,\ldots,m\)} \Comment{Reconstructing the hybrids}
                \State
                Run \cref{alg:meta_recon} on \(P_{\zeta,\bv}\).
                \If{the algorithm aborts}
                \State\label{line:bad_hybrid_reconstruction} Move to the next sample.
                \Else
                \State Let \(\{(\mu_j,g_j)\}_{j=1}^r\) be its output.
                \State\label{line:indv alignment}Using a PIT algorithm for $\Sigma^2\wedge^d \cC$ find a unique $\pi \in S_r$ so that for every $i \in [r]$: $$
                \lambda_i\ell_i(\x_S)^d
                    \equiv\mu_{\pi(i)}g_{\pi(i)}(u_{\zeta},\x_S)^d.
                $$
                \If{No such permutation exists or if there is more than one} \label{line:indv-recon-failed-alignment}
                \State Continue to the next sample.
                \EndIf
                \EndIf
                \For{\(i=1,\ldots,r\)} \Comment{Calculating the normalizing factors}
                    \State\label{line:g_def} Set \(g\gets g_{\pi(i)}\). Search
                    \(\cH_r\) for \(\bw\) with \(g(u_{\zeta},\bw)\neq0\).
                    \State\label{line:rec_poly_not_in_cC} Reject the sample if no such \(\bw\) exists.
                    \State\label{line:unified_normalization}
                    Set \(\displaystyle\beta_i\gets
                        \beta_i\cdot\frac{\ell_i(\bw)}{g(u_{\zeta},\bw)}\).
                    \State Set \(\ell_i(\x_S)\gets g(v_{\zeta},\x_S)\)
                    and \(\lambda_i\gets\mu_{\pi(i)}\).
                \EndFor
            \EndFor
            \State\label{line:good_sample}
            For every \(i\), append
            \((\gamma,\bv,\beta_i\ell_i(\bv_S))\) to \(\cE_i\). \Comment{Adding $c_if_i(\bv)$ to $\cE_i$}
        \EndFor
        \State If \(|\cE_1|<(1-\epsilon)|\hre(\epsilon)|\),
        \textbf{return} \(\bot\).
        \State\label{line:unified_final_reconstruction}
        For each \(i\), reconstruct \(H_i\) from \(\cE_i\)
        using the reconstruction algorithm from Property~\eqref{item:hre_recon_prop} in \cref{def:hre}.
        \State\Return \((\lambda_i^0,H_i)_{i=1}^r\).
    \end{algorithmic}
\end{algorithm}

\begin{theorem}[Reconstruction for small individual degree]
    \label{thm:small_individual_degree_reconstruction}
    Let $r,d, \delta \in \N$, $0<\epsilon<1$, $\cC$ some polynomial class and $P \in \Sigma^r\wedge^d\cC$. Moreover, let \(\Sigma_{i=1}^r\alpha_i f_i^d\) be a minimal representation of $P$, where each \(f_i\in\cC\) has individual degree at most \(\delta\). Assume that $d > (r+1)^4r\delta$, that $|\F|>\kappa\delta$
    and that \(\F\) has characteristic \(0\) or greater than
    \(2r^2\delta d\).

    Let \(\cH_{\mathrm{Alt}}\) satisfy the requirements of
    \cref{lem:recon_ext_proj_to_r_var_HS}, let \(\cH_r\) be a
    hitting set for the \(r\)-variate restrictions of
    \(f_1,\ldots,f_r\), and let \(\hre(\epsilon)\) satisfy
    \cref{def:hre}. Moreover, let \(\hp^{(r)}\) and \(\hr^{(r)}\)
    be the sets required to apply \cref{alg:meta_recon} to these
    \(r\)-variate restrictions.

    Then, given these sets and parameters as input,
    \cref{alg:meta_recon_ext} succeeds and returns
    \((\lambda_i^0,H_i)_{i=1}^r\) satisfying
    \begin{equation*}
        P\equiv\sum_{i=1}^r\lambda_i^0H_i^d,
        \qquad
        H_i\in\cC
        \quad\text{for every }i\in[r].
    \end{equation*}
    Let \(T_{\mathrm{meta}}\) bound the running time of a single
    application of \cref{alg:meta_recon} to an \(r\)-variate
    restriction of $P$, let \(T_{\mathrm{pit}}\) bound the cost of a single test in \Cref{line:indv alignment}, and
    let \(T_{\mathrm{rec}}\) bound the running time of the
    reconstruction procedure from
    Property~\eqref{item:hre_recon_prop} of \cref{def:hre}.
    Then the running time of \cref{alg:meta_recon_ext} is
    \begin{equation*}
        \poly\!\left(
            n^r,r, 
            |\cH_{\mathrm{Alt}}|,
            |\cH_r|,
            |\hre(\epsilon)|,
            T_{\mathrm{meta}},
            T_{\mathrm{pit}},
            T_{\mathrm{rec}}
        \right).
    \end{equation*}
\end{theorem}

\begin{proof}

    The claimed running time follows from the fact that each step either reconstructs an $r$-variate polynomial using \cref{alg:meta_recon}, executes a PIT algorithm between two $d$-th powers of $r$-variate polynomials in $\cC$, performs basic operations on the reconstructed polynomials or points in one of the input sets or reconstructs the output polynomials using the reconstruction algorithm from Property~\eqref{item:hre_recon_prop} of $\hre(\epsilon)$. 
    We next prove the correctness of the algorithm.
    
    \begin{claim}[Finding an anchor pair]
        \label{claim:unified_anchor_correctness}
        The search for an anchor pair in \cref{line:anchor_reconstruction} succeeds and returns an anchor pair and an ordered representation \((\lambda_i^0,h_i^0)_{i=1}^r\) for which, after relabeling the original summands and for any $\bv$,
        \begin{equation}
            \label{eq:ext_recon_step_3_norm_start}
            h_i^0=c_if_i({\bu_{-S},\x_S}) = c_iE_{i,0,\bv},
            \qquad \lambda_i^0c_i^d=\alpha_i,
            \qquad c_i\in\F^{\times}.
        \end{equation}
    \end{claim}

    \begin{proof}
        Consider any accepted pair. If one of the original restricted
        polynomials were zero, or if two were associates, deleting or
        combining the corresponding summands would give a representation
        with fewer than \(r\) terms. In this case, \cref{alg:meta_recon} would abort and the pair would be rejected. The pair that is returned is therefore an anchor pair. By uniqueness (\cref{cor:uniq repres of high powers}) its
        reconstructed summands match to the original restricted summands,
        up to permutation and nonzero scalars. Fixing this ordering gives for some $\{c_i\}_{i=1}^r \subseteq \F$,
        \(h_i^0=c_iE_{i,0,\bv}\) and
        \(\lambda_i^0(h_i^0)^d=\alpha_iE_{i,0,\bv}^d\).
        Since the $E_{i,0,\bv}$ are  nonzero, we also have
        \(\lambda_i^0c_i^d=\alpha_i\).
        
        By \cref{lem:recon_ext_proj_to_r_var_HS} there exist $S \subset [n]$ of size $|S|=r-1$ and a point $\bu \in \cH_{\mathrm{Alt}}$ which form an anchor pair. Such a pair is accepted and satisfies the two assertions, as was already proved.
    \end{proof}

    Using the above claim, we get that $(S,\bu)$ satisfy \cref{def:anchor_pair} and that the algorithm does not abort in \cref{line:no_anchor}. From now on, we use the ordering fixed in the above claim, the constants $c_i$ and the $\lambda_i^0$-s. Next, we argue that any hybrid sequence built using a good point $\bv$ and the anchor pair results in a correct evaluation of $c_i f_i(\bv)$. 
    
    \begin{claim}[Correct evaluations of retained samples]
        \label{claim:unified_sample_soundness}
        Every sample $(\gamma,\bv)$ that gets to \cref{line:good_sample} contributes the evaluation of \(c_if_i(\bv)\) to $\cE_i$, for each $i \in [r]$.
    \end{claim}

    \begin{proof}
        Fix a sample $\bv$ that reached \cref{line:good_sample}. Let
        \(\ell_i^{(\zeta)},\lambda_i^{(\zeta)},\beta_i^{(\zeta)}\) be the
        stored values after step \(\zeta\), with superscript
        zero denoting the initialization values. We prove the following
        invariants
        \begin{align}
            \lambda_i^{(\zeta)}\bigl(\ell_i^{(\zeta)}\bigr)^d
                &\equiv\alpha_iE_{i,\zeta,\bv}^d,
                \label{eq:unified_alignment_invariant}\\
            \beta_i^{(\zeta)}\ell_i^{(\zeta)}
                &\equiv c_iE_{i,\zeta,\bv}
                \label{eq:unified_normalization_invariant}
        \end{align}
        using induction on $\zeta$.
        
        They hold at \(\zeta=0\) by the parameters initialization and the identities in \cref{claim:unified_anchor_correctness}.

        Suppose these invariants hold for \(\zeta-1\). Since the sample was not rejected in \cref{line:bad_hybrid_reconstruction}, the original polynomials
        \(q_{1,\zeta,\bv},\ldots,q_{r,\zeta,\bv}\) are nonzero and
        pairwise non-associate, otherwise their representation could
        be shortened and \cref{alg:meta_recon} aborts. \cref{cor:uniq repres of high powers} gives a bijection between these original polynomials and
        the reconstructed pairs. This and the correctness of the PIT algorithm used in \Cref{line:indv alignment} imply that the unique permutation $\pi \in S_r$ is found, and that \begin{equation*}
            \lambda_i^{(\zeta-1)} \ell_i^{(\zeta-1)}(\x_S)^d \equiv \mu_{\pi(i)} g_{\pi(i)}(u_{\zeta},\x_S)^d, \qquad i \in [r].
        \end{equation*}
        
        In particular, for every \(i \in [r]\) we get in \cref{line:g_def}
        \begin{equation} \label{eq:identifying_g}
            g=a_{i,\zeta}q_{i,\zeta,\bv},
            \qquad
            \mu_{\pi(i)} g^d=\alpha_iq_{i,\zeta,\bv}^d,
            \qquad a_{i,\zeta}\in\F^{\times}.
        \end{equation}
        Substituting $x_\zeta=v_\zeta$ in \eqref{eq:identifying_g} proves \eqref{eq:unified_alignment_invariant} for step \(\zeta\).

        By \eqref{eq:identifying_g} $0 \not \equiv g \in \cC$. Therefore, by its definition, $\cH_r$ contains a point $\bw$ with
        \begin{equation*}
            g(u_{\zeta},\bw)=a_{i,\zeta}E_{i,\zeta-1,\bv}(\bw)\neq0.
        \end{equation*}
        Using \eqref{eq:unified_normalization_invariant} at
        step \(\zeta-1\) we get
        \begin{equation*}
            \beta_i^{(\zeta)}
            =\frac{\beta_i^{(\zeta-1)}\ell_i^{(\zeta-1)}(\bw)}
                   {g(u_{\zeta},\bw)}
            =\frac{c_iE_{i,\zeta-1,\bv}(\bw)}
                   {a_{i,\zeta}E_{i,\zeta-1,\bv}(\bw)}
            =\frac{c_i}{a_{i,\zeta}}.
        \end{equation*}
        Since the new endpoint is
        \(\ell_i^{(\zeta)}=a_{i,\zeta}E_{i,\zeta,\bv}\), this proves
        \eqref{eq:unified_normalization_invariant} at step
        \(\zeta\).

        At \(\zeta=m\), evaluating the second invariant at
        \(\x_S=\bv_S\) gives
        \begin{equation*}
            \beta_i^{(m)}\ell_i^{(m)}(\bv_S)
            =c_iE_{i,m,\bv}(\bv_S)=c_if_i(\bv).
        \end{equation*}
    \end{proof}

    \begin{claim}
        \label{claim:unified_sample_completeness}
        Let $\bv \in \hre(\epsilon)$ be a good point, then $\bv$ passes the tests in \cref{line:bad_hybrid_reconstruction}, \cref{line:indv-recon-failed-alignment} and \cref{line:rec_poly_not_in_cC} for every $\zeta \in [m]$ and $i \in [r]$, that is $\bv$ gets to \cref{line:good_sample}.
        In particular, for every $i \in [r]$, $|\cE_i| \geq (1-\epsilon)|\hre(\epsilon)|$.
    \end{claim}

    \begin{proof}
        Fix a good point \(\bv\). By \autoref{def:good_point} at every $0 \leq \zeta \leq m$, the \(E_{i,\zeta,\bv}\) are nonzero and pairwise
        non-associate. Since these are restrictions of the hybrid polynomials, the hybrid polynomials are also nonzero and pairwise non-associate. This implies using \cref{thm:gen recon algo} and \cref{cor:uniq repres of high powers} that for every $\zeta \in [m]$ \cref{alg:meta_recon} does not abort and the reconstructed polynomials satisfy for some permutation $\pi \in S_r$
        \begin{equation*}
            \alpha_i q_{i,\zeta,\bv}^d(x_\zeta,\x_S) \equiv \mu_{\pi(i)} g_{\pi(i)}(x_\zeta,\x_S)^d, \qquad i \in [r].
        \end{equation*}
        Since by assumption $\cC$ is closed under substitutions, the $g_i$-s are all nonzero members of $\cC$. Therefore the identity test used in \Cref{line:indv alignment} matches the polynomials correctly, and $\bv$ passes the tests in \cref{line:bad_hybrid_reconstruction}. Moreover, this implies that $\cH_r$ contains a nonzero point for each one of the $g_i$-s, hence $\bv$ passes the test in \cref{line:rec_poly_not_in_cC}. This, the fact that the polynomials are pairwise non-associate, and thus different and that $\pi$ exists, imply that $\bv$ passes the tests in \cref{line:indv-recon-failed-alignment}.

        Therefore, all tests are passed, $\bv$ reaches \cref{line:good_sample} and $(\gamma,\bv,\beta_i\ell_i(\bv_S))$ is added to $\cE_i$.

        The number of good points is at least
        \((1-\epsilon)|\hre(\epsilon)|\), by
        \cref{prop:ext_recon_step_2_key_prop}, this implies the second assertion.
    \end{proof}

    By combining \cref{claim:unified_sample_soundness} and \cref{claim:unified_sample_completeness} we get that for every $i \in [r]$ \(\cE_i\) contains the evaluations of $c_if_i$ on at least a $1-\epsilon$ fraction of $\hre$. By Property~\eqref{item:hre_recon_prop} this suffices for reconstructing $c_i f_i$ using the robust reconstruction algorithm for the underlying class. Hence \(H_i=c_if_i\). Using the second equality in \cref{claim:unified_anchor_correctness} we get
    \begin{equation*}
        \sum_{i=1}^r\lambda_i^0H_i^d
        =\sum_{i=1}^r\lambda_i^0c_i^df_i^d
        =\sum_{i=1}^r\alpha_if_i^d=P.\qedhere
    \end{equation*}
\end{proof}

\begin{remark}
    To make the algorithm non-adaptive, we query \(P\) in advance at all points required for every possible anchor. Enumerating the variable sets introduces an \(O(n^r)\) factor in the reconstruction-set size.
\end{remark}

\subsection{From Reconstruction Set to Robust Reconstruction Set} \label{subsec:recon_to_robust}

The proofs of the  lifting theorems  relied on the existence of robust hitting and reconstruction sets (recall \Cref{def:hp}, \Cref{def:hr} and \Cref{def:hre}). In this section we show how to obtain such sets from suitable hitting and reconstruction sets for the base class $\cC_\Delta$.

\subsubsection{Making hitting sets robust}

We first show how to convert an interpolating set into an ordered set
satisfying the requirements of \Cref{def:hp}. This is done by simple
coordinatewise univariate interpolation.

\begin{proposition}
    \label{prop:interpolation_set_satisfies_hp}
    Let \(r,d,\Delta\in\N\), and let \(\cC\) be a polynomial class.
    Let \(\cH\subseteq\F^n\) of size $|\cH| = s$ and \(H:\F\to\F^n\) be as in
    \autoref{thm:bb pit for high powers}.

    For any set \(S\subseteq\F\) of size
    \begin{equation*}
        |S|
        =
        \left(r+\binom{r}{2}d\right)\Delta(s-1)+1,
    \end{equation*}
    the ordered set \(H(S)\), satisfies the requirements of \autoref{def:hp}. 
    Moreover, if the points in $\cH$ and in $S$ require bit length at most $B_{\text{eval}}$, the points in $H(S)$ have bit complexity at most $\poly \left(\log r,\log d,\log \Delta,s,B_{\text{eval}} \right)$
\end{proposition}

\begin{proof}
    Throughout the proof, we count the elements of \(H(S)\) with
    multiplicity. Thus \(|H(S)|=|S|\), even if $H$ maps distinct points
    in \(S\) to the same image. Define
    \begin{equation*}
        \epsprod
        \coloneqq
        \frac{r\Delta(s-1)}{|S|},
        \qquad
        \epsst
        \coloneqq
        \frac{d\Delta(s-1)}{|S|}.
    \end{equation*}
    We show that these parameters satisfy the requirements of
    \autoref{def:hp}.

    First, let
    \begin{equation*}
        0\not\equiv g=\prod_{i=1}^r g_i,
        \qquad g_i\in\cC_\Delta.
    \end{equation*}
    Since \(\cH\) hits \(\cC_\Delta\), each nonzero factor \(g_i\)
    has a nonzero evaluation on \(\cH\). The curve \(H\) passes
    through every point of \(\cH\), so \(g_i\circ H\not\equiv0\)
    for every \(i\in[r]\). Consequently,
    \begin{equation*}
        g\circ H
        =
        \prod_{i=1}^r(g_i\circ H)
        \not\equiv0.
    \end{equation*}
    Every coordinate of \(H\) has degree at most \(s-1\), and
    therefore
    \begin{equation*}
        \deg(g\circ H)\leq r\Delta(s-1).
    \end{equation*}
    It follows that \(g\circ H\) vanishes at at most
    \(r\Delta(s-1)\) points in \(S\). Equivalently, \(g\)
    vanishes on at most an \(\epsprod\)-fraction of the multiset
    \(H(S)\), proving Property~\eqref{item:hs prod prop of hp}.

    Next, let
    \(0\not\equiv g\in\Sigma^2\wedge^d\cC_\Delta\).
    The argument for $r \leq 2$ in the proof of
    \autoref{thm:bb pit for high powers} gives
    \(g\circ H\not\equiv0\) (this argument does not impose restrictions on $d$ or on the field's characteristic). Moreover,
    \begin{equation*}
        \deg(g\circ H)\leq d\Delta(s-1).
    \end{equation*}
    Thus \(g\circ H\) vanishes at at most \(d\Delta(s-1)\)
    points in \(S\), so \(g\) vanishes on at most an
    \(\epsst\)-fraction of \(H(S)\), proving Property~\eqref{item:hs s2 prop of hp}.

    Finally, by the choice of \(|S|\),
    \begin{equation*}
        \epsprod+\binom{r}{2}\epsst
        =
        \frac{\left(r+\binom{r}{2}d\right)\Delta(s-1)}{|S|}
        =
        \frac{|S|-1}{|S|}
        <1.
    \end{equation*}
    Since \(|H(S)|=|S|\), we also obtain
    \begin{equation*}
        \left(
            1-\epsprod-\binom{r}{2}\epsst
        \right)|H(S)|
        =
        \frac{|H(S)|}{|S|}
        =
        1.
    \end{equation*}
    These bounds establish all the requirements of
    \autoref{def:hp}.

    For the bit-complexity bound, write the coordinates of \(H(\alpha)\), for \(\alpha\in S\), using Lagrange interpolation:
\[
H_j(\alpha)=
\sum_{i=1}^s u_{i,j}
\prod_{\substack{k\in[s]\\k\ne i}}
\frac{\alpha-a_k}{a_i-a_k},
\qquad j\in[n].
\]
Each Lagrange basis polynomial evaluates to a scalar of bit complexity \(O(sB_{\text{eval}})\). Multiplying by the coordinates of \(\mathbf u_i\) and summing over \(i\) gives the claimed polynomial bound.
\end{proof}


\subsubsection{Making reconstruction sets robust}

We now show how to transform a reconstruction set into a robust one and then we modify it to satisfy the other requirements of  \Cref{def:hr}. The construction is again via interpolation as in the proof of \Cref{prop:interpolation_set_satisfies_hp}.

It will be useful to work with the slightly more general notion of a robust reconstruction map.

\begin{definition}[{Robust Reconstruction maps}] \label{def:robust_recon_map}
Let $\mathcal{C}$ be a class of algebraic circuits, let
$0 \leq \epsilon < 1$ and let $D,m \in \N$. A map $\phi:\F\to\F^n$ is called a $(D,m,\epsilon)$-robust reconstruction-map for $\cC$ if the degree of each coordinate $\phi_i(t)$ is at most $D$, and for every $E\subseteq\F$ of size $m$ the image $\phi(E)$, viewed as a multiset, is an $\epsilon$-robust reconstruction set for $\cC$. 
\end{definition}

We're now ready for the construction itself.

\begin{theorem} \label{thm:rec-set-to-robust-map}
Let $\cR=\{\bu_i\}_{i=1}^s$ be a reconstruction set of size $s$ for
$\cC_{\Delta}$. Let $E=\{\alpha_1,\ldots,\alpha_s\}\subseteq\F$ be a set of distinct field elements. Let
\begin{equation*}
\phi(t)\colon \F\longrightarrow\F^n \quad \text{s.t.} \quad \forall i \in [s] : \phi(\alpha_i) = \bu_i
\end{equation*}
be the unique polynomial map of degree at most $s-1$ that evaluates to $\bu_i$ on $\alpha_i$.

Let $T\subseteq\F$ be a set of distinct field elements such that
$ m:=|T|\geq \Delta(s-1)+1$.
Then $\phi$ 
is an $(s-1,m,\epsilon)$-robust reconstruction map for $\cC_{\Delta}$, where
$
    1-\epsilon
    =
    \frac{\Delta(s-1)+1}{|T|}
$.
Moreover, whenever $\cR$ is explicit and the required field elements can be
efficiently enumerated, the resulting robust reconstruction map is
explicit. Denote the bit complexity of points in $\cR, T$ and $E$ by $B_{\text{eval}}$. Then the bit complexity of points in $\phi(T)$ is at most $\poly(s, B_{\text{eval}})$.
\end{theorem}

\begin{proof}
It is clear that $\phi$ is explicit when $R$ and $E$ are explicit, by simple Lagrange interpolation.
    

Fix a polynomial $C\in\cC_{\Delta}$ and consider the univariate polynomial
$F(t) := (C\circ \phi)(t)$. Since $C$ has total degree at most $\Delta$ and each $\phi_j$ has degree at
most $s-1$, we have
$\deg(F)\leq\Delta(s-1)
$. Therefore, the values of $F$ on any $\Delta(s-1)+1$ distinct elements of
$T$ uniquely determine it.

Now let $T'\subseteq T$ satisfy
$|T'|\geq (1-\epsilon)|T| = \Delta(s-1)+1$.
Clearly we can interpolate $F$ from the (ordered) evaluations of $C$ on the (ordered) set $\phi(T')$. 

Since
$C(\bu_i) = C\big(\phi(\alpha_i)\big) =F(\alpha_i)$
we can recover all evaluations of $C$ on the original reconstruction set
$\cR$. Applying the reconstruction algorithm associated with $\cR$ then
yields a circuit computing $C$. Hence, every $(1-\epsilon)$-fraction of
$\phi(T)$ is a reconstruction set for $\cC_{\Delta}$.

The bit complexity bound follows by the same calculation as in \cref{prop:interpolation_set_satisfies_hp}.
\end{proof}

Using the notion of robust reconstruction maps and \Cref{thm:rec-set-to-robust-map} we can easily show how to transform any reconstruction set to one that satisfies the properties in \Cref{def:hr}.

\begin{proposition} \label{prop:recon_set_satisfies_hr}
    Let $r,d,\Delta \in \N$, let $\cC$ be a polynomial class closed under scalar multiplication and let $\cR$ be a reconstruction set of size $s$ for the class $\cC_{\Delta}$. There exists a set of size at most $O(r^2 \cdot (s-1) \cdot \Delta)$ that  satisfies the requirements in \autoref{def:hr} for the class $\cC_{\Delta}$, with $\epsrcn = \frac{1}{2}$. Moreover, if $B_{\text{eval}}$ bounds the bit complexity of points in $\cR$ then the points in the new set have bit complexity $\poly(s, \log r, \log \Delta, B_{\text{eval}})$.
\end{proposition}

\begin{proof}
    Since $\cR$ is a reconstruction set it is also a hitting set for $\Sigma^2\cC_\Delta$.     
    Let $\phi(t)$ be the polynomial map interpolating through the set $\cR$, as in the proof of \Cref{thm:rec-set-to-robust-map}.
    
    Pick any $S \subseteq \F$ of size 
    \[
    |S| =
    \begin{cases}
        r^2 \cdot (s-1)\cdot \Delta, & r>1, \\
        3\cdot\big((s-1)\cdot \Delta+1\big), & r=1,
    \end{cases} \, .
    \] 
    We argue that $\phi(S)$ satisfies the requirements of \autoref{def:hr} with $\epsrcn = \frac{1}{2}$, where we index $S$ in an arbitrary way.

    When $r=1$ we get that \autoref{def:hr}~\eqref{item:recon prop of hr} holds since by \autoref{thm:rec-set-to-robust-map} $\phi(S)$ is an $\epsilon$-robust reconstruction set with
    \[
    \epsilon = 1 - \frac{\Delta \cdot (s - 1) + 1}{3 \cdot\left( \Delta \cdot (s-1)  +1\right)} > \frac{1}{2} \, .
    \] 
    By \autoref{rem:hr_for_r_eq_one}, this completes the proof when $r = 1$. 
    Henceforth, assume that $r \geq 2$. \autoref{def:hr}~\eqref{item:recon prop of hr} holds since by \autoref{thm:rec-set-to-robust-map} $\phi(S)$ is an $\epsilon$-robust reconstruction set with
    \[
    \epsilon = 1 - \frac{\Delta \cdot (s - 1) + 1}{r^2 \cdot \Delta \cdot (s-1) } > \frac{1}{2} \, .
    \] 

    Let $0 \not \equiv g \in \Sigma^2\cC_{\Delta}$.
    Since a reconstruction set for $\cC$ must distinguish any two different polynomials in it, we must have that $g \circ \phi(t) \not \equiv 0$. Considering its degree as a univariate polynomial, we see that it vanishes on at most a
    \[
        \frac{(s-1) \cdot \Delta }{r^2 \cdot (s-1) \cdot \Delta } = \frac{1}{r^2}
    \]
    fraction of the points in $\phi(S)$. Since $\epsrcn = \frac{1}{2} > \frac{1}{r^2}\cdot {r\choose2}$, it follows that $\phi(S)$ also satisfies \autoref{def:hr}~\eqref{item:hs s2 prop of hr}.
\end{proof}

\subsubsection{Constructing $\hre(\epsilon)$} \label{sssec:constr_hre}

Here we provide a construction for the set $\hre(\epsilon)$ defined in \cref{def:hre}. To do so, we combine the ideas from the previous two parts of this subsection.

\begin{construction}[Construction of \(\hre(\epsilon)\)]
    \label{cnstr:hre}
    Let $r,\Delta \in \N$, define $\kappa$ as in Equation~\eqref{eq:definition_of_kappa} and let $\cC$ be a polynomial class. Let \(S_{\text{pit}}\) hit \(\sum^2\prod^2\cC_{\Delta}\), and let
    \(S_{\text{recon}}\) be a reconstruction set for \(\cC_{\Delta}\). Denote
    \(L=|S_{\text{pit}}|+|S_{\text{recon}}|\) and choose
    \begin{equation*}
        \frac{1}{2}\leq\epsilon\leq\frac{2\kappa}{1+2\kappa}.
    \end{equation*}
    Let \(M:\F\to\F^n\) be a Lagrange interpolation curve
    passing through all points of \(S_{\text{pit}}\) and \(S_{\text{recon}}\) at
    prescribed points of $\F$. Each coordinate of \(M\)
    has degree at most \(L-1\).

    Choose a set \(A\subseteq\F\) of size
    \begin{equation*}
        |A|=N=\left\lfloor
            \frac{2\Delta\kappa(L-1)}{\epsilon}
          \right\rfloor+1,
    \end{equation*}
    and set
    \begin{equation*}
        \hre(\epsilon)=((\gamma,M(\gamma)))_{\gamma\in A}.
    \end{equation*}
\end{construction}

\begin{lemma}[Correctness of the construction]
    \label{lem:size_bound_of_hre}
    The indexed set in \cref{cnstr:hre} satisfies \cref{def:hre}, and
    \begin{equation*}
        N\leq4\Delta\kappa(L-1)+1.
    \end{equation*}
    If $B_{\text{eval}}$ bounds the bit complexity of points in $S_{\text{pit}}$ and $S_{\text{recon}}$, then each point in the new set has bit complexity $\poly(L, \log \Delta, \log n, \log r, B_{\text{eval}})$.
\end{lemma}

\begin{proof}
    Let \(0 \not \equiv g \in \Pi^{\kappa}\Sigma^2\Pi^2\cC_{\Delta}\).
    Every factor of $g$ remains nonzero after composition with \(M\),
    since \(M\) passes through \(S_{\text{pit}}\). Each factor has total
    degree at most \(2\Delta\), so
    \begin{equation*}
        0\neq g\circ M,
        \qquad
        \deg(g\circ M)\leq2\Delta\kappa(L-1)<\epsilon N.
    \end{equation*}
    Property~\eqref{item:hre_pit_prop} follows.

    For reconstruction, let \(f \in \cC_{\Delta}\), then \(f \circ M\) has degree at most
    \(\Delta(L-1)\). Any set \(A'\subseteq A\) of at least
    \((1-\epsilon)N\) indices satisfies
    \begin{equation*}
        |A'|\geq(1-\epsilon)N
        >\frac{1-\epsilon}{\epsilon}
            2\Delta\kappa(L-1)
        \geq\Delta(L-1),
    \end{equation*}
    where the last inequality follows from the fact that $\epsilon \leq \frac{2\kappa}{1+2\kappa}$.
    Thus its evaluations determine \(f \circ M\) by univariate
    interpolation and the evaluations of $f$ on $S_{\text{recon}}$ can be found. This satisfies Property~\eqref{item:hre_recon_prop}.

    Finally, \(\kappa \geq 1\), so the interval for \(\epsilon\)
    is nonempty. The size bound follows from \(\epsilon\geq1/2\).

    The bit complexity bound follows by the same calculation as in \cref{prop:interpolation_set_satisfies_hp}.
\end{proof}

\subsection{Sufficient conditions for reconstruction}
\label{subsec:suff_cond_for_main_thms}

We now give sufficient conditions for applying each of our two
reconstruction meta-algorithms. For the first algorithm, a reconstruction
set for \(\cC_\Delta\), as defined in \autoref{def:recon_set}, suffices
whenever \(d>(r+1)^4\Delta\). The second algorithm removes the
dependence on the total degree, but requires additional properties from \(\cC\) and also a hitting set
for \(\Sigma^2\Pi^2\cC_\Delta\).

We begin with the sufficient conditions for the first algorithm.

\begin{theorem}[Lifting reconstruction sets]
    \label{thm:recon_sets_imply_high_power_recon}
    Let \(r,d,\Delta\in\N\) satisfy
    \(d>(r+1)^4\Delta\), and let \(\cC\) be a polynomial class
    closed under scalar multiplication. Assume that \(\F\) has characteristic \(0\) or greater than \(2\Delta rd\). 
    
    Suppose that \(\cC_\Delta\) has a reconstruction set \(\cH\) of size
    \(s_{\mathrm{recon}}\), with a corresponding reconstruction
    algorithm running in time \(T_{\mathrm{recon}}\). 
    
    Then every \(P\in\Sigma^r\wedge^d\cC_\Delta\)
    can be reconstructed in time
    \begin{equation*}
        \poly\!\left(
            n,r,\Delta,d,p,
            s_{\mathrm{recon}},T_{\mathrm{recon}}
        \right).
    \end{equation*}
\end{theorem}

\begin{proof}
    Since \(\cH\) is a reconstruction set for \(\cC_\Delta\), it
    is also an interpolating set for this class. Applying
    \autoref{prop:interpolation_set_satisfies_hp} to \(\cH\)
    therefore yields a multiset \(\hp\) of size
    \begin{equation*}
        |\hp|
        =
        \left(r+\binom{r}{2}d\right)
        \Delta(s_{\mathrm{recon}}-1)+1
    \end{equation*}
    satisfying \autoref{def:hp}.

    Likewise, applying \autoref{prop:recon_set_satisfies_hr} to $\cH$ provides a
    set \(\hr\) satisfying \autoref{def:hr}, with
    \begin{equation*}
        |\hr|
        =
        O\!\left(
            1+r^2\Delta(s_{\mathrm{recon}}-1)
        \right).
    \end{equation*}
    Using these sets as the inputs \(\hp\) and \(\hr\) to
    \autoref{thm:gen recon algo} gives a reconstruction algorithm
    for \(P\). Substituting their size bounds and the running time of the univariate reconstruction algorithm yields the claimed running time.
\end{proof}

\begin{remark}
    The construction of \(\hp\) need not use the reconstruction
    set \(\cH\). A separate hitting set for
    \(\Sigma^2\cC_\Delta\) also suffices and may yield a smaller
    \(\hp\). Thus, in applications, it can be advantageous to use
    different sets for the construction of $\hp$ and $\hr$.
\end{remark}

We next give sufficient conditions for the small-individual-degree
algorithm of \cref{thm:small_individual_degree_reconstruction}.

\begin{theorem}
    \label{thm:suff_cond_for_high_power_recon_ext}
    Let \(n,r,d,\delta, \Delta\in\N\). Assume that \(d>(r+1)^4r\delta\), and
    let \(\cC\) be a polynomial class closed under scalar
    multiplication and substitutions. Assume that \(\F\) has characteristic \(0\) or greater than \(2r^2\delta d\).
    
    Suppose that \(\cC_{\Delta,\delta}\) admits a reconstruction set
    \(S_{\mathrm{recon}}\) of size \(s_{\mathrm{recon}}\), and
    that \(\Sigma^2\Pi^2\cC_{\Delta,\delta}\) admits a hitting set
    \(S_{\mathrm{pit}}\) of size \(s_{\mathrm{pit}}\).
    Let \(T_{\mathrm{recon}}\) be the running time of the
    reconstruction algorithm associated with
    \(S_{\mathrm{recon}}\). 

    Then every \(P\in\Sigma^r\wedge^d\cC_{\Delta,\delta}\) 
    can be
    reconstructed in time
    \begin{equation*}
        \poly\!\left(
            n^r,r,\delta,d, p,
            s_{\mathrm{pit}},
            s_{\mathrm{recon}},
            T_{\mathrm{recon}}
        \right).
    \end{equation*}
\end{theorem}

\begin{proof}
    We construct the sets and bound the subroutine costs required
    by \cref{thm:small_individual_degree_reconstruction}.

    First, applying \cref{cnstr:hre} to
    \(S_{\mathrm{pit}}\) and \(S_{\mathrm{recon}}\), with
    \(\epsilon=\frac12\), gives an ordered set
    \(\hre(1/2)\) satisfying \cref{def:hre}, of size
    \begin{equation*}
        |\hre(1/2)|
        =
        O\!\left(
            n\delta(n-r+1)\binom{r}{2}
            (s_{\mathrm{pit}}+s_{\mathrm{recon}})
        \right).
    \end{equation*}
    In particular, this size is polynomial in
    \(n,r,\delta,s_{\mathrm{pit}},s_{\mathrm{recon}}\).
    The associated robust reconstruction algorithm first performs
    univariate interpolation to recover the evaluations (of the polynomial under reconstruction) on
    \(S_{\mathrm{recon}}\), and then applies the reconstruction
    algorithm for \(\cC_\Delta\). Its running time is therefore
    polynomial in these parameters and \(T_{\mathrm{recon}}\).

    Next, \(S_{\mathrm{pit}}\) also hits $r$-variate restrictions of \(\cC_\Delta\), hence it can serve as $\cH_{r}$.

    To construct \(\cH_{\mathrm{Alt}}\), let
    \(M:\F\to\F^n\) be a curve passing through
    \(S_{\mathrm{pit}}\). Such a curve can be built with coordinate degrees at most
    \(s_{\mathrm{pit}}-1\). Every nonzero polynomial in
    \(\Sigma^2\Pi^2\cC_\Delta\) remains nonzero after composition
    with \(M\), and hence so does every product of such
    polynomials. A product of \(\binom{r}{2}\) such factors has
    total degree at most \(2\Delta\binom{r}{2}\), so evaluating
    \(M\) at
    \begin{equation*}
        2\Delta\binom{r}{2}(s_{\mathrm{pit}}-1)+1
    \end{equation*}
    distinct points yields a hitting set of the required
    type. Thus
    \begin{equation*}
        |\cH_{\mathrm{Alt}}|
        =
        O\!\left(1+r^2\Delta s_{\mathrm{pit}}\right).
    \end{equation*}

    We now bound the cost of reconstructing the restricted
    polynomials. Every restriction of a polynomial in $\cC_{\Delta}$ to at most \(r\) variables
    belongs to \(\cC\) and has total degree at most \(r\delta\).
    Moreover, \(S_{\mathrm{recon}}\) can obviously serve as a reconstruction set for the corresponding
    restricted class. 

    Using the bound $d > (r+1)^4(r\delta)$ and 
    \cref{thm:recon_sets_imply_high_power_recon} with \(S_{\mathrm{recon}}\) as the reconstruction set gives us
    \begin{equation} \label{eq:bounding_T_meta_run_time}
        T_{\mathrm{meta}}
        =
        \poly\!\left(
            n,r,d,\delta,  p,
            s_{\mathrm{recon}},T_{\mathrm{recon}}
        \right).
    \end{equation}

    The set \(\hp^{(r)}\) constructed in \cref{thm:recon_sets_imply_high_power_recon} can serve as a hitting set for the differences of two \(d\)-th powers from $\cC$, which can be used in the tests that occur in \Cref{line:indv alignment} for alignment. This implies that $T_{\mathrm{pit}}$ is also bounded by the right hand side of Equation~\eqref{eq:bounding_T_meta_run_time}.

    Substituting these set-size  and subroutine bounds into
    \cref{thm:small_individual_degree_reconstruction} yields
    the running time
    \begin{equation*}
        \poly\!\left(
            n^r,r,\delta,d,  p,
            s_{\mathrm{pit}},
            s_{\mathrm{recon}},
            T_{\mathrm{recon}}
        \right),
    \end{equation*}
    as claimed.
\end{proof}


\section{Sparse Polynomials}\label{sec:sparse}

In this section we give our results for sums of powers of sparse polynomials. We treat sparse inner polynomials separately from the other bounded-depth models as in this case we obtain both PIT and reconstruction for every $d$.  For the more general bounded-depth classes, our approach yields PIT but not reconstruction.

We prove the following unconditional PIT and reconstruction results.

\begin{theorem}[PIT for sums of powers of sparse polynomials] \label{thm:sum_pow_sparse_poly_pit}
   Let $n,d,r,s,\Delta\in \N$ and let $\F$ be a field of characteristic $p=0$ or $p > rd\cdot s^3\Delta^3n^2$. There is a hitting set for the class of $\Sigma^{r}\!\wedge^{d}\!\Sigma^{s} \Pi^{\Delta}$ of size
   \[
   \begin{cases}
     \poly(s,n,\Delta,d) & d +1 \geq (r-1)^2\\
     \poly(s^{r^2},r,n,\Delta) & d +1< (r-1)^2
   \end{cases}.
   \]
   In particular, when $r=O(1)$ the size of the hitting set is polynomial in $n$ and when $r=\poly\log(n)$ it is of quasi-polynomial size, regardless of $d,\Delta$.
\end{theorem}

\begin{theorem}[Reconstruction for sums of powers of sparse polynomials] \label{thm:sum_pow_sparse_poly_recon}
   Let $n,d,r,s,\Delta, \delta\in \N$ and let $\F$ be a field of characteristic $p$. There is a reconstruction algorithm for the class of $\Sigma^{r}\!\wedge^{d}\!\Sigma^{s} \Pi^{\Delta}$ according to the following table \begin{table}[H]
    \centering
    \small
    \renewcommand{\arraystretch}{1.5}
    \setlength{\tabcolsep}{4pt}

    \begin{threeparttable}
        \caption{Deterministic reconstruction for
        \(\Sigma^r\wedge^d\Sigma^s\Pi^\Delta\)}
        \label{tab:sparse_reconstruction_results}

        \medskip

        \begin{tabularx}{\textwidth}{@{}
            >{\centering\arraybackslash\hsize=1\hsize}X
            >{\centering\arraybackslash\hsize=1.20\hsize}X
            >{\centering\arraybackslash\hsize=0.80\hsize}X
            >{\centering\arraybackslash\hsize=0.6\hsize}X
        @{}}
            \toprule
            \textbf{Power regime}
            &
            \textbf{Running time}
            &
            \textbf{Characteristic}
            &
            \textbf{Reconstruction}
            \\
            \midrule

            \(d>(r+1)^4\Delta\)
            &
            \(\poly\!\bigl(n,r,s,d,\Delta , p\bigr)\)
            &
            \(p=0\) or
            \(p>2\Delta rd\)
            &
            Proper
            \\[1ex]

            \(d>(r+1)^4r\delta\)
            &
            \(\poly\!\bigl(n^r,r,s,d,\delta , p\bigr)\)
            &
            \(p=0\) or \(p>2r^2\delta d\)
            &
            Proper
            \\[1ex]

            otherwise
            &
            \(\poly\!\bigl(n,r,s^{d},d,\Delta,   p\bigr)\)
            &
            Any characteristic
            &
            Non-proper
            \\

            \bottomrule
        \end{tabularx}

    \end{threeparttable}
\end{table}
\end{theorem}


\subsection{PIT and reconstruction set for sparse polynomials}
\label{sec:KS}

We recall the generator of Klivans and Spielman~\cite{KlivansSpielman01}, which yields a hitting set as well as a reconstruction set for sparse polynomials.
While these results are known, we repeat them here for the sake of explicitness.

For a prime number $q$, and an integer $k<q$ let $\Psi_{k,q}:\F[\vx]\mapsto \F[y]$ be defined as 
\begin{equation}\label{eq:psi-k-q}
\Psi_{k,q}(x_i)=
y^{(k^{i-1}\bmod q)},    
\end{equation}
with the natural extension to monomials and polynomials. Namely,  for an exponent vector $\ve=(e_1,\ldots,e_n)$
\[
\Psi_{k,q}(\vx^\ve)= y^{\Sigma_{i} e_i\left(k^{(i-1)}\bmod q\right)}.
\]
Clearly, 
\begin{equation}\label{eq:deg-psi}
 \deg(\Psi_{k,q}(\vx^\ve))\leq  (q-1)\Sigma_{i}e_i=(q-1)\deg(\vx^\ve) \ .   
\end{equation}

\begin{theorem}[Corollary of Klivans–Spielman~{\cite[Lemma~1]{KlivansSpielman01}}]\label{thm:PIT:sparse:KS}
Let $f\in\F[\vx]$ be a nonzero $n$-variate polynomial, such that $f$ has at most~$s$ monomials and  $\deg(f)\leq \Delta$.
Let $q>\Delta$ be a prime.
Then, for all but at most $\binom{s}{2}(n-1)$ values of $k\in[q-1]$, the univariate polynomial $\Psi_{k,q}(f)$
has the same number of monomials as $f$. Furthermore, $\deg(\Psi_{k,q}(f))\leq \Delta q$. Moreover, for every prime $q>\Delta$, the polynomial
$\Psi_{k,q}(f)$ is nonzero for all but at most
$(s-1)(n-1)$ values of $k\in[q-1]$.
\end{theorem}

\begin{corollary} [Hitting Set for Sparse Polynomials] \label{cor:hs_sparse_poly}
There exists an explicit hitting set for $\Sigma^{s} \Pi^\Delta$ circuits of size  \[O\big(\big((s-1)(n-1) + 1\big)(\Delta q + 1)\big) \leq O\big(s^2n^2\Delta^2\big) .\]
\end{corollary}

\begin{proof}
    By Bertrand's postulate there exists a prime number  $q \in \big[ sn + \Delta + 1, \; 2(sn + \Delta + 1) \big]$.
    Let $f \in \Sigma^s\Pi^\Delta$. Let $T \subseteq [q-1]$ of size $|T| = (s-1)(n-1)+1$. \autoref{thm:PIT:sparse:KS} implies that there exists $k \in T$ such that $\Psi_{k,q}(f) \not \equiv 0$ is a polynomial of degree at most $\Delta \cdot q$. Pick $S \subseteq \F$ of size $|S| = \Delta \cdot q + 1 \leq 2sn\Delta + 2\Delta^2 + 3\Delta$. Then,
    \[
        \Psi_{k,q}(f)|_{S} \neq \0.
    \]
    Hence, \[\cH = \bigcup_{k\in T} \Psi_{k,q}(\x)|_{S} = \Big\{\big(\Psi_{k,q}(x_1),\ldots,\Psi_{k,q}(x_n)\big)\big|_{y=s} \mid s \in S,\ k\in T \Big\}\] is a hitting set for $\Sigma^s\Pi^\Delta$ of size \[|\cH| =|T| \cdot |S| = O(\Delta^2s^2n^2).\qedhere\]
\end{proof}

We next use the well-known generator, which follows from the work of Klivans-Spielman, in order to create a reconstruction set for sparse polynomials.  
For the construction we shall consider the modified generator
\begin{equation}\label{eq:ks-gen-modified}
\Psi_{j,k,q,\lambda}(x_i)=
 \begin{cases}
  \lambda \cdot y^{(k^{j-1}\bmod q)} & i=j\\
  y^{(k^{i-1}\bmod q)} & i\ne j
  \end{cases}.    
\end{equation}

\begin{construction}[Reconstruction set for sparse polynomials]\label{construction:rec-ks}
Let $n,s,\Delta\in \N$.
Let 
\[2\Delta s^2n^2 <q < 4\Delta s^2n^2\] be a prime number, and let $\cA_{\Delta,q}\subset\F$ be a set of size
\[|\cA_{\Delta,q}| =  \Delta q.\]  
Let $\lambda\in \F$ (or $\lambda\in \mathbb{E}$ if needed) have multiplicative order larger than $\Delta $.

Denote 
\begin{equation*} \cS_{n,\Delta,q,\lambda}:=\bc{\Psi_{j,k,q,\lambda}[\alpha] \mid j\in [n],  k\in [q-1], \alpha\in\cA_{\Delta,q}}\ \cup \  \bc{\Psi_{k,q}[\alpha] \mid k\in [q-1], \alpha\in\cA_{\Delta,q}} \ .
\end{equation*}
\end{construction}

\begin{theorem}\label{thm:sparse-rec}
    Let $n,s,\Delta\in \N$.
    \Cref{construction:rec-ks} yields a reconstruction set for $n$-variate polynomials in $\Sigma^s\Pi^\Delta$  of size $O\left(\Delta^3 n^5s^4  \right)$.
\end{theorem}

\begin{proof}
    We first observe that 
    \[| \cS_{n,\Delta,q,\lambda}|=(n+1) \cdot (q-1) \cdot |\cA_{\Delta,q}|= O\left(\Delta nq^2  \right)= O\left(\Delta^3 n^5s^4  \right)\ .\]

    Note that for any polynomial $f$ of degree $\Delta$, $\deg(f\circ \Psi_{j,k,q,\lambda})\leq (q-1)\Delta$.
    We further observe that each monomial $\x^{\e}$ is mapped to
 \[
 \Psi_{j,k,q,\lambda}(\x^{\e})= \lambda^{e_j}\cdot y^{\Sigma_{i} e_i\left(k^{(i-1)}\bmod q\right)}.
 \]
 Since the order of $\lambda$ is larger than $\Delta\geq e_j$, if we know both $\Psi_{k,q}(\x^{\e})$ and $\Psi_{j,k,q}(\x^{\e})$ then we can recover the exponent of $x_j$ from each of the monomials by simple interpolation. 

    Let $f\in \Sigma^s\Pi^\Delta$. Since $|\cA_{\Delta,q}|=q\Delta>\deg(f\circ \Psi_{k,q})$, it follows that from the evaluations of $f$ on the set $ \  \bc{\Psi_{k,q}[\alpha] \mid k\in [q-1], \alpha\in\cA_{\Delta,q}} $  we can interpolate  $f\circ \Psi_{k,q}$. Similarly, knowing the values of $f$ on the set $\bc{\Psi_{j,k,q,\lambda}[\alpha] \mid j\in [n],  k\in [q-1], \alpha\in\cA_{\Delta,q}}$ we can interpolate $f\circ \Psi_{j,k,q}$.

 For any monomial $\vx^\ve$ in $f$, denote $m_\ve:=\Psi_{k,q}(\x^{\ve})$ be its image under $\Psi_{k,q}$.
 Comparing the coefficient of $m_\ve$ in the different interpolations, we can  recover the exponent of $x_j$ in $\vx^\ve$, as well as the coefficient of $\vx^\ve$ in $f$. 
\end{proof}


\subsection{PIT for sums of powers of sparse polynomials} \label{sec:pit_sparse_poly}

In this section, we prove \Cref{thm:sum_pow_sparse_poly_pit} that gives a hitting set for  $\Sigma^r\wedge^d\Sigma^s\Pi^\Delta$ circuits.

The proof for the high degree case is based on \Cref{thm:bb pit for high powers}. In the case of small powers we write a   $\Sigma^r\wedge^d\Sigma^s\Pi^\Delta$ circuit as a sparse polynomial (with much larger sparsity) and apply \Cref{cor:hs_sparse_poly}.

\begin{proof}[Proof of \Cref{thm:sum_pow_sparse_poly_pit}]
    Assume first that $d + 1 \geq (r-1)^2$. 
    In this case the claim follows by applying \Cref{thm:bb pit for high powers} with the hitting set for $2s$-sparse polynomials given in \Cref{cor:hs_sparse_poly}.

    Assume now that $d + 1 < (r-1)^2$. It is clear that any polynomial in $\wedge^d\Sigma^s\Pi^\Delta$ is also an $\binom{s+d-1}{d}$-sparse polynomial of degree at most $d\Delta$. Thus, $\Sigma^r\wedge^d\Sigma^s\Pi^\Delta \subseteq \Sigma^{r\cdot \binom{s+d-1}{d}}\Pi^{d\Delta}$. \Cref{cor:hs_sparse_poly} gives a hitting set of size 
    \[\br{r\cdot \binom{s+d-1}{d}}^3n^2d^2\Delta^2 < r^7n^2\Delta^2s^{3r^2}
    \]
    for the larger class, where we have used the assumption that $d<r^2$.
\end{proof}

\subsection{Reconstruction of sums of powers of sparse polynomials}

We now give reconstruction algorithms for
\(\Sigma^r\wedge^d\Sigma^s\Pi^\Delta\) circuits. As in
\cref{sec:pit_sparse_poly}, we distinguish between large and small values of the powering parameter. For large values, when \(d>(r+1)^4\Delta\), we apply
our total-degree reconstruction meta-theorem and when \(d > (r+1)^4r\delta\), we apply our individual-degree reconstruction meta-theorem. For small values, we reconstruct the sparse expansion of the input polynomial, getting a polynomial of the form $\Sigma^{s'}\Pi^{d\Delta}$. 
The hitting set and reconstruction set in \cref{cor:hs_sparse_poly} and \cref{thm:sparse-rec} can be constructed with points of bit complexity \(\operatorname{poly}(n,s,\Delta)\) over \(\mathbb Q\). Over finite fields, the bound is \(\operatorname{poly}(n,s,\Delta,\log|\mathbb F|)\).

\paragraph{Reconstruction using the total-degree bound.}
Our first result follows by applying
\cref{thm:recon_sets_imply_high_power_recon} using the sparse polynomials reconstruction set.

\begin{proposition}
    \label{prop:sparse_recon_high_total_degree}
    Let \(n,r,s,d,\Delta \in\N\) satisfy \(d>(r+1)^4\Delta\). Assume
    \(\F\) has characteristic \(0\) or greater than
    \(2\Delta rd\). Then, there is a deterministic reconstruction algorithm for
    \(\Sigma^r\wedge^d\Sigma^s\Pi^\Delta\)
    running in time
    \begin{equation*}
        \poly(n,r,s,d,\Delta, p).
    \end{equation*}
\end{proposition}

\begin{proof}
    By \cref{thm:sparse-rec} there is
    a reconstruction set of size $s_{\mathrm{recon}}$ and reconstruction algorithm with running time $T_{\mathrm{recon}}$ satisfying
    \begin{equation*}
        s_{\mathrm{recon}}=O(\Delta^3n^5s^4),
        \qquad
        T_{\mathrm{recon}}=\poly(n,s,\Delta).
    \end{equation*}
    The theorem follows by applying
    \cref{thm:recon_sets_imply_high_power_recon} with these parameters.
\end{proof}

\paragraph{Reconstruction using the individual-degree bound.}
To reconstruct in the case of small-individual-degree, we also need a
hitting set for $\Sigma^2\Pi^2\Sigma^s\Pi^{\Delta}$ in order to apply \Cref{thm:suff_cond_for_high_power_recon_ext}.
For sparse polynomials, this follows using the simple observation that the product of two $s$-sparse polynomials has at most $s^2$ monomials and degree $2\Delta$ and a sum of two $s^2$-sparse polynomials has at most \(2s^2\) monomials. Then use \cref{cor:hs_sparse_poly}.

\begin{observation}
    \label{lem:sparse_partial_alternant_hs}
    There is an explicit hitting set for \(\Sigma^2\Pi^2\Sigma^s\Pi^\Delta\) of size \(O(s^4n^2\Delta^2)\).
\end{observation}

\begin{proposition}
    \label{prop:sparse_recon_small_individual_degree}
    Let \(n,r,s,d,\Delta,\delta \in\N\) and suppose that \(d > (r+1)^4r\delta\).
    Assume that \(\F\) has characteristic \(0\) or greater than \(2\delta r^2 d\).
    Then, there is deterministic reconstruction algorithm for \(\Sigma^r\wedge^d\Sigma^s\Pi^\Delta\) 
    whose inner sparse polynomials have individual degree at most $\delta$, running in time 
    \begin{equation*}
        \poly(n^r,r,s,d,\delta, p).
    \end{equation*}
\end{proposition}

\begin{proof}
    Let \(\cC\) be the class of \(s\)-sparse polynomials with
    total degree at most \(\Delta\) and individual degree at most
    \(\delta\). Let \(S_{\mathrm{pit}}\) be the hitting set from \cref{lem:sparse_partial_alternant_hs} and \(S_{\mathrm{recon}}\) the reconstruction set from \cref{thm:sparse-rec}, then these sets have sizes
    \[
         s_{\mathrm{pit}}=O(s^4n^2\Delta^2) \qquad \text{and} \qquad
        s_{\mathrm{recon}}=O(\Delta^3n^5s^4),
    \]
    respectively. Moreover, the reconstruction algorithm has running time \[
    T_{\mathrm{recon}}=\poly(n,s,\Delta).
    \]
    Apply
    \cref{thm:suff_cond_for_high_power_recon_ext} to \(\cC\),
    using these sets. Since we may replace $\Delta$ by $\min\{\Delta,n\delta\}$, the running time of the algorithm is bounded by \[
    \poly(n^r,r,s,d,\delta,\Delta,  p)
        =\poly(n^r,s,d,\delta , p),    
    \] as claimed.
\end{proof}

\paragraph{General non-proper reconstruction.}

In the general case, as observed in \cref{sec:pit_sparse_poly}, every polynomial computed by a
\(\Sigma^r\wedge^d\Sigma^s\Pi^\Delta\) circuit can be also computed by a
\(\Sigma^{r\cdot \binom{s+d-1}{d}}\Pi^{d\Delta}\) circuit. Therefore, by applying the reconstruction algorithm from \cref{thm:sparse-rec}, we obtain a non-proper reconstruction of the original polynomial. This immediately implies the following.

\begin{proposition} \label{cor:sparse_recon_gen_non_proper}
    Let \(n,r,s,d,\Delta \in\N\). There is a deterministic non-adaptive reconstruction algorithm for \(\Sigma^r\wedge^d\Sigma^s\Pi^\Delta\) 
    running in time
    \begin{equation*}
        \poly(n,r,s^d,d,\Delta).
    \end{equation*}
    This reconstruction algorithm returns a \(\Sigma^{r\cdot \binom{s+d-1}{d}}\Pi^{d\Delta}\) circuit, hence it is non-proper.
\end{proposition}

\medskip

To prove \cref{thm:sum_pow_sparse_poly_recon} we apply \cref{prop:sparse_recon_high_total_degree}, \cref{prop:sparse_recon_small_individual_degree} and \cref{cor:sparse_recon_gen_non_proper} with the corresponding bounds on $d$.


\section{Algebraic Branching Programs}\label{sec:roabp}

Algebraic branching programs form a well studied model, whose power lies between the power of general circuits and formulas. 
In this section we give our results for this model. 

We first state our main PIT result. This is a formal version of \Cref{thm:intro:pit-abp}.
To ease the reading we introduce the following notation.
The next function is used throughout this section
\begin{equation}
\label{eq:roabp_intro_consts}
    \lambda(x) = \frac{\log n}{\max\left(1,\log\log n - \log \log (x)\right)}.
\end{equation}

 \begin{theorem}[PIT for sums of powers of $\roabp$s]\label{thm:main-pit-abp}
 Let $n,r,d,\delta\in \N$. Let $\F$ be a field of characteristic $p$ such that $p=0$ or $p$ is sufficiently large.
 Then, there are explicit hitting sets as in \Cref{tab:roabp-hitting-set-sizes} below, where we also list the requirement on positive characteristic.
\begin{table}[h]
\centering

\setlength{\arraycolsep}{12pt}
\[
\begin{array}{lll} 
    \toprule
    \multirow{2}{*}{\textbf{\(\roabp\) type}} &
    \multicolumn{2}{c}{
        \hspace{-3cm}\textbf{Hitting set size } T
    }
    \\
    &
    \textbf{\(d+1 \geq (r-1)^2\)}
    &
    \textbf{\(d+1 < (r-1)^2\)}
    \\
    \midrule
  
    \text{arbitrary orders}
        &
        \displaystyle
        d\cdot
        \left(n\delta w\right)^{
            O\left(\log (n\delta w)\right)
        }
        &
        \displaystyle
        \left(n\delta w\right)^{
            2^{O(r)}\log(n\delta w)
        }
        \\

    \indent\charac(\F)
        & p>rT
        & p>d\delta
        \\

    \textbf{quasi-poly regime}
        &  
        \text{always}
        & r=O(\log\log n)
        \\[2.5ex]

    \text{same variable order}
        &
        d\cdot \poly\left(
            \delta,
            (2nw)^{\lambda(2w)}
        \right)
        &
        O\left(
            \delta\cdot
            (nw)^{
                O\left(r^2\lambda(w^{r^2})\right)
            }
        \right)
        \\

    \indent\charac(\F)
        & p>rT
        & p>d\delta
        \\

    \textbf{quasi-poly regime}
        & \text{always}
        & r=\poly(\log n)
        \\[2.5ex]

    \text{any-order}
        &
        d\cdot
        \poly\left(
            n\delta,\,
            w^{\log\log w}
        \right)
        &
        \begin{cases}
            \poly\left(
                n\delta,\,
                r^{\log\log r}
            \right),
                & w=1, \\[1ex]
            \poly\left(
                n\delta,\,
                w^{r^2(\log r+\log\log w)}
            \right),
                & w>1
        \end{cases}
        \\

    \indent\charac(\F)
        & p>rT
        & p>n^4(d\delta+1)^2
        \\

    \textbf{quasi-poly regime}
        & \text{always}
        & r=\poly(\log n)
        \\

    \bottomrule
\end{array}
\]\caption{Hitting set sizes for sums of powers of $\roabp$s
}
\label{tab:roabp-hitting-set-sizes}
\end{table}
\end{theorem}

Our reconstruction result builds on a deterministic non-adaptive
algorithm for a single \(\roabp\) given in \cref{sec:non_adaptive_algo_roabp}.
Using this algorithm and its associated reconstruction set, we apply
our two reconstruction meta-algorithms in the high-power regime.
In the low-power regime, we convert a sum of powers of \(\roabp\)s
sharing a common variable order into a single \(\roabp\) with larger
width and individual degree. This yields non-proper reconstruction.

\begin{theorem}[Reconstruction for sums of powers of $\roabp$s] \label{thm:main-rec-abp}
   Let $n,d,r,w,\delta,\Delta\in \N$ and let $\F$ be a field of characteristic $p=0$ or sufficiently large. Consider \(P=\sum_{i=1}^{r'}\alpha_i f_i^d\), 
   where \(r'\le r\), each \(f_i\) has an $\roabp$ of width at most \(w\) and individual degree at most \(\delta\), and \(\deg f_i\le\Delta\). The required variable-order information is specified for each case in \cref{tab:roabp_reconstruction_results}, where we also list the requirement on positive characteristic. The non-proper algorithms return a single
    \(\roabp(W,D)\), where \begin{equation*}
        W=r\binom{w+d-1}{d} 
        ,\qquad D= d\delta \leq(r+1)^4r\delta^2.
    \end{equation*}
\end{theorem}

When $w=O(1)$, $W \le r(d+1)^{w-1}$ is polynomial in $r$ and $d$; for any-order inner $\roabp$s with $r, \delta = O(1)$ this makes the non-proper reconstruction polynomial-time.

\begin{table}[t]
    \centering
    \small
    \renewcommand{\arraystretch}{1.35}
    \resizebox{\textwidth}{!}{%
    \begin{tabular}{@{}lll@{}}
        \toprule
        \textbf{Underlying \(\roabp\)s}
        & \textbf{Running time}
        & \textbf{Positive characteristic} \\
        \midrule
        \multicolumn{3}{@{}l}{\textbf{Proper reconstruction:
        \(d>(r+1)^4\Delta\)}} \\[0.5ex]
        arbitrary orders
        & \(\poly\!\left((n\delta w)^{O(\log(n\delta w))},r,d , p\right)\)
        & \(p>2\Delta rd\) \\
        same variable order
        & \(\poly\!\left((nw)^{\lambda(w)},\delta,r,d , p\right)\)
        & \(p>2\Delta rd\) \\
        any-order
        & \(\poly\!\left(n,(w)^{\log\log(w)},\delta,r,d , p\right)\)
        & \(p>\max\{2\Delta rd,n^4(\delta+1)^2\}\) \\
        \midrule
        \multicolumn{3}{@{}l}{\textbf{Proper reconstruction:
        \(d > (r+1)^4r\delta\)}} \\[0.5ex]
        same variable order
        & \(\poly\!\left(n^r,(2nw^2)^{\lambda(2w^2)},\delta,r,d , p\right)\)
        & \(p>2 r^2\delta d\) \\
        any-order
        & \(\poly\!\left(n^r,(2w)^{\log\log(2w)},\delta,r,d , p\right)\)
        & \(p>\max\{2 r^2\delta d,n^4(2\delta+1)^2\}\) \\
        \midrule
        \multicolumn{3}{@{}l}{\textbf{Non-proper reconstruction:
        \(d\le (r+1)^4r\delta\)}} \\[0.5ex]
        same variable order
        & \(\poly\!\left(n,D,(nW)^{\lambda(W)} \right)\)
        & \(p>D\) \\
        any-order
        & \(\poly\!\left(n,D,(W)^{\log\log(W)} \right)\)
        & \(p>n^4(D+1)^2\) \\
        \bottomrule
    \end{tabular}%
    }
    \caption{Deterministic reconstruction for sums of powers of $\roabp$s}
    \label{tab:roabp_reconstruction_results}
\end{table}

\paragraph{Organization:}
The section is organized as follows. We begin by defining the branching-program models used in this section. We then give PIT results for sums of powers of some of these models.
Next, we give the non-adaptive reconstruction algorithm for a single
\(\roabp\). Finally, we combine these ingredients with our reconstruction
meta-theorems to prove \cref{thm:main-rec-abp}.

\subsection{Models of ABPs}\label{sec:abp-def}

We begin with definitions, most of which are taken from \cite{AndersonFSSV18}.

\begin{definition} [{Algebraic Branching Program \cite{NisanABPDef}}]
An Algebraic Branching Program ($\abp$) is a directed acyclic graph with one vertex $s$ of in-degree zero (the source) and one vertex $t$ of out-degree zero (the sink). The vertices of the graph are partitioned into layers labeled $0,1,\ldots,L$. Edges in the graph can only go from layer $\ell - 1$ to layer $\ell$, for $L \geq \ell \geq 1$. The source is the only vertex at layer $0$ and the sink is the only vertex at layer $L$. Each edge is labeled with a polynomial over a field $\F$ in the input variables $\x$. Each path from $s$ to $t$ computes the polynomial that is the product of the labels of the path edges, and the $\abp$ computes the sum, over all $s$ to $t$ paths, of such polynomials.
The \emph{width} of an $\abp$ is the maximum number of nodes in any layer, and the \emph{size} of an $\abp$ is the number of vertices in the $\abp$.
\end{definition}

In this work we only consider restricted types of $\abp$s. 

\begin{definition} [{Read-$k$ Oblivious $\abp$s}]
    An algebraic branching program is said to be \emph{oblivious} if for every layer $n \geq \ell \geq 1$, all the edge labels entering that layer are labeled with univariate polynomials in a variable $x_{i_\ell}$. Such a branching program is said to be a \emph{read-once oblivious $\abp$ ($\roabp$)} if the $x_{i_{\ell}}$'s are distinct variables, that is, each $x_i$ appears in the edge labels of at most one layer. An oblivious $\abp$ is said to be a \emph{read-$k$} if each variable $x_i$ appears in the edge labels of at most $k$ layers. The \emph{individual degree} of an $\roabp$ is defined to be the maximum degree of the polynomial edge labels.

    We denote by $\roabp(w,\delta)$ the class of polynomials computed by a width-$w$ and individual degree $\delta$ read-once oblivious $\abp$s.
\end{definition}

\begin{definition} [{$k$-Pass $\abp$s}] \label{def:k-pass-abp}
    An oblivious $\abp$ is said to be a \emph{$k$-pass $\abp$} if there exists a permutation $\pi$ on $[n]$ such that the $\abp$ reads variables in the order
    $$
       \underbrace{x_{\pi(1)}, x_{\pi(2)},\ldots,x_{\pi(n)},x_{\pi(1)},x_{\pi(2)},\ldots,x_{\pi(n)},\ldots,x_{\pi(1)},x_{\pi(2)},\ldots,x_{\pi(n)}}_{k \text{ repetitions}} \, .
    $$
\end{definition}

\begin{definition} [{Any-order}]
A polynomial $f$ is computed by a $\roabp$, width $w$ and individual degree $\delta$ in \emph{any-order} if for every permutation $\pi$ of $[n]$, we can find a $\roabp$ with the variable order $\pi$ and these width and individual degree that computes $f$. 

We denote by $\aoroabp(w,\delta)$ the class of polynomials computed by a width-$w$ and individual degree $\delta$ $\roabp$s in any-order.
\end{definition}

In many cases, it is useful to view $\abp$s as a product of matrices. We will state this explicitly only for $\roabp$s.

\begin{lemma} \cite[Lemma~3.1]{roabp_recon_adaptive_fs13} \label{lem:roab_as_prod_of_matrices}
Let $f \in \roabp(w,\delta)$ be with variable order $x_1,x_2,\ldots,x_n$. There exists $1 = w_0, w_1,\ldots,w_n = 1$ such that $\forall i :w_i \leq w$ and matrices $M_i \in \F[x_i]^{w_{i-1} \times w_i}$ of degree at most $\delta$ such that
\[
    f(\x) = \prod_{i=1}^n M_i(x_i) .
\]
Conversely, any such function can be computed by a $\roabp(w,\delta)$ with variable order $x_1,x_2,\ldots,x_n$.
\end{lemma}

The following simple lemma will also be useful for us.

\begin{lemma}
\label{lem:product-roabps}
Let $f,g$ be width-$w$, individual-degree-$\delta$ $\roabp$s in the same
variable order. Then $fg$ is a width-$w^2$, individual-degree-$2\delta$
$\roabp$ in the same variable order.
\end{lemma}

\begin{proof}
Write
\[
f=A_1(x_1)\cdots A_n(x_n),
\qquad
g=B_1(x_1)\cdots B_n(x_n).
\]
Then
\[
fg
=
(A_1(x_1)\otimes B_1(x_1))
\cdots
(A_n(x_n)\otimes B_n(x_n)).
\]
Each tensor-product matrix has width at most $w^2$, and every entry has
degree at most $2\delta$ in the corresponding variable.
\end{proof}

\begin{corollary}\label{cor:sum-2-prod-2-roabp}
\[
\Sigma^2\Pi^2\ROABP(w,\delta)
\subseteq
\ROABP(2w^2,2\delta),
\]
where all the involved $\roabp$s have the same order.
\end{corollary}

\subsection{PIT for sums of powers of $\roabp$s}

The proof of \Cref{thm:main-pit-abp} for the case of $d+1\ge (r-1)^2$ is based on \Cref{thm:bb pit for high powers}. To apply the theorem we require appropriate hitting sets for the base class. We give these results in the next subsection and then give the full proof of \Cref{thm:main-pit-abp}.

\subsubsection{Known PIT for models of Algebraic Branching Programs}\label{sec:known-pit-abp}

We state the currently best black-box PIT results for various types of $\roabp$s.

\begin{theorem} [{PIT for $\roabp$s \cite[Theorem 1]{guo-gurjar-roabps-pit}, \cite[Theorem~3.6]{GurjarKS17}}] \label{thm:roabp_hs}
    Denote $\charac(\F) = p$. There exists an explicit hitting set for $\roabp(w,\delta)$ whenever $p = 0$ or $p > \delta$. 
    The size of the hitting set is 
    \[
    \poly\left(\delta, (nw)^{\lambda(w)}\right).
    \]
    If the order is known and $p=0$ or $p>\delta n^{\log w}$, then there is a hitting set of size \[
    \poly\left(\delta, n^{\log w}\right).
    \]
\end{theorem}

For any-order $\roabp$s a better size bound can be given.

\begin{theorem} [{PIT for any-order $\roabp$s \cite[Theorem 3]{guo-gurjar-roabps-pit}}] \label{thm:ao_roabp_hs}
    Denote $\charac(\F) = p$. There exists an explicit hitting set for $\aoroabp(w,\delta)$ whenever $p = 0$ or $p > n^4(\delta+1)^2$. 
    The size of the hitting set is 
    $$
    \poly\left(n\delta, w^{\log \log w}\right) \, .
    $$
\end{theorem}

We will also use the next result for sum of $c$ $\roabp$s.

\begin{theorem} [{PIT for sums of $\roabp$s \cite[Theorem 4.9]{gurjar_et_al_sum_roabps}}] \label{thm:sum_roabp_hs}
    Denote $\charac(\F) = p$. There exists an explicit hitting set for $\Sigma^{c} \roabp(w,\delta)$ whenever $p = 0$ or $p > \delta$. 
    The size of the hitting set is 
    \[
    (n\delta w)^{O(c2^c \cdot \log (n\delta w))}.
    \]
\end{theorem}




\subsubsection{PIT for high powers} \label{sec:pit_results_roabps}

\begin{proof}[{Proof of \Cref{thm:main-pit-abp} when $d+1\ge (r-1)^2$}]

For each of the models under consideration we apply \Cref{thm:bb pit for high powers}. 

Recall that the requirement in \Cref{thm:bb pit for high powers} is the existence of a hitting set $\cH$ for $\Sigma^2 \cC_\Delta$, depending on the model $\cC_\Delta$ under consideration. Furthermore, if the size of that hitting set is $|\cH|=A$ then we obtain a hitting set of size $T=A\Delta d$ and we require that the characteristic of $\F$ is either $p = 0$ or $p > (r-1) d \Delta A$ so it is enough to require $p>rT$. 
In our case, since the individual degrees are bounded by $\delta$, the total degree is at most $\Delta\le n\delta$ so the requirement from the characteristic is that either $p = 0$ or $p > (r-1) d \delta n A$.

\begin{itemize}
    \item {\bf Arbitrary orders:} According to \Cref{thm:sum_roabp_hs} there is a hitting set of size  $(n \delta w)^{O(\log (n\delta w))}$ for the sum of two $\roabp$s of arbitrary orders. 

    \item {\bf Same variable order:}
    Since the sum of two $\roabp(w,\delta)$ in the same variable order is a $\roabp(2w,\delta)$ in the same variable order, \Cref{thm:roabp_hs} gives a hitting set of size $\poly\left(\delta \cdot (2nw)^{\lambda(2w)}\right)$ for this case.

\item {\bf Any order:} As in the previous case, the sum of two $\aoroabp(w,\delta)$ is an $\aoroabp(2w,\delta)$, \Cref{thm:ao_roabp_hs} gives a hitting set of size  $\poly\left(n\delta, w^{\log \log w}\right)$ in this case.
 
\end{itemize}

    This concludes the proof for the case $d+1\ge (r-1)^2$, where the requirement on the characteristic is as explained above for each of the cases.
\end{proof}

\subsubsection{PIT for low powers}\label{sec:pit-roabp-low}

To prove \Cref{thm:main-pit-abp} in the regime \(d+1<(r-1)^2\), we show that sums of powers of \(\roabp\)s can be represented by a single \(\roabp\) of larger size.

\smallskip

Let
\begin{equation*}
    \cW_d(w) \coloneqq \binom{w+d-1}{d}.
\end{equation*}

\begin{lemma}
\label{lem:power-roabp}
Let $f\in \roabp(w,\delta)$ be computed in the variable order
$x_1,\ldots,x_n$ and let $B_{\text{in}}$ be a bound on the total bit complexity of the coefficients in the edge polynomials.
Then, for every $d\geq 1$, the polynomial $f^d$ is computed by an
$\roabp$ in the same variable order, of individual degree at most
$d\delta$, width at most $\cW_d(w)$ and total bit complexity $O\left(nd^2\delta\cW_d(w)^2 B_{\text{in}}\right)$.
\end{lemma}

\begin{proof}
Let $w_i$ be the width of the original $\roabp$ at layer $i$, with
$w_0=w_n=1$, and let $q_{i,j}$ denote the partial polynomial computed
at the $j$-th vertex of layer $i$. In particular, $q_{0,1}=1$ and
$q_{n,1}=f$.
Write the given $\roabp$ in matrix-product form as
\begin{equation*}
    f(\vx)=M_1(x_1)M_2(x_2)\cdots M_n(x_n),
\end{equation*}
where
$M_i(x_i)\in \F[x_i]^{w_{i-1}\times w_i}$ and every entry of
$M_i(x_i)$ has degree at most $\delta$.

For $m\geq 1$, let
\begin{equation*}
    I_{m,d}
    =
    \left\{
        \e=(e_1,\ldots,e_m)\in\mathbb{Z}_{\geq 0}^{m}
        :
        \sum_{j=1}^{m}e_j=d
    \right\},
    \qquad
    |I_{m,d}|=\binom{m+d-1}{d}=\cW_d(m).
\end{equation*}
For every $0\leq i\leq n$, define
\begin{equation*}
    v_i=(q_{i,1},\ldots,q_{i,w_i})
    \qquad\text{and}\qquad
    v_i^{[d]}
    =
    \left(v_i^{\e}\right)_{\e\in I_{w_i,d}},
    \qquad
    v_i^{\e}
    \coloneqq
    \prod_{j=1}^{w_i}q_{i,j}^{e_j}.
\end{equation*}
That is, $v_i$ is the row vector of partial polynomials at the $i$-th layer of the $\roabp$ computing $f$ and $v_i^{[d]}$ contains all products of exactly $d$ entries of $v_i$,
with repetition.

Observe that $v_n^{[d]}=(f^d)$. It therefore suffices to construct
an $\roabp$ whose vector of partial polynomials at layer $i$ is
$v_i^{[d]}$.
The transition rule of the original $\roabp$ is
\begin{equation*}
    v_i=v_{i-1}M_i(x_i).
\end{equation*}
Our goal is to obtain a similar transition rule for the vectors
$v_i^{[d]}$.
Specifically, for each $i\in[n]$, we construct a matrix
\begin{equation*}
    \widehat M_i(x_i)
    \in
    \F[x_i]^{\cW_d(w_{i-1})\times\cW_d(w_i)}
\end{equation*}
whose entries are univariate polynomials in $x_i$ that have degree at most $d\delta$ and 
\begin{equation*}
    v_{i-1}^{[d]}\widehat M_i(x_i)=v_i^{[d]}.
\end{equation*}

Fix $i\in[n]$ and $\e\in I_{w_i,d}$.
Expanding the product below and collecting terms according to
the multiplicities of the factors $q_{i-1,t}$ gives
\begin{equation*}
\begin{aligned}
    \left(v_i^{[d]}\right)_{\e}
    &=
    \prod_{j=1}^{w_i}q_{i,j}^{e_j}
    \\
    &=
    \prod_{j=1}^{w_i}
    \left(
        \sum_{t=1}^{w_{i-1}}
        q_{i-1,t}\,M_i(x_i)[t,j]
    \right)^{e_j}
    \\
    &=
    \sum_{\ba\in I_{w_{i-1},d}}
    c_{\ba,\e}(x_i)
    \prod_{t=1}^{w_{i-1}}q_{i-1,t}^{a_t},
\end{aligned}
\end{equation*}
for some coefficients $c_{\ba,\e}(x_i)\in\F[x_i]$.
Since $\sum_{j=1}^{w_i}e_j=d$, each coefficient
$c_{\ba,\e}(x_i)$ obtained in this expansion is a sum of products
of exactly $d$ entries of $M_i(x_i)$.
Consequently, every such coefficient has degree at most $d\delta$ and bit-complexity with the same blowup.

Index the rows of $\widehat M_i$ by $\ba\in I_{w_{i-1},d}$ and
its columns by $\e\in I_{w_i,d}$, and define
\begin{equation*}
    \widehat M_i(x_i)[\ba,\e]
    \coloneqq
    c_{\ba,\e}(x_i).
\end{equation*}
Let $u_0=(1)$, and for every $i\in[n]$, define
\begin{equation*}
    u_i=u_{i-1}\widehat M_i(x_i).
\end{equation*}

\begin{claim}
    For every $0\leq i\leq n$, we have $u_i=v_i^{[d]}$.
\end{claim}

\begin{proof}[Proof of the claim]
We prove the claim by induction on $i$.
For $i=0$, we have $u_0=(1)=v_0^{[d]}$.
Now let $i\geq 1$ and assume that $u_{i-1}=v_{i-1}^{[d]}$.
For every $\e\in I_{w_i,d}$,
\begin{equation*}
\begin{aligned}
    (u_i)_{\e}
    &=
    \sum_{\ba\in I_{w_{i-1},d}}
    (u_{i-1})_{\ba}\,c_{\ba,\e}(x_i)
    \\
    &=
    \sum_{\ba\in I_{w_{i-1},d}}
    c_{\ba,\e}(x_i)
    \prod_{t=1}^{w_{i-1}}q_{i-1,t}^{a_t}
    \\
    &=
    \prod_{j=1}^{w_i}q_{i,j}^{e_j}
    =
    \left(v_i^{[d]}\right)_{\e}.
\end{aligned}
\end{equation*}
The first equality follows from the definitions of $u_i$ and
$\widehat M_i$, the second uses the induction hypothesis, and
the third follows from the expansion defining the coefficients
$c_{\ba,\e}$.
This proves the claim.
\end{proof}
Taking $i=n$, the claim gives $u_n=v_n^{[d]}=(f^d)$, and hence
\begin{equation*}
    f^d
    =
    \widehat M_1(x_1)\widehat M_2(x_2)\cdots\widehat M_n(x_n).
\end{equation*}
These matrices define an $\roabp$ in the same variable order,
with width $\cW_d(w_i)\leq\cW_d(w)$ at layer $i$ and individual degree at most $d \delta$. The bit complexity bound follows by the fact that there are at most $n \cdot \cW_d(w)^2 \cdot (d\delta + 1)$ scalars in the new representation each having bit complexity at most $O(d B_{\text{in}})$.
\end{proof}

\begin{corollary}
\label{cor:sum-powers-roabp}
Let \(P=\sum_{j=1}^r \alpha_j f_j^d\), where $\alpha_j\in\F$ and each $f_j \in \roabp(w,\delta)$ and all the $f_j$'s have the same
variable order. Then $P$ can be computed by an $\roabp\left(r\cW_d(w), d\delta\right)$ in the same variable order. If the inner $\roabp$s are any-order, then the $P$ can be computed by an any-order $\roabp$ with the stated parameters.
\end{corollary}

We now combine this with the PIT results from \Cref{sec:known-pit-abp} to prove \Cref{thm:main-pit-abp} in the low powering degree case.

\begin{proof}[{Proof of \Cref{thm:main-pit-abp} when $d+1<(r-1)^2$}]

    By \cref{lem:power-roabp}, any $d$-th power of an $\roabp(w,\delta)$ is an $\roabp(\cW_d(w),\delta d)$ arbitrary order/any-order $\roabp$.
    Let $\cW_d(w)=\binom{w+d-1}{d}$ be the width of each term and by $D=d\delta$ its degree.
    
\begin{itemize}
    \item {\bf Arbitrary orders:} According to \Cref{thm:sum_roabp_hs} there is a hitting set of size  $(nD \cW_d(w))^{O(r2^r\log nD\cW_d(w))}$ for the sum of $r$ $\roabp(\cW_d(w),D)$ of arbitrary orders. Plugging in the values for $\cW_d(w),D$, and using the assumption that $d<(r-1)^2-1<r^2$ we obtain\footnote{This holds under the assumption $w>1$. When $w=1$ we are in the any-order case.\label{fn:w>1}}
    \begin{align*}
        (nD \cW_d(w))^{O(r2^r\log nD\cW_d(w))} &\le (nd\delta  w^{d})^{O(r2^r\log (nd\delta w^{d}))}\\ &\le (nr^2\delta  w^{r^2})^{O(r2^r\log (nr^2\delta w^{r^2}))}\le (n\delta w)^{2^{O(r)}\log (n\delta w)}.
    \end{align*}
    The requirement on the characteristic as given in \Cref{thm:sum_roabp_hs} is $p=0$ or $p>d\delta$.

    \item {\bf Same variable order:}
    Since the sum of $r$ $\roabp(\cW_d(w),D)$ in the same variable order is a $\roabp(r\cW_d(w),D)$ (in the same variable order), \Cref{thm:roabp_hs} gives a hitting set of size\footref{fn:w>1}
    \begin{align*}
    \poly\left(D \cdot (nr\cW_d(w))^{\lambda(r\cW_d(w))}\right)&\le \poly\left(d\delta \cdot (nrw^{d})^{\lambda(rw^{d})}\right)\\&\le\poly\left(r^2\delta \cdot (nrw^{r^2})^{\lambda(rw^{r^2})}\right) \le \poly\left(\delta \cdot (nw)^{O(r^2 \lambda(w^{r^2}))}\right).
    \end{align*}
    The requirement on the characteristic as given in \Cref{thm:roabp_hs} is $p=0$ or $p>d\delta$.

\item {\bf Any order:} As in the previous case, the sum of $r$ $\aoroabp(\cW_d(w),D)$is an $\aoroabp(r\cW_d(w),D)$, \Cref{thm:ao_roabp_hs} gives a hitting set of size 
\begin{align*}
\poly\left(nD, (r\cW_d(w))^{\log \log (r\cW_d(w))}\right)
&\le\poly\left(nr^2\delta, (rw^{r^2})^{\log \log (rw^{r^2})}\right)
\\&\le\begin{cases}
    \poly\left(n\delta, r^{\log \log r}\right) &  w =1  \\
    \poly\left(n\delta, w^{r^2(\log r+ \log \log w)}\right)  & w >1
\end{cases}
\end{align*}
    The requirement on the characteristic as given in \Cref{thm:ao_roabp_hs} is $p=0$ or $p>n^4(d\delta+1)^2$.
\end{itemize}
This completes the proof of \Cref{thm:main-pit-abp}.
\end{proof}

\subsection{Reconstruction} \label{ssec:recon-set-roabps}

\subsubsection{Non-Adaptive Reconstruction of $\roabp$} \label{sec:non_adaptive_algo_roabp}

In \cite{roabp_recon_adaptive_fs13, klivansShpilkaRecon06}, deterministic adaptive reconstruction algorithms were given for several polynomial classes, including $\roabp$s. To apply our results on reconstructing sums of high powers of certain polynomial classes (\autoref{thm:recon_sets_imply_high_power_recon}), we need to convert this adaptive algorithm into a non-adaptive one. This is the goal of this section and is achieved in \autoref{thm:non_adaptive_recon_algo_for_roabp}.

In \autoref{thm:non_adaptive_recon_algo_for_roabp_relaxed} we relax the requirement of knowing the exact order in advance to having a set of several possible orders, one of which is the correct order of the inner $\roabp$. This allows us to reconstruct in the next section $\Sigma^r\wedge^d\roabp$s with varying orders for the inner $\roabp$s. 

The main idea for the next algorithm is to reconstruct each layer of the $\roabp$ separately. To isolate a given layer, we use a hitting set for $\roabp$s, which guarantees that the relevant prefix and suffix computations retain their full rank. This isolation can be carried out non-adaptively, since all required evaluations are fixed in advance. Finally, we combine the reconstructed layers using suitable gluing matrices. For each layer, the corresponding gluing matrix consists of evaluations of the $\roabp$ we are trying to reconstruct on points formed by independently combining a prefix and a suffix of the same hitting set.

\begin{theorem} \label{thm:non_adaptive_recon_algo_for_roabp}
    Let $\delta,w \in \N$, let $\cH$ be a hitting set for $\roabp(w,\delta)$ (respectively $\aoroabp(w,\delta)$). 
    Moreover, denote $\charac (\F) = p$ and assume $p = 0$ or $p > \delta$. There is an $  O\left(n \left(|\cH|^3 + w^\omega + w^2 (\delta+1) \right) \right)$-time non-adaptive deterministic reconstruction algorithm for $\roabp(w,\delta)$ (respectively $\aoroabp(w,\delta)$) of known variable order, 
    that uses a reconstruction set of size $n(\delta+1)|\cH|^2$.  
\end{theorem}

\begin{proof}
    First, we check whether $f \equiv 0$ using known PIT algorithms. If so, we return the zero polynomial. Otherwise we proceed as described below.

    Let $f\in \roabp(w,\delta)$ with variable order $x_1,x_2,\ldots,x_n$ and let $S \subseteq \F$ be a set of size $|S| = \delta + 1$.  
    
    Define the prefix and suffix projections of $\cH$ by
    \[
        L_i= \left\{ (a_1,\ldots,a_i) \mid \va \in \cH \right\}\subseteq \mathbb F^i,
    \qquad
    R_i= \left\{ (a_{i+1},\ldots,a_n) \mid \va \in \cH \right\} \subseteq \mathbb F^{n-i}.
    \]
    with the conventions
    $L_0=R_n=\{\epsilon\}$. Set   
\begin{equation} \label{eq:roabp_recon_set_size}
J
    =
    \bigcup_{i=1}^n
    L_{i-1}\times S\times R_i, \quad \text{so that }\quad
    |J|
    \le
    \sum_{i=1}^n |L_{i-1}|\, |S|\, |R_i|
    \le
    n(\delta+1)|\cH|^2.
\end{equation}

We first note that $J$ is a hitting set for
$\roabp(w,\delta)$. Indeed, since $f\not\equiv0$, then some
$\va=(a_1,\ldots,a_n)\in\cH$ satisfies $f(\va)\neq0$. Hence the
univariate polynomial
\[
f(x_1,a_2,\ldots,a_n)
\]
is nonzero and has degree at most $\delta$, so $f$ must be nonzero at some point of
$S\times R_1\subseteq J$. 

Fix a width-$w$ $\roabp$
representation in the known order:
\[
    f(\vx)
    =
    M_1(x_1)M_2(x_2)\cdots M_n(x_n),
\]
where
\[
    M_j(x_j)\in\F[x_j]^{w_{j-1}\times w_j},
    \qquad
    w_0=w_n=1,
    \qquad
    w_j\le w.
\]
We may assume that the representation has minimal width at every
layer. 
For $i\in[n-1]$, write
\[
    A_i(\vx_{[i]})
    =
    M_1(x_1)\cdots M_i(x_i)
\qquad \text{and}\qquad
    B_i(\vx_{[i+1:n]})
    =
    M_{i+1}(x_{i+1})\cdots M_n(x_n).
\]
Thus
\[
    f(\va,\vb)=A_i(\va)B_i(\vb) \quad \text{for every} \quad \va\in\F^i, \, \vb\in\F^{n-i}.
\]

Consider the full evaluation matrix
\[
    M_i^{\mathrm{full}}
    =
    \bigl(f(\va,\vb)\bigr)_{
        \va\in\F^i,\,
        \vb\in\F^{n-i}},
\]
where its rank means the maximum rank of a finite submatrix. The above
factorization implies
\[
    r_i
    :=
    \rank(M_i^{\mathrm{full}})
    \le w_i
    \le w.
\]
We next show that $M_i^{\mathrm{full}}$ is spanned by a submatrix corresponding to the queries in $L_{i-1}\times S\times R_i$.
Define
\[
    M_i
    =
    \bigl(f(\va,\vb)\bigr)_{
        \va\in L_i,\,
        \vb\in R_i}.
\]
We note that, since we can interpolate $f(\ba_{[i-1]},z,\bb)$ using $S$, each entry of $M_i$ is determined by the values of $f$ on $L_{i-1}\times S\times R_i$.
We next show that
\begin{equation}
\label{eq:roabp-restricted-rank}
    \rank(M_i)=r_i.
\end{equation}
Let $\cA_i$ be the matrix whose rows are the vectors
$A_i(\va)$, for $\va\in L_i$, and let
$\cA_i^{\mathrm{full}}$ be the analogous matrix containing all
$\va\in\F^i$. We claim that
\[
    \rowspan(\cA_i)
    = \rowspan(\cA_i^{\mathrm{full}}).
\]
As $(\cA_i)$ is a submatrix of $\cA_i^{\mathrm{full}}$
if there is no equality there is a vector $\bv \in\F^{w_i}$ such that
\[
    A_i(\va)\bv=0
    \quad\text{for every }\va\in L_i,
\quad \text{but} \quad    A_i(\vx_{[i]})\bv\not\equiv0.
\]

\begin{claim}
     \(A_i(\x_{[i]})\bv \in \roabp(w,\delta)\),
    if \(f\in\aoroabp(w,\delta)\), then
    \(A_i(\x_{[i]})\bv\in\aoroabp(w,\delta)\).
\end{claim}

\begin{proof}
    The first part of the claim is trivial, we continue to the any-order case.
    
    Write \(g(\x_{[i]})=A_i(\x_{[i]})\bv\). Now suppose that \(f\in\aoroabp(w,\delta)\).
    By minimality, the entries of \(B_i\) are linearly independent.
    Since their individual degrees are at most \(\delta\) and
    \(|\F|>\delta\), the vectors
    \(\{B_i(\bb):\bb\in\F^{n-i}\}\) span \(\F^{w_i}\).
    Hence there exist points
    \(\bb_1,\ldots,\bb_{w_i}\in\F^{n-i}\) and scalars
    \(\gamma_1,\ldots,\gamma_{w_i}\in\F\) such that
    \begin{equation*}
        \bv=\sum_{j=1}^{w_i}\gamma_jB_i(\bb_j),
        \qquad
        g(\x_{[i]})
        =
        \sum_{j=1}^{w_i}\gamma_j f(\x_{[i]},\bb_j).
    \end{equation*}

    Let \(\pi\) be an arbitrary ordering of the first \(i\) variables.
    Since \(f\) is any-order width-$w$, there exists a factorization of $f$ according to this order
    \begin{equation*}
        f(\x)
        =
        \widetilde A_\pi(\x_{[i]})
        \widetilde B_\pi(\x_{[i+1:n]}).
    \end{equation*}
    Then
    \begin{equation*}
        g(\x_{[i]})
        = \sum_{j=1}^{w_i} \gamma_j f(\x_{[i]}, \bb_j) =
        \widetilde A_\pi(\x_{[i]})
        \left(
            \sum_{j=1}^{w_i}
            \gamma_j\widetilde B_\pi(\bb_j)
        \right).
    \end{equation*}
    Since the order is now fixed, we may absorb the constant column vector $\left(
            \sum_{j=1}^{w_i}
            \gamma_j\widetilde B_\pi(\bb_j)
        \right)$ into the last layer and get a width-$w$ $\roabp$ for $g$ with the order $\pi$. The claim follows since $\pi$ was chosen arbitrarily.
\end{proof}

By the above claim and since $\cH$ is a hitting set for $\roabp(w,\delta)$, there is a point
$\va\in\cH$ such that
\[
    A_i(a_1,\ldots,a_i)\bv\neq0,
\]
in contradiction.
The same argument, applied to the suffix vectors $B_i(\vb)$, shows
that their evaluations on $R_i$ span all suffix evaluations.
Consequently, there are matrices $X_i,Y_i$ such that
\[
    M_i^{\mathrm{full}}
    =
    X_iM_iY_i.
\]
Hence
\[
    \rank(M_i^{\mathrm{full}})
    \le
    \rank(M_i).
\]
The reverse inequality follows because $M_i$ is a submatrix of
$M_i^{\mathrm{full}}$, proving
\eqref{eq:roabp-restricted-rank}.

    For every $i\in[n-1]$, choose subsets
    \[
        U_i \subseteq L_i,
        \qquad
        V_i \subseteq R_i,
        \qquad
        |U_i| = |V_i| = r_i,
    \]
    such that
    \[
        K_i
        =
        \bigl(f(\ba,\bb)\bigr)_{\ba\in U_i,\,\bb\in V_i}
    \]
    is nonsingular. Such subsets can be found by Gaussian elimination on $M_i$. We also set $U_0=V_n=\{\epsilon\}$ and $K_0 = K_n = (1)$.

In what follows, for every $A\subseteq \F^i,B\subseteq\F^{n-i}$ denote by $f(A,B)$ the submatrix of $M_i^{\mathrm{full}}$ indexed by the rows in $A$ and columns in $B$.

\begin{claim} \label{claim:roabp-hankel-identity}
For every $\vu\in\F^i$ and $\bv\in\F^{n-i}$ it holds that
\begin{equation*}
    f(\vu,\bv)
    =
    f(\vu,V_i)K_i^{-1}f(U_i,\bv).
\end{equation*}
\end{claim}

\begin{proof}
    The rows of $M_i^{\mathrm{full}}$ indexed by $U_i$ form a
basis of its row space. Therefore, for some $\bw$,
\begin{equation}\label{eq:zu}
    f(\vu,\F^{n-i})
    =
    \bw f(U_i,\F^{n-i}).
\end{equation}
Restricting this equality to the columns indexed by $V_i$ gives
\[
    f(\vu,V_i)=\bw K_i,
\]
and hence
\[
    \bw =f(\vu,V_i)K_i^{-1}.
\]
Using this equality in \eqref{eq:zu} and fixing $\bv$ as the second coordinate proves the claim.
\end{proof}

For $i\in[n]$, define
\[
    C_i(x_i)
    =
    f(U_{i-1},x_i,V_i).
\]
This is an $r_{i-1}\times r_i$ polynomial matrix, where
$r_0=r_n=1$, and each entry has degree at most $\delta$. Furthermore, observe, as before, that the matrices $C_i$ are determined by
the queries in $J$. 

\begin{claim} \label{claim:roabp-recon-main-poly-identity}
    \[
    f(U_{i-1},x_i,x_{i+1},\ldots,x_n) =
    C_i(x_i)K_i^{-1}
    f(U_i,x_{i+1},\ldots,x_n).
\]
\end{claim}

\begin{proof}
Applying \Cref{claim:roabp-hankel-identity} row by row gives
\[
    f(U_{i-1},\alpha_i,\alpha_{i+1},\ldots,\alpha_n) =
    C_i(\alpha_i)K_i^{-1}
    f(U_i,\alpha_{i+1},\ldots,\alpha_n),
\]
for all possible values of $\alpha_i,\ldots,\alpha_n \in \F$. 
Since we assume $|\F|>\delta$, and both sides of the claimed identity have individual degrees bounded by $\delta$, this proves the claim.
\end{proof}

Iterating the identity in \Cref{claim:roabp-recon-main-poly-identity} yields
\[
    f(x_1,\ldots,x_n)
    =
    C_1(x_1)K_1^{-1}
    C_2(x_2)K_2^{-1}
    \cdots
    K_{n-1}^{-1}C_n(x_n).
\]
Since $r_i\le w$, this is a width-$w$ $\roabp$ in the order
$x_1,\ldots,x_n$.

\medskip 

\paragraph{Running time.} For every $i$, recovering $M_i$
from the values on $J$ and performing Gaussian elimination costs
$O(h^3)$ field operations. The matrix $K_i$ has
size at most $w\times w$, and its inverse can be computed in
$O(w^\omega)$ operations. Finally, the matrix $C_i$ has at most
$w^2$ entries, each determined by $\delta+1$ evaluations. 
Summing over all $n$ layers gives
\[
    O\left(
        n\left(
            |\cH|^3
            +
            w^\omega
            +
            w^2(\delta+1)
        \right)
    \right)
\]
field operations, which implies the claim. 
\end{proof}

\subsubsection{$\roabp$ Reconstruction using Multiple Possible Orders} \label{subsec:multiple_orders_roabp_recon}

In this section we show that we can reconstruct a $\roabp$ non-adaptively, even when the exact order is not known in advance, but is one of several possible orders.

The idea is to try and reconstruct a given $f\in\roabp$ using all possible orders, each reconstruction results in a polynomial $g_i$. Then, we can check which of these orders is the correct one by checking if $f - g_i$ is identically zero, using a hitting set for $\Sigma^2\roabp(w,\delta)$. This can be done non-adaptively by adding the interpolating set to the reconstruction set.

\begin{theorem}
    \label{thm:non_adaptive_recon_algo_for_roabp_relaxed}
    Let \(n,w,\delta,r\in\N\), and assume that
    \(p=\charac(\F)\) satisfies \(p=0\) or
    \(p>\delta\). Let \(S_{\mathrm{ord}}\) be a list
    of at most \(r\) permutations of \([n]\). Let \(\cH\)
    be a hitting set for \(\roabp(w,\delta)\) and let
    \(H\) be a hitting set for \(\Sigma^2\roabp(w,\delta)\). 

    Let $P \in \roabp(w,\delta)$ 
    and assume the order of $P$ is given by some permutation in $S_{\mathrm{ord}}$. Then, there is a deterministic non-adaptive reconstruction algorithm that reconstructs $P$. It uses at most
    \begin{equation*}
        r n(\delta+1)|\cH|^2+|H|
    \end{equation*}
    evaluation points.
    Its running time is \[
    O\Big(r\cdot \big(n  \left(|\cH|^3 + w^\omega + w^2 (\delta+1) \right) + rnw^2(\delta+1)|H|\big) \Big).
    \]
\end{theorem}


\begin{proof}
    Let $f \in \roabp(w,\delta)$ in a variable order that appears in $S_{\text{ord}}$.
    Denote $S_{\text{ord}} = \{\sigma_i\}_{i\in [r]}$ and let $J_i$ be the reconstruction set used in \autoref{thm:non_adaptive_recon_algo_for_roabp} whenever the variable order of the $\roabp$ is given by $\sigma_i$. Let $\mathcal{J} = \left\{ H \cup \bigcup_{i \in [r]} J_i \right\}$, by the size bound in \autoref{thm:non_adaptive_recon_algo_for_roabp} we get that $|\mathcal{J}| \leq r \cdot n\cdot (\delta+1)|\cH|^2 + |H|$. We use $\mathcal{J}$ as our reconstruction set, the last statement implies the size bound claimed.

    Evaluate $f$ on $\mathcal{J}$, and attempt reconstructing $f$ using \autoref{thm:non_adaptive_recon_algo_for_roabp} and the reconstruction set $J_i$ (assuming variable order $\sigma_i$). Let $g_i$ be the polynomial resulting in reconstructing $f$ using $J_i$. If in the process of reconstruction of $g_i$, any of the matrices constructed has rank greater than $w$, discard $g_i$ and continue to the next order. This can be checked using the rank of the flattening matrices $M_i$.

    \begin{observation} \label{obs:rem_g_is_are_good_roabp}
        Every remaining $g_i$ can be computed by a $\roabp(w,\delta)$.
    \end{observation} 

    \begin{proof}
    Let $\{K_j\}$ and $\{C_j\}$ be the gluing matrices and polynomial matrices constructed in \autoref{thm:non_adaptive_recon_algo_for_roabp} when reconstructing $g_i$.

    Since $g_i$ was not discarded, every $K_j$ has rank at most $w$. Moreover, the polynomial matrices $C_j$ are of degree at most $\delta$ since this is the resulting polynomials in the interpolation process used. This implies the claim.
    \end{proof}
    
    For each such $g_i$, check if $f(H) - g_i(H) \equiv \0$. If so, return $g_i$. This can be done using the evaluations of $f$ on $\mathcal{J}$ prior to the reconstruction of the $g_i$'s and by evaluating $g_i$ on $H$ at this point. By the definition of $H$ and \autoref{obs:rem_g_is_are_good_roabp} we get that $f-g_i$ is identically zero if and only if $f \equiv g_i$.

    It remains to analyze the running time. The algorithm can be divided into three sections.
    \begin{enumerate}
        \item \label{item:diff_var_recon_evaluate} Evaluate $f$ on $\mathcal{J}$.
        \item \label{item:diff_var_recon_reconstruct} Reconstruct $r$ candidates $\{g_i\}_{i \in [r]}$.
        \item \label{item:diff_var_recon_interpolate} Find a candidate that computes $f$.
    \end{enumerate}

Part~\eqref{item:diff_var_recon_evaluate} is the preceding ``evaluation'' part of the algorithm. This part takes time relative to the size of the reconstruction set, $\mathcal{J}$, that is $O(r \cdot n\cdot (\delta+1)|\cH|^2 + |H|)$.

Part~\eqref{item:diff_var_recon_reconstruct} consists of at most $r$ independent executions of the reconstruction algorithm described in \autoref{thm:non_adaptive_recon_algo_for_roabp}. This takes $O\big(r \cdot n \cdot \left( |\cH|^3 + w^{\omega} + w^2(\delta + 1) \right)\big)$.

Part~\eqref{item:diff_var_recon_interpolate} consists of evaluating the $r$ candidates $\{g_i\}_{i \in [r]}$ on $H$, then subtracting $f(H)$ from these values. This takes $O(rnw^2(\delta+1) |H|)$ field operations.

By combining these bounds and the fact that $|\cH| \geq\delta+1$ we get an overall number of field operations of \[
O\Big(r\cdot \big(n  \left(|\cH|^3 + w^\omega + w^2 (\delta+1) \right) + rnw^2(\delta+1) |H|\big) \Big).
\]
\end{proof}

\subsection{Reconstruction for sums of powers of $\roabp$s}

We now prove \cref{thm:main-rec-abp}, using the reconstruction
sets from \cref{ssec:recon-set-roabps}. We treat the three power
regimes in \cref{tab:roabp_reconstruction_results} separately.
When \(d>(r+1)^4\Delta\), the total-degree meta-theorem gives
proper reconstruction in all three variable-order settings.
When \(d > (r+1)^4r\delta\), the small-individual-degree
meta-theorem additionally requires a hitting set for $\Sigma^2\Pi^2\roabp$, which we give in \Cref{lem:hs_alternants_same_ord_roabp}.
We construct such a set for same-order and any-order
\(\roabp\)s. Finally, for these two models, we also obtain
non-proper reconstruction for any $d$ by expressing the input as a single
\(\roabp\) of larger width and individual degree.

In all of the reconstruction algorithms, the variable orders are given to us. In the
arbitrary-orders setting, a list of the \(r\)
orders of the inner $\roabp$s is supplied as in
\cref{thm:non_adaptive_recon_algo_for_roabp_relaxed}.





\paragraph{Reconstruction using the total-degree bound.}

\begin{proof} [{Proof of \Cref{thm:main-rec-abp} when $d > (r+1)^4\Delta$}]
For each underlying class, we apply
    \Cref{thm:recon_sets_imply_high_power_recon}. 

    To apply \Cref{thm:recon_sets_imply_high_power_recon} we need a reconstruction set for the underlying class. We also require that $p=0$ or $p > 2\Delta rd$. Write \(s_{\mathrm{rec}}\) and
    \(T_{\mathrm{rec}}\) for the size and running time of the
    reconstruction set and algorithm for the underlying class. 
\begin{itemize}
    \item {\bf Arbitrary orders:} 
    We use \cref{thm:non_adaptive_recon_algo_for_roabp_relaxed}
    with the supplied list of orders. Its two required
    hitting sets are \(\cH\) for \(\roabp(w,\delta)\) and
    \(H\) for \(\sum^2\roabp(w,\delta)\). By
    \cref{thm:roabp_hs,thm:sum_roabp_hs}, such sets exist and satisfy 
    \begin{equation*}
        |\cH|=\poly\!\left(\delta(nw)^{\lambda(w)}\right),
        \qquad
        |H|=(n\delta w)^{O(\log(n\delta w))}.
    \end{equation*}
    Substituting these bounds into
    \cref{thm:non_adaptive_recon_algo_for_roabp_relaxed}
    gives
    \begin{equation*}
        s_{\mathrm{rec}},T_{\mathrm{rec}}
        \leq
        \poly\!\left((n\delta w)^{O(\log(n\delta w))},r\right).
    \end{equation*}
    \cref{thm:recon_sets_imply_high_power_recon} now gives the
    arbitrary-orders entry in the first column of
    \cref{tab:roabp_reconstruction_results}.
    
    \item {\bf Same variable order:}
    We use \cref{thm:non_adaptive_recon_algo_for_roabp} with the same
    hitting set \(\cH\) as above. This gives
    \begin{equation*}
        s_{\mathrm{rec}},T_{\mathrm{rec}}
        \leq\poly\!\left((nw)^{\lambda(w)},\delta\right).
    \end{equation*}
    Substituting into the meta-theorem yields the stated result.
    
    \item {\bf Any order:}  Fix any variable order and use \cref{thm:non_adaptive_recon_algo_for_roabp}. By \cref{thm:ao_roabp_hs}, we can choose a smaller set $\cH$ that satisfies
    \begin{equation*}
        |\cH|=\poly\!\left(n\delta,w^{\log\log(w)}\right),
    \end{equation*}
    with the additional requirement that $p=0$ or \(p>n^4(\delta+1)^2\).
    Thus
    \begin{equation*}
        s_{\mathrm{rec}},T_{\mathrm{rec}}
        \leq\poly\!\left(n,w^{\log\log(w)},\delta \right).
    \end{equation*}
    Combining this with the meta-theorem gives the stated
    running time and the characteristic condition
    \begin{equation*}
        p=0\qquad\text{or}\qquad
        p>\max\{2\Delta rd,n^4(\delta+1)^2\}.
    \end{equation*}
\end{itemize}
\end{proof}

\paragraph{Reconstruction using the individual-degree bound.}

To apply \Cref{thm:suff_cond_for_high_power_recon_ext} we need a hitting set for the class \(\Sigma^2\Pi^2\roabp(w,\delta)\).
When all inner $\roabp$s share the same variable order, this class can
be represented by an \(\roabp(2w^2,2\delta)\) in that same order, and we can use known hitting sets.




\begin{lemma} \label{lem:hs_alternants_same_ord_roabp}
    Let $n,w,\delta \in \N$ and let $\F$ be a field of characteristic $p$ such that $p = 0$ or sufficiently large. Then, there exists an explicit hitting set for $\Sigma^2\Pi^2 \roabp(w,\delta)$ given that the inner $\roabp$s have the same order. The size of the hitting set and the bound on the positive characteristic are given in the next table
    \begin{table}[H]
    \centering
    \small
    \renewcommand{\arraystretch}{1.4}
    \setlength{\tabcolsep}{10pt}

    \begin{tabular}{ccc}
        \toprule
        \textbf{Variable order}
        &
        \textbf{Hitting set size}
        &
        \textbf{Positive characteristic bound}
        \\
        \midrule

        same order
        &
        \(\displaystyle
        \poly\!\left(
            \delta\left(2nw^2\right)^{\lambda(2w^2)}
        \right)\)
        &
        \(p>2\delta\)
        \\[1ex]

        any order
        &
        \(\displaystyle
        \poly\!\left(n
            \delta,(2w)^{\log\log w}
        \right)\)
        &
        \(p>n^4(2\delta+1)^2\)
        \\

        \bottomrule
    \end{tabular}
\end{table}
\end{lemma}

\begin{proof}
 Let
    \begin{equation*}
        P=F_1F_2+F_3F_4,
        \qquad F_1,\ldots,F_4\in\roabp(w,\delta),
    \end{equation*}
    where the four programs share the same variable order. By \Cref{cor:sum-2-prod-2-roabp}, $P$ has an
    \(\roabp(2w^2,2\delta)\) in the same order.

    Applying \cref{thm:roabp_hs} with these parameters gives the
    same-order bound. If the four inner $\roabp$s are
    any-order, the same construction works in every variable
    order, so \(P\in\aoroabp(2w^2,2\delta)\). Applying
    \cref{thm:ao_roabp_hs} yields
    \begin{equation*}
        \poly\!\left(n\delta,
            (2w^2)^{\log\log(2w^2)}\right)
        =\poly\!\left(n\delta,(2w)^{\log\log(2w)}\right).
    \end{equation*}
\end{proof}

We're now ready to prove the results in the second column of \cref{tab:roabp_reconstruction_results}.

\begin{proof} [{Proof of \Cref{thm:main-rec-abp} when $d > (r+1)^4r\delta$}]
We apply the lifting theorem in
\Cref{thm:suff_cond_for_high_power_recon_ext} to same-order and
any-order \(\roabp\)s.
To apply this theorem, for common order and any order, in addition to the two sets used in the total degree case, we need a hitting set for \(\Sigma^2\Pi^2\roabp(w,\delta)\). Such a hitting set, for both models, is given in \Cref{lem:hs_alternants_same_ord_roabp}.
Substituting the bounds on the sets that we use yields the stated results.
\end{proof}

\paragraph{General non-proper reconstruction.}

For a common variable order, the conversion used in the low-power PIT
argument also gives a non-proper reconstruction algorithm. We view
\(P\) as a single \(\roabp\) with larger width and individual degree and reconstruct that
representation.
Let
\begin{equation}
    \label{eq:roabp_recon_const_def}
    W\coloneqq r\binom{w+d-1}{d},\qquad D\coloneqq d\delta.
\end{equation}
These are the width and individual-degree bounds of the reconstructed
$\roabp$s. The next proposition has no restriction on \(d\).

\begin{proposition}
    \label{prop:recon_low_deg_sum_power_roabp}
    Let \(n,r,d,w,\delta\in\N\), and let
    \(P\in\Sigma^r\wedge^d\roabp(w,\delta)\), where the individual $\roabp$s share a given variable order. Let
    \(\cH\) be a hitting set for \(\roabp(W,D)\) of size
    \(s\). Assume that \(p=\operatorname{char}(\F)\) satisfies
    \(p=0\) or \(p>D\). 

    There is a deterministic non-adaptive reconstruction
    algorithm that returns an \(\roabp(W,D)\) computing \(P\).
    Its running time and reconstruction-set size are, respectively,
    \begin{equation*}
        \poly(D,W) \cdot O\!\left(n\left(s^3+W^\omega+W^2(D+1)\right)\right)
        \qquad\text{and}\qquad
        n(D+1)s^2.
    \end{equation*}
    When the individual $\roabp$s are any-order, $\cH$ can be a hitting set for
    \(\aoroabp(W,D)\).
\end{proposition}

\begin{proof}
    Let
    \begin{equation*}
        P=\sum_{i=1}^r\alpha_i G_i^d,
        \qquad G_i\in\roabp(w,\delta),
    \end{equation*}
    with all \(G_i\) in the given order. By \Cref{lem:power-roabp}
    each \(G_i^d\) has an \(\roabp\) representation of
    width \(\binom{w+d-1}{d}\) and individual degree \(D\), in the same
    variable order. Adding the $\roabp$ in parallel, each multiplied by its respective weight gives a single \(\roabp(W,D)\) computing \(P\).

    Apply \cref{thm:non_adaptive_recon_algo_for_roabp} with
    width \(W\), individual degree \(D\) and hitting set
    \(\cH\). This gives the stated results, we note that if the inner $\roabp$s are
    any-order, the above argument applies in every
    order, hence the single $\roabp$ computing $P$ is itself any-order.
\end{proof}

\begin{proof} [{Proof of \Cref{thm:main-rec-abp} when $d \leq (r+1)^4r\delta$}]
    In the given power regime, $D \leq (r+1)^4r\delta^2$.
    
    When the underlying class is same-order $\roabp$, apply
    \cref{prop:recon_low_deg_sum_power_roabp} with the hitting
    set from \cref{thm:roabp_hs}. Its size is
    \begin{equation*}
        s=\poly\!\left(D(nW)^{\lambda(W)}\right),
    \end{equation*}
    and requires \(p=0\) or \(p>D\).
    Substitution into the proposition gives the same-order
    running time in the table.

    For any-order inner polynomials, use
    \cref{thm:ao_roabp_hs} instead. It gives
    \begin{equation*}
        s=\poly\!\left(nD,(W)^{\log\log(W)}\right),
    \end{equation*}
    under \(p=0\) or \(p>n^4(D+1)^2\), and yields the
    corresponding any-order bound.
\end{proof}


\section{Small Depth Circuits}\label{sec:dmall-depth}

In this section we give our results for small depth circuits. 

\begin{definition}[Bounded-depth circuit]
A bounded-depth circuit is a layered arithmetic circuit with a bounded number of layers. Each layer consists of gates of a single type. We denote a layer of addition gates by $\Sigma$, a layer of multiplication gates by $\Pi$, and a layer of power gates by $\wedge$.

When the fan-in of a layer is bounded in the circuit class under consideration, we indicate this bound by a superscript. Thus, for example, $\Sigma^s$ denotes a layer of addition gates of fan-in at most $s$, and $\Pi^\Delta$ denotes a layer of multiplication gates of fan-in at most $\Delta$. For power gates, the superscript denotes the exact power to which the input polynomial is raised; thus $\wedge^d$ denotes a layer of gates computing $g \mapsto g^d$.
\defsymbol
\end{definition}

We first state our main PIT result. This is a formal version of \Cref{thm:intro:pit-bounded-depth}.
In what follows we denote
\begin{equation}\label{eq:top-fan-in}
    S:=r\binom{d+s-1}{d} \le r\min(s^d,(d+1)^{s-1}).
\end{equation} 

 \begin{theorem}[PIT for small depth powering circuits]\label{thm:main-pit-small-depth}
 Let $n,r,d,\delta\in \N$. Let $\F$ be a field of characteristic $p$ such that $p=0$ or $p$ is sufficiently large.
 Then, there are explicit hitting sets as in \Cref{tab:hitting-set-sizes-small-depth}, where we also list the requirement on positive characteristic.
\begin{table}[h]
\centering
\setlength{\arraycolsep}{12pt}
\[
\begin{array}{lll} 
    \toprule
    \multirow{2}{*}{\textbf{Circuit Model}} &
    \multicolumn{2}{c}{\hspace{-3cm}\textbf{Hitting set size } T} 
     \\
    & \textbf{\(d+1 \geq (r-1)^2\)} &
      \textbf{\(d+1 < (r-1)^2\)}\\
    \midrule
    


    
    \Sigma^r\!\wedge^d\!\Sigma^s\!\wedge^\Delta\!\Sigma
        &
        d\cdot\poly(n,\Delta,s)
        &
        \poly\left(n,(s\Delta + r)^{r^2}\right)
        \\

    \indent\charac(\F)
        &p> \max\{rT, O(ns^2\Delta^2)\} 
        & p>O\left(n(s\Delta + r^2)^{2r^2}\right)
        \\

    \textbf{quasi-poly regime}
        & \text{always poly}
        & r=\poly(\log n)
        \\[2ex]

    \Sigma^r\!\wedge^d\!\Sigma^s\Pi^\Delta\Sigma
        &
        d \cdot \poly(n)(\Delta+1)^{2s+1}
        &
        \poly(n)\cdot (d\Delta+1)^{S}
        \\

    \indent\charac(\F)
        & p>rT
        & \text{always}
        \\

    \textbf{quasi-poly regime}
        & s=\poly(\log n)
        & r=O(1),\,s=\poly(\log n)
        \\[2ex]

    \Sigma^r\!\wedge^d\!\Sigma^s
    \Pi^\Delta\Sigma^m\!\wedge^\delta
        &
        d
        \left(s\Delta(m+\delta)\right)^{
            O\left(
                s\cdot\log\log(s\Delta(m+\delta))
            \right)
        }
        &
        \left(Sd\Delta(m+\delta)\right)^{
            O\left(
                S\cdot\log\log(Sd\Delta(m+\delta))
            \right)
        }
        \\

    \indent\charac(\F)
        & p>rT
        & p>(d\Delta\delta)^S
        \\

    \textbf{quasi-poly regime}
        & s=\poly(\log n)
        & r=O(1),\,s=\poly(\log n)
        \\[2ex]

    \Sigma^r\!\wedge^d\!\Sigma^s
    \Pi^\Delta\Sigma^m\Pi^\delta
        &
        d
        \left(s\Delta (m+\delta)\right)^{
            O\left(
                \delta^2s\cdot\log(s\Delta (m+\delta))
            \right)
        }
        &
        \left(Sd\Delta(m+\delta)\right)^{
            O\left(
                \delta^2S\cdot\log(Sd\Delta(m+\delta))
            \right)
        }
        \\

    \indent\charac(\F)
        & p>rT
        & p>(d\Delta\delta)^S
        \\

    \textbf{quasi-poly regime}
        & \delta,s=\poly(\log n)
        & r=O(1),\,\delta,s=\poly(\log n)
        \\

    \bottomrule
\end{array}
\]
\caption{Hitting sets sizes for small depth powering circuits
}
\label{tab:hitting-set-sizes-small-depth}

\end{table}
\end{theorem}




As in \Cref{sec:roabp} we first state known PIT results and then separate the proof of \Cref{thm:main-pit-small-depth} to the low and high powering degree cases.

\subsection{{Proof of \Cref{thm:main-pit-small-depth}}}

\subsubsection{Known PIT results}\label{sec:known-pit-small-depth}

We state known PIT results for the classes whose sums of powers we consider.

\begin{theorem} [{\cite{BS26HSSmallDer}}] \label{thm:hs_small_pd_space}
    Let $n,R,\Delta \in \N$ and let $\F$ be a field of characteristic $p$ such that $p=0$ or $p > \max\{\Delta,nR^2 + R\}$. Then, there exists an explicit hitting set for the class of polynomials with partial derivative space dimension at most $R$ and total degree at most $\Delta$. The size of the hitting set is $$\poly(n,\Delta,R).$$
\end{theorem}

In the same paper the authors give the following corollary.

\begin{corollary}
     [{\cite{BS26HSSmallDer}}] \label{thm:hs_diagonal}
    Let $n,s,\Delta \in \N$ and let $\F$ be a field of characteristic $p$ such that $p=0$ or $p > O\left(n\big(s(\Delta+1)\big)^2\right)$. Then, there exists an explicit hitting set for $\Sigma^{s}\! \wedge^{\Delta}\! \Sigma$ circuits of size  $$\poly(n,\Delta,s).$$
\end{corollary}

\begin{remark}
    In \cite{BS26HSSmallDer}, the authors establish the result over
    fields of characteristic zero. A slight modification of their proof
    extends the result to fields of characteristic greater than
    \(\max\{\Delta,nR^2+R\}\).
    The stated positive-characteristic bound in the corollary then follows from the fact that polynomials computed by
    \(\Sigma^s\wedge^\Delta\Sigma\) circuits have partial derivative spaces of dimension \(s(\Delta+1)\).
\end{remark}


\begin{theorem} [{\cite{SaxenaSesh12pit}}] \label{thm:depth-3-hs}
    There exists an explicit hitting set for $\Sigma^{s}\Pi^\Delta\Sigma$ of size
    \[
     \poly(n)(\Delta+1)^s\, ,
    \]
    regardless of the characteristic of the field.
\end{theorem}


\begin{theorem} [{\cite[Theorem 2(a)]{DDS-depth4-pit}}] \label{thm:hs_depth4_prod_wedge}
    Denote $\charac(\F) = p$. There exists an explicit hitting set for $\Sigma^{s}\Pi^\Delta\Sigma^m\wedge^\delta$ whenever $p = 0$ or $p > (\delta\cdot \Delta)^{s}$. 
    The size of the hitting set is 
    $$
     \left(s \cdot \Delta \cdot (m+\delta)\right)^{O(s \cdot \log \log (s \Delta (m+\delta)))} \, .
    $$
\end{theorem}

\begin{theorem} [{\cite[Theorem 2(b)]{DDS-depth4-pit}}] \label{thm:hs_depth4_prod_prod}
    Denote $\charac(\F) = p$. There exists an explicit hitting set for $\Sigma^{s}\Pi^\Delta\Sigma^m\Pi^\delta$ whenever $p = 0$ or $p > (\delta\cdot \Delta)^{s}$. 
    The size of the hitting set is 
    $$
     \left(s \cdot \Delta \cdot (m+\delta)\right)^{O(\delta^2 s \cdot \log (s \Delta (m+\delta)))} \, .
    $$
\end{theorem}


\subsubsection{PIT for high powers}
\label{sec:pit_results_small-depth-high}

\begin{proof}[{Proof of \Cref{thm:main-pit-small-depth} when $d+1\ge (r-1)^2$}]
As before, \Cref{thm:bb pit for high powers} shows how to construct a hitting set for $\Sigma^r\wedge^d\cC_\Delta$, with size blowup of factor $d\Delta$, from  a hitting set $\cH$ for $\Sigma^2 \cC_\Delta$.

Furthermore, if the size of that hitting set is $|\cH|=A$ then we construct a hitting set of size $T=A\Delta d$ and the characteristic of the field can be either $0$ or at least $rT$. 

\begin{itemize}

    \item {\bf $\Sigma^r\!\wedge^d\!\Sigma^s\!\wedge^\Delta\!\Sigma$:} According to \Cref{thm:hs_diagonal} there is a hitting set of size $\poly(n,\Delta,s)$  for the sum of two $\Sigma^s\!\wedge^\Delta\!\Sigma$ circuits.

    \item {\bf $\Sigma^r\!\wedge^d\!\Sigma^s\Pi^\Delta\Sigma$:}
    \Cref{thm:depth-3-hs} gives a hitting set of size $\poly(n)(\Delta+1)^{2s}$ for the sum of two $\Sigma^s\Pi^\Delta\Sigma$ circuits.

\item {\bf $\Sigma^r\!\wedge^d\!\Sigma^s    \Pi^\Delta\Sigma^m\!\wedge^\delta$:} \Cref{thm:hs_depth4_prod_wedge} gives a hitting set of size $\left(s \cdot \Delta \cdot (m+\delta)\right)^{O(s \cdot \log \log (s \Delta (m+\delta)))}$ for the sum of two $\Sigma^s    \Pi^\Delta\Sigma^m\!\wedge^\delta$
circuits. 

\item {\bf $\Sigma^r\!\wedge^d\!\Sigma^s    \Pi^\Delta\Sigma^m\!\Pi^\delta$:} \Cref{thm:hs_depth4_prod_prod} gives a hitting set of size 
$
     \left(s \cdot \Delta \cdot (m+\delta)\right)^{O(\delta^2 s \cdot \log (s \Delta (m+\delta)))} \, .
    $ for sum of two circuits from the base class.
\end{itemize}
    
    This concludes the proof for the case $d+1\ge (r-1)^2$.
\end{proof}

\begin{remark}
    We also note that by using \cref{thm:bb pit for high powers} with the interpolation set $\cH = \{e_i\}_{i \in [n]} \cup \{\0\}$ for affine functions, one gets a hitting set of size $dn+1$ for $\Sigma^r\wedge^d\Sigma$ circuits whenever $d +1 \geq (r-1)^2$.
\end{remark}


\subsubsection{PIT for low powers}
\label{sec:pit_results_small-depth-low}

As in \Cref{sec:pit-roabp-low} we show that for each of the base classes, powers of circuits from it belong to a somewhat larger class to which PIT is known. 

For this we shall use the well-known multinomial identity.

\begin{fact}[Multinomial Identity] \label{fact:multinomial identity}
    For any $g_1,g_2,\ldots,g_s\in\F[\x]$,
    \[
        \left(\sum_{i=1}^s g_i\right)^d
        =
        \sum_{\substack{e_1+e_2+\cdots+e_s=d\\e_1,e_2,\ldots,e_s\geq 0}}
        \binom{d}{e_1,e_2,\ldots,e_s}
        \prod_{i=1}^s g_i^{e_i}.
    \]
    Note that the number of terms in the sum equals $\binom{d+s-1}{d}$.
\end{fact}


As corollary we obtain the following containment results.

\begin{corollary}\label{cor:closure-small-depth}
    Let $S=r\binom{d+s-1}{d}$ as in \eqref{eq:top-fan-in}. The following containment results hold:
    \begin{align*}
        \sum^r\bigwedge^d\sum^s\prod^\Delta\sum & \subseteq \sum^S\prod^{d\Delta}\sum \\
        \sum^r\bigwedge^d\sum^s\prod^\Delta\sum^m\bigwedge^\delta
        &\subseteq  \sum^S\prod^{d\Delta}\sum^m\bigwedge^\delta\\
        \sum^r\bigwedge^d\sum^s\prod^\Delta\sum^m\prod^\delta &\subseteq \sum^S\prod^{d\Delta}\sum^m\prod^\delta.
    \end{align*}
\end{corollary}

\begin{proof}
    The result follows immediately from \Cref{fact:multinomial identity} since it allows us to replace the first occurrence of $\wedge^d\Sigma^s\cC$ by $\Sigma^{\binom{d+s-1}{d}}\Pi^d\cC$. Thus the upper bound on the top fan-in comes from collapsing the top $\Sigma^r\Sigma^{\binom{d+s-1}{d}}$ layers into one after the transformation. The upper bounds on the fan-in of the first product gate is obtained in a similar manner.
\end{proof}

\begin{lemma} [{Small derivative space for depth-5 powering circuits}] \label{lem:der_space_depth_5_powering}
    Let $P \in \Sigma^r\wedge^d\Sigma^s\wedge^\Delta\Sigma$. Then the dimension of the derivative space of $P$ is at most $r \cdot \binom{s(\Delta + 1) + d - 1}{d}$.
\end{lemma}

\begin{proof}
    For a polynomial $f \in \F[\x]$ denote by $\dim\partial^* f$ the dimension of the partial derivative space of $f$.

    \begin{claim}
        Let $f \in \Sigma^s\wedge^\Delta\Sigma$, then $\dim\partial^* (f^d) \leq \binom{s(\Delta + 1) + d - 1}{d}$.
    \end{claim}

    \begin{proof}
        It is a well-known fact that every $f \in \Sigma^s\wedge^\Delta\Sigma$ admits $\dim\partial^* f \leq s(\Delta + 1)$. Next, every derivative of $f^d$ can be represented as a linear combination of products of exactly $d$ elements from $ \partial^* f$. This proves the claim.
    \end{proof}

    The lemma follows from the above claim and the fact that any derivative of $\Sigma^s_{i=1}f_i$ can be presented as a sum of elements from $\{ \partial^* f_i\}_{i=1}^s$.
\end{proof}


We now conclude the result for the case of small powering degree.

\begin{proof}[Proof of \Cref{thm:main-pit-small-depth} when $d+1<(r-1)^2$]
    The proof follows from combining \cref{lem:der_space_depth_5_powering}, \Cref{cor:closure-small-depth} and the results from \Cref{sec:known-pit-abp} and \Cref{sec:known-pit-small-depth}.

    \begin{itemize}
    
    \item {\bf $\Sigma^r\!\wedge^d\!\Sigma^s\!\wedge^\Delta\!\Sigma$:} Combining \cref{lem:der_space_depth_5_powering} with \Cref{thm:hs_small_pd_space} we obtain a hitting set of size 
    $\poly\big(n,r(s\Delta + d)^d\big) \leq \poly\left(n,(s\Delta + r)^{r^2}\right)$.

    \item {\bf $\Sigma^r\!\wedge^d\!\Sigma^s\Pi^\Delta\Sigma$:}
    \Cref{cor:closure-small-depth} combined with
    \Cref{thm:depth-3-hs} give a hitting set of size $\poly(n)(d\Delta+1)^{S}\leq \poly(n)(d\Delta+1)^{rs^{r^2}}$.

\item {\bf $\Sigma^r\!\wedge^d\!\Sigma^s    \Pi^\Delta\Sigma^m\!\wedge^\delta$:} Combining \Cref{cor:closure-small-depth} with 
\Cref{thm:hs_depth4_prod_wedge} gives a hitting set of size 
\[\left(S \cdot d\Delta \cdot (m+\delta)\right)^{O(S \cdot \log \log (S d\Delta (m+\delta)))}\]

\item {\bf $\Sigma^r\!\wedge^d\!\Sigma^s    \Pi^\Delta\Sigma^m\!\Pi^\delta$:} Here we get from \Cref{thm:hs_depth4_prod_prod} a hitting set of size 
\[   \left(S \cdot d\Delta \cdot (m+\delta)\right)^{O(\delta^2 S \cdot \log (S d\Delta (m+\delta)))} \, .
    \]
\end{itemize}

    This concludes the proof for the case $d+1< (r-1)^2$.
\end{proof}

\section{Bounded-Read Formulas}\label{sec:rof}

 In this section we give our results for sums of powers of bounded-read formulas. Since we have very few PIT results for bounded-read formulas, and as this model is not closed under taking sums, the results that we obtain hold in the high powering degree regime. We start the section by giving some definitions and stating known PIT results. Then we state and prove our results.

 \subsection{Models of bounded read formulas}

\begin{definition} [{Bounded read formula}]
 Arithmetic formula is a binary tree where each leaf is labeled with a variable $x_i \in \{x_{1}, \ldots, x_{n}\}$ and each internal node, called a gate, has an operation $+,\times$.
Additionally, each node $v$ is labeled with some $\alpha_{v}, \beta_{v} \in \F$.

The formula is evaluated recursively as follows: 
Each leaf $\ell$ labeled with $x_i$ computes $q_{\ell} = \alpha_{\ell} x_i + \beta_\ell$.
Each gate $v$ with an operation $\op \in \{+, \times\}$,
computes $\alpha_{v} (q_{(v)_\tl} \hspace{2pt} \op \hspace{3pt} q_{(v)_\tr}) + \beta_{v}$, where $q_{(v)_\tl}$ and $q_{(v)_\tr}$ are the polynomials computed by its left and right children respectively.

A read-$k$ formula is an arithmetic formula in which each variable $x_{i} \in \x$ labels at most $k$ leaves, i.e., $\fread_{x_i}(F)\leq k$. A polynomial is a read-$k$ polynomial if it can be computed by a read-$k$ formula. We denote by $\rkf$ the class of read-$k$ formulas. For read-once formulas, we use the standard notation $\rof$. A polynomial computed by an $\rkf$ is called a read-$k$ polynomial, or an $\mathsf{RkP}$ (similarly a read-once polynomial, or an $\rop$).

     The class $\Sigma^m \rkf$ consists of all the polynomials that can be decomposed into a sum of $m$ $\mathsf{RkP}$ polynomials. 

     The class $\Pi^m \rkf$ consists of all polynomials that can be expressed as a product of $m$ $\mathsf{RkP}$ polynomials. 
     
     We similarly define the classes $\Sigma^s \Pi^m \rkf$ and $\Sigma^s \wedge^d \rkf$.
\defsymbol

\end{definition}

For $x_i \in \x$ and $\alpha \in \F$, we write $F|_{x_i = \alpha}$ to denote the formula obtained by substituting $x_i = \alpha$ in~$F$. This extends naturally to any subset $I \subseteq [n]$, as in the polynomial case.

\begin{definition}[{Preprocessed Arithmetic Read-Once Formula, \cite[Definition 3.15]{Shpilka2015-fd}}] \label{def:prof}
    A \emph{preprocessing} is a transformation
    $
        T(\vec{x}) : \F^n \to \F^n
    $
    of the form
    $
        T(\vec{x})
        \coloneqq
        \bigl(
            T_1(x_1),
            T_2(x_2),
            \ldots,
            T_n(x_n)
        \bigr),
    $
    where each $T_i$ is a non-constant univariate polynomial. 
    
    A \emph{preprocessed arithmetic read-once formula} (PROF for short) is the result of composing a $\rof$ $F$ with a preprocessing $T(\vec{x})
        \coloneqq
        \bigl(
            T_1(x_1),
            T_2(x_2),
            \ldots,
            T_n(x_n)
        \bigr)$, that is, it is $F\big(T(\x)\big)$. We denote the class of $\prof$s of individual degree at most $\delta$, that is $\forall i\in [n]:\deg(T_i) \leq \delta$, by $\prof_{\delta}$. 
        A polynomial computed by a $\prof$ is called a $\prop$.
\end{definition}


\subsection{Known PIT and reconstruction results} \label{ssec:known_pit_and_recon_bounded_read_formulas}

Here we state the results for bounded read formulas. We state all of the results we'll use in this subsection.

\begin{lemma} [{\cite[Fact 3.9]{bisht_et_al}}] \label{lem:hs_s2_rof}
    There exists an explicit hitting set of size  $(n+1)^6$ for $\Sigma^{2} \rof$.
\end{lemma}

\begin{theorem} [{\cite[Theorem 2]{MinahanVolkovich18}}] \label{lem:hs_s2_sum_prof}
    There exists an explicit hitting set of size $(n \cdot \delta +1)^{2\cdot\left(1+6s\right)}$ for $\Sigma^{2s} \prof$ of individual degrees at most $\delta$.
\end{theorem}

\begin{lemma} [{\cite[Implicit in the proof of Theorem 1.1]{r4fPIT}}] \label{lem:hs_s2_rtf}
    There exists an explicit hitting set of size  $\left(2n+1\right)^{2 \cdot \left(\log n + c_{\ref{lem:hs_s2_rtf}} \right)}$ where $c_{\ref{lem:hs_s2_rtf}}$ is some absolute constant for $\Sigma^2 \rtf$.
\end{lemma}

For this type of algebraic circuits, there is almost a complete match between the PIT results for the base class and the results for the orbit classes, as was shown in \cite{r4fPIT}.

\begin{theorem} [{\cite[Theorem 1.5]{r4fPIT}}] \label{thm:hs_orbit_sum_rof}
    There exists an explicit hitting set of size  $(n+1)^{2 \cdot (\log n + 6s)}$ for $(\Sigma^{2s} \rof)^{\gla}$. If $s = 1$ then the size bound is improved to $(n+1)^{2 \cdot (\log n + 3)}$.
\end{theorem}

The next theorem can be easily improved using the fact that the polynomial we're considering is more structured than a general $\mathsf{mlR(2k)F}$.

\begin{theorem} [{\cite[Theorem 1.1]{multilinear_formulas_pit} and \cite[Theorem 1.6]{r4fPIT}}] \label{lem:hs_orbit_s2_mlrkf}
    There exists an explicit hitting set of size $(n + 1)^{2\cdot\left(r_{2k}+2k \log n\right)}$ where $r_{2k} = (2k)^{O(k)}$ for $(\Sigma^{2} \mlrkf)^{\gla}$. 
\end{theorem}

\begin{lemma} [{\cite[Implicit in the proof of Theorem 1.9]{r4fPIT}}] \label{lem:hs_orbit_s2_rtf}
    There exists an explicit hitting set of size  $\left(2n+1\right)^{2 \cdot (3\log n + c_{\ref{lem:hs_orbit_s2_rtf}})}$ where $c_{\ref{lem:hs_orbit_s2_rtf}}$ is some absolute constant, for $(\Sigma^2 \rtf)^{\gla}$.
\end{lemma}

Lastly, for the class of preprocessed read-once formulas, we have a non-adaptive, deterministic reconstruction algorithm. We use it to reconstruct sums of powers of such formulas.

\begin{theorem} [{\cite[Theorem 1.2]{sv14_rec_rof}} and \cite{MinahanVolkovich18}] \label{thm:rof recon set} There exists a reconstruction set, of size $2 n^5\delta^5$ for $\prof$ of individual degrees bounded by $\delta$. The running time of the algorithm is polynomial in $n$ and $\delta$.
\end{theorem}

To use the individual degree bound, we need to apply \Cref{thm:suff_cond_for_high_power_recon_ext}. For that we give the following theorem.

\begin{theorem} [{\cite[Theorem 1]{bisht_et_al}}] \label{thm:hs_alternants_rofs}
    There exists an explicit hitting set of size $\left(2n^2+1\right)^{8}$ for $\Sigma^2\Pi^2 \rof$.
\end{theorem}


\subsection{PIT and reconstruction for sum of powers of bounded-read formulas} \label{sec:pit_and_recon_bounded_read_formulas}

All the results in this subsection follow from applying our meta-algorithms to the interpolation and reconstruction sets given in \cref{ssec:known_pit_and_recon_bounded_read_formulas}. First, we apply \autoref{thm:bb pit for high powers} to various models, the results are summarized in the following table. They hold whenever $d+1\geq (r-1)^2$ and over fields of sufficiently large characteristic\footnote{As explained in previous sections.}.

\begin{theorem}\label{thm:hs-powers-rof}
    We can explicitly construct hitting sets given in \Cref{tab:high_pow_pit_rofs} when $d+1\ge (r-1)^2$.
\begin{table}[H]
    \centering
    \begin{threeparttable}
    \renewcommand{\arraystretch}{1.4}

    \begin{tabular}{lll}
        \toprule
        \textbf{Model} & \textbf{Hitting-set size} & \textbf{Proof} \\
        \midrule

        $\Sigma^r\wedge^d\rof$
        & $d\cdot(n+1)^7$ & \Cref{lem:hs_s2_rof} \\

        $\Sigma^r\wedge^d\Sigma^s\prof_{\delta}$
        & $d\cdot(n\delta+1)^{2(1+6s)+1}$ &         \Cref{lem:hs_s2_sum_prof} \\

         $\Sigma^r\wedge^d\rtf$
        & $d\cdot(2n+1)^{2(\log n+c_{\ref{lem:hs_s2_rtf}}+1)}$ & \Cref{lem:hs_s2_rtf}\\[3.5ex]

        $(\Sigma^r\wedge^d\rof)^{\gla}$
        & $d\cdot(n+1)^{2\log n+7}$ &  \Cref{thm:hs_orbit_sum_rof} \\

         $(\Sigma^r\wedge^d\Sigma^s\rof)^{\gla}$
        & $d\cdot(n+1)^{2(\log n+6s)+1}$ & \Cref{thm:hs_orbit_sum_rof} \\

         $(\Sigma^r\wedge^d\rtf)^{\gla}$
        & $d\cdot(2n+1)^{2(3\log n+c_{\ref{lem:hs_orbit_s2_rtf}})+1}$ & \Cref{lem:hs_orbit_s2_rtf}\\

         $(\Sigma^r\wedge^d\mlrkf)^{\gla}$
        & $d\cdot(n+1)^{2(2k\log n+r_{2k})+1}$ & \Cref{lem:hs_orbit_s2_mlrkf}\\
        \bottomrule
    \end{tabular}    \caption{Black-Box PIT for Sums of powers of bounded-read formulas and their orbits, when $d+1\ge (r-1)^2$}    \label{tab:high_pow_pit_rofs}

    \end{threeparttable}
\end{table}
\end{theorem}

\begin{proof}
    As before the proof follows from combining known results with \Cref{thm:bb pit for high powers}. For each model the relevant PIT result is given in the third column of \Cref{tab:high_pow_pit_rofs}.
\end{proof}

Next, we prove a reconstruction result when the inner polynomials are $\prop$.

\begin{theorem}[Reconstruction for sums of powers of $\prop$s] \label{thm:sum_pow_prop_recon}
   Let $n,d,r,\delta,\Delta\in \N$, assume $d > (r+1)^4\Delta$ and let $\F$ be a field of characteristic $p$ such that $p=0$ or $p > 2\Delta r d$. Then, there exists a deterministic reconstruction algorithm for $\Sigma^{r}\!\wedge^{d}\!\prop$ whose inner preprocessed read-once polynomials have total degree at most $\Delta$ and individual degree at most $\delta$, that runs in time \[
   \poly(n,r,d,\delta,p).
   \]
\end{theorem}

\begin{proof}
    By \cref{thm:rof recon set} there is a reconstruction set for $\prop$ of size \(2 n^5\delta^5\) that corresponds to a reconstruction algorithm that runs in time $\poly(n,\delta)$. Using this set and algorithm in \cref{thm:recon_sets_imply_high_power_recon} gives us the claimed result.
\end{proof}

When the inner polynomials are $\rop$s, we have a hitting set for $\Sigma^2\Pi^2\rop$. This allows us to use \cref{alg:meta_recon_ext} and get a relaxed bound on $d$ (for $n > r$).

\begin{theorem} [Reconstruction for sums of powers of $\rop$s] \label{thm:sum_pow_read_once_recon}
   Let $n,d,r\in \N$, assume $d > (r+1)^4r$ and let $\F$ be a field of characteristic $p$ such that $p=0$ or $p > 2 n r d$. Then, there exists a deterministic reconstruction algorithm for $\Sigma^{r}\!\wedge^{d}\!\rop$
   that runs in time \[
   \poly(n^r,r,d,p).
   \]
\end{theorem}

\begin{proof}
    By \cref{thm:rof recon set} there is a reconstruction set for $\rop$s of size \(2 n^5\) that corresponds to a reconstruction algorithm that runs in time $\poly(n)$. By \cref{thm:hs_alternants_rofs} there is a hitting set for $\Sigma^2\Pi^2\rop$ of size $\left(2n^2+1\right)^{8}$. Using these two sets and the reconstruction algorithm's running time in \cref{thm:small_individual_degree_reconstruction} gives us the claimed result.
\end{proof}

\bibliographystyle{alpha}
\bibliography{Ref}

\appendix
\section{Linear Independence and Wronskian}\label{app:wronskian}

We prove \autoref{thm:wronskian-lin-ind}. We recall its statement.

\thmwronskilinind*

\begin{proof}
The 'only if' direction is immediate.  
For the converse, assume that $g_1, \dots, g_n$ are linearly independent.  
Since the determinant is invariant under elementary column operations, we may assume, without loss of generality, that $g_1, \dots, g_n$ have distinct degrees.

Let their degrees be $d_1 > d_2 > \cdots > d_n \ge 0$, and write  
$g_j = c_j x^{d_j} + \text{(lower terms)}$ with $c_j \ne 0$.  
By multilinearity of the determinant, the leading term of the Wronskian equals the Wronskian of the leading monomials, provided its coefficient is nonzero:
\[
W(\mathrm{LM}(g_1), \dots, \mathrm{LM}(g_n))
  = W(c_1 x^{d_1}, \dots, c_n x^{d_n}).
\]

The determinant of the matrix of derivatives of $c_j x^{d_j}$ is
\[
\det\!\big( c_j (d_j)_{i}\, x^{d_j - i} \big)_{\substack{i\in [[n-1]] \\ j\in [n]}},
\]
where $(d_j)_i = d_j(d_j-1)\cdots(d_j - i + 1)$ is the falling factorial (defined to be $1$ for $i=0$).
Factoring out $c_j$ from each column and collecting powers of $x$ yields
\[
W(c_1 x^{d_1}, \dots, c_n x^{d_n})
  = \left(\prod_{j=1}^n c_j\right)
    \left(\prod_{1 \le i < j \le n} (d_j - d_i)\right)
    x^{\sum_j d_j - \binom{n}{2}}.
\]

Because $p=0$ or $p > d$, all integers $d_j$ and differences $(d_j - d_i)$ are nonzero in $\F$.  
Since each $c_j \ne 0$, the product above is nonzero, so $W(g_1, \dots, g_n)$ is not identically zero.
\end{proof}

\end{document}